\documentclass[showpacs, amsmath, amssymb, twocolumn, aps, prl, superscriptaddress,notitlepage,floatfix]{revtex4-2} 

\usepackage{lipsum,mwe}
\usepackage{soul}
\usepackage{floatrow}
\usepackage{float}   
\usepackage{braket}
\usepackage{graphicx}
\usepackage{dcolumn} 
\usepackage{bm}      
\usepackage{comment} 
\usepackage{xcolor} 
\usepackage{color}
\usepackage[colorlinks=true, linkcolor=blue]{hyperref} 
\usepackage{physics}

\newcommand{\diag}[0]{\text{diag}}
\newcommand{\IMR}[0]{\mathcal{M}} 
\newcommand{\M}[0]{\mathbb{M}}    
\newcommand{\R}[0]{\mathcal{R}}   
\newcommand{\T}[0]{\mathrm{T}}    

\newcommand{\comments}[1]{}

\begin{document}

\title{Invested and Potential Magic Resources in Measurement-Based Quantum Computation}

\author{Gong-Chu Li}
\author{Lei Chen}
\author{Si-Qi Zhang}
\author{Xu-Song Hong}
\author{Huaqing Xu}
\author{Yuancheng Liu}
\affiliation{CAS Key Laboratory of Quantum Information, University of Science and Technology of China, Hefei, Anhui 230026, China.}
\affiliation{Anhui Province Key Laboratory of Quantum Network, Hefei, Anhui 230026, China.}
\affiliation{CAS Center For Excellence in Quantum Information and Quantum Physics, University of Science and Technology of China, Hefei, Anhui 230026, China.}
\affiliation{Hefei National Laboratory, Hefei 230088, China}
\author{You Zhou}
\email{you\_zhou@fudan.edu.cn}
\affiliation{Key Laboratory for Information Science of Electromagnetic Waves (Ministry of Education), Fudan University, Shanghai 200433, China}
\author{Geng Chen}
\email{chengeng@ustc.edu.cn}
\author{Chuan-Feng Li}
\email{cfli@ustc.edu.cn}
\author{Guang-Can Guo}
\affiliation{CAS Key Laboratory of Quantum Information, University of Science and Technology of China, Hefei, Anhui 230026, China.}
\affiliation{Anhui Province Key Laboratory of Quantum Network, Hefei, Anhui 230026, China.}
\affiliation{CAS Center For Excellence in Quantum Information and Quantum Physics, University of Science and Technology of China, Hefei, Anhui 230026, China.}
\affiliation{Hefei National Laboratory, Hefei 230088, China}
\author{Alioscia Hamma}
\email{alioscia.hamma@unina.it}
\affiliation{Dipartimento di Fisica ‘Ettore Pancini’, Università degli Studi di Napoli Federico II, Via Cintia 80126, Napoli, Italy}
\affiliation{INFN, Sezione di Napoli, Italy}
\affiliation{Scuola Superiore Meridionale, Largo S. Marcellino 10, 80138 Napoli, Italy}

\begin{abstract}
    Magic states and magic gates are crucial for achieving universal quantum computation, but important questions about how magic resources should be implemented to attain maximal quantum advantage have remained unexplored, especially in the context of measurement-based quantum computation (MQC). 
    This work bridges the gap between MQC and the resource theory of magic by introducing the key concepts of ``\textit{invested}'' and ``\textit{potential}" magic resources. The former quantifies the magic cost associated with MQC, serving as both a resource witness and a feasible upper bound for the practical realization, and is gate-order independent; The latter represents the maximal achievable magic resource in a given graph structure defining MQC. 
    We utilize both concepts to analyze the quantum Fourier transform (QFT) and provide a fresh perspective on the universality of MQC, highlighting the crucial role of non-Pauli measurements in injecting magic. In particular, we theoretically prove that high-dimensional graphs can generate an exponential advantage of MQC compared to classical computing. 
    We demonstrate experimentally our theoretical findings in a high-fidelity four-photon setup, surpassing conventional magic state injection (MSI) methods in both qubit efficiency and resource utilization.
    Our findings pave the way for future research exploring magic resource optimization and novel distillation schemes within the MQC framework, advancing fault-tolerant universal quantum computation.
    
    Published in \textit{Physical Review Letters} 135, 160203 (2025).
    
    DOI: \href{https://doi.org/10.1103/4yyv-hggz}{10.1103/4yyv-hggz}
\end{abstract}

\maketitle

\textit{Introduction ---}
The ultimate goal of quantum computation is to realize large-scale, fault-tolerant, universal quantum computation \cite{campbell2017roads}.
To this end, quantum error correction codes are used to encode logical qubits and introduce sets of fault-tolerant universal gates \cite{postler2022demonstration, google2023suppressing, gupta2024encoding}.
Within such a framework, T states/gates, often referred to as magic states/gates \cite{bravyi2005universal}, have been recognized as crucial resources, 
which can be quantified by the corresponding resource theory \cite{veitch2014resource, leone2022stabilizer, bittel}. 

Magic resources connect closely with magic distillation \cite{bravyi2005universal, gupta2024encoding, souza2011experimental}, magic synthesis \cite{knill2005quantum,howard2017application, regula2021fundamental} tasks, and classical simulation complexity \cite{aaronson, bravyi2016improved, bravyi2016trading}. 
At the core of these discussions is the celebrated Gottesman-Knill theorem \cite{aaronson, bravyi2016improved, bravyi2016trading}, which moves beyond the first layer of computation, i.e., injecting $k$ magic states to Clifford circuits exponentially increases the complexity of classical simulation as $2^{O(k)}\text{poly}(n)$ of an $n$-qubit system. From such a perspective, universal quantum computation is characterized by an exponential increase in resource requirements for classical computation, quantified by the magic resource present. 

A reliable measure of magic resource should satisfy three key properties: faithfulness (zero for Clifford gates and stabilizer states), non-increasingness under free operations (i.e., Clifford operations), and (sub-)additivity (for independent components). These ensure the consistency and practicality of the measure in real quantum computations. 
While measures like mana \cite{veitch2014resource}, robustness \cite{howard2017application}, thauma \cite{wang2020efficiently}, and nullity \cite{beverland2020lower} satisfy faithfulness and Clifford invariance, they lack the additivity property. 
Recent approaches based on stabilizer norm \cite{howard2017application}, stabilizer Rényi entropy\cite{leone2022stabilizer, bittel}, and Gottesman-Kitaev-Preskill codes \cite{hahn2022quantifying} have successfully addressed this limitation, providing measures that fulfill all three desired criteria. 

Recently, the issue of the magic resource cost of non-Pauli measurements has risen to attention \cite{oliviero2021transitions, niroula2023phase}. However, a significant gap remains in our understanding of their fundamental role within the framework of measurement-based quantum computation (MQC) \cite{briegel2009measurement}. MQC, a prominent paradigm in quantum computing, relies on the preparation of cluster states, which are stabilizer states associated with a graph and thus inherently lack magic resources. Universality in MQC is achieved through consecutive non-Pauli single-qubit measurements on these graph stabilizer states. This highlights a crucial insight that non-Pauli measurements are responsible for generating magic resources, whereas the source of MQC's universality has traditionally been attributed to the rich entanglement structure of graph states \cite{van2007fundamentals, van2006universal}. 

While magic resource distillation/synthesis is a well-established approach for achieving universal quantum computation, the role of non-Pauli measurements has been less explored, and their critical role in achieving quantum advantage is not well understood. 
The current work introduces two new measures for magic resources—\textit{invested} and \textit{potential}—that address the gap in current quantum resource theories, shedding light on how measurement-based processes contribute to achieving quantum advantage, particularly in the context of MQC, where non-Pauli measurements play a crucial role.

\begin{figure}[t]
    \centering
    \includegraphics[width=0.7\textwidth]{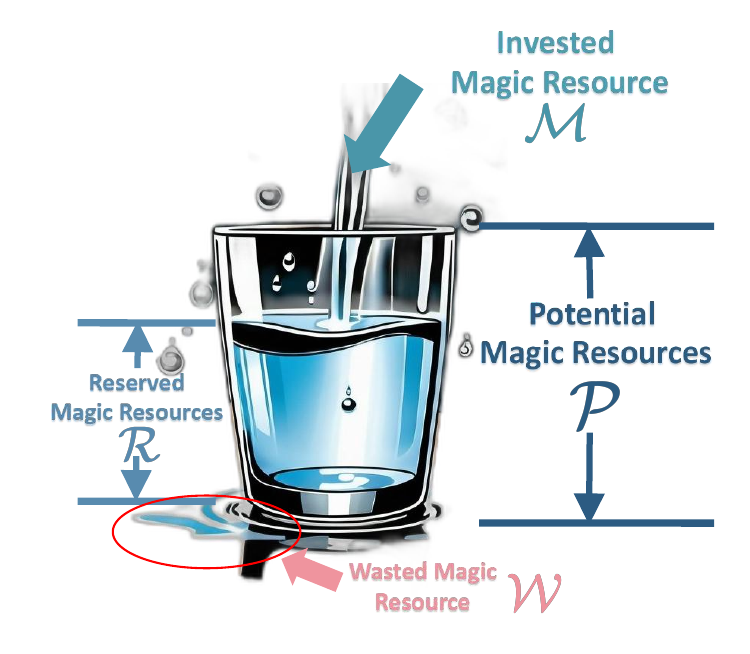}
    \caption{
    \textit{Cartoon showing the relationship between invested magic resources ($\IMR$), potential magic resources ($\mathcal{P}$), and reserved magic resources ($\R$)} using a water-pouring analogy.
    }
    \label{fig:Resources_Relationship}
\end{figure}

This work consists of two parts: Theory and Experiment. In the first part, we establish the theory of {\em invested magic resources} ($\mathcal{M}$) and {\em potential magic resources} ($\mathcal{P}$) as sketched in Fig.~\ref{fig:Resources_Relationship}, and will be addressed later. 
In the second part, we experimentally demonstrate the effectiveness of injecting magic resources through MQC in a four-photon experiment, involving a single-qubit rotation and Quantum Fourier Transform (QFT) tasks. 
We show that non-Pauli measurements, combined with the entanglement structure of graph states, are essential for the universality of MQC.
Our framework provides a novel way of quantifying the magic resources needed for these measurements, thereby advancing our understanding of MQC's computational power.

\ 

\textit{Review of MQC --- }
The setup of MQC starts with a graph state $|G\rangle$, which is an entangled state of n qubits defined by a graph $G = (V, E)$ with n vertices at $|+\rangle=(|0\rangle+|1\rangle)/\sqrt{2}$ and edges representing controlled-Z (CZ) gates. 
Graph states are a special type of stabilizer state, meaning they possess a set of stabilizing operators that leave the state unchanged. 
MQC relies on preparing a suitable graph state and performing a series of adaptive single-qubit measurements in the X-Y plane of the Bloch sphere. The final state of the computation is locally Pauli equivalent to the desired target state. 
Based on the stochastic outcomes of the measurements and additional Pauli corrections, MQC becomes a universal quantum computation platform. 
This process can be summarized as the CME pattern \cite{danos2005parsimonious, danos2007measurement}, consisting of Entanglement (preparing the graph state), Measurement (adaptive single-qubit measurements), and Correction (Pauli corrections).

CME relies on the fact that any unitary operation can be decomposed into a sequence of CZ gates and J gates, $J(\theta)=\frac{1}{\sqrt{2}}\begin{pmatrix}1& e^{i\theta}\\ 1&-e^{i\theta}\end{pmatrix}$. This decomposition process is known as J-decomposition. Each J gate in the decomposition corresponds to a small CME structure. For example, $J(\theta)$ on the qubit 2 can be implemented as $J(\theta) := X_2^{s_1}M_1^{-\theta}E_{12}$, where $E_{12}$ represents a CZ gate between qubits 1 and 2, $M_1^{-\theta}$ denotes a measurement on the first qubit in the basis $(|0\rangle\pm e^{-i\theta}|1\rangle)/\sqrt{2}$, $s_1$ is the signal of measurement outcome, and $X_2^{s_1}$ is a Pauli X correction applied to the second qubit based on $s_1$. This equation illustrates how the action of $J(\theta)$ on a state $|\psi\rangle$ can be achieved using an auxiliary qubit, i.e., the second qubit, initialized in the $|+\rangle$ state and an appropriate MQC procedure. For more complex circuits, by replacing all the J gates with a small CME pattern, we could move all the entanglement to the beginning of the circuit and all the corrections to the final part, given the measurement calculus \cite{danos2007measurement}. We obtain a practical MQC plan where qubits are connected in the graph representing CZ gates, and measurements are performed sequentially, adapting the measurement basis based on previous outcomes. 
Overall, the CME pattern could be summarized as:
\begin{equation}
    U := [C][M][E].
    \label{eq: CME}
\end{equation}

\textit{Invested Magic Resources and Upper Bound for Quantum Fourier Transform (QFT) --- }
From a resource-theoretic perspective, introducing auxiliary qubits in the $|+\rangle$ state, entangling them with CZ gates, and applying final Pauli corrections do not require any magic resources. As such, during the whole MQC procedure, the key source of magic resource investment lies in the non-Pauli measurements.  
These measurements lead to the generation of states with higher magic content, which is ultimately reflected in the final output state of the computation.

\begin{figure*}[htbp]
    \centering
    \includegraphics[width=0.9\textwidth]{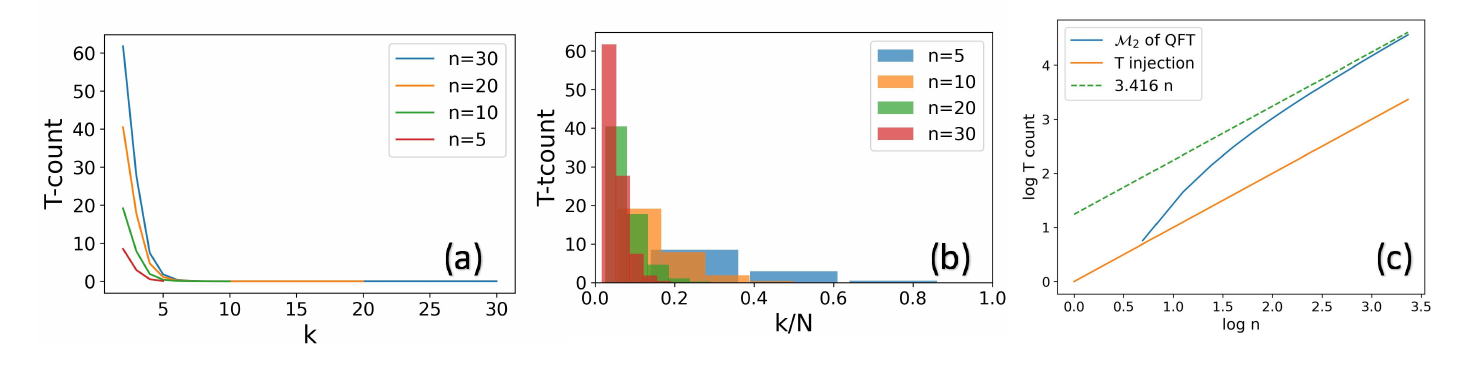}
    \caption{
    \textit{Demonstration of invested magic resources in QFT.}  
    (a) and (b) show the distribution of invested magic resources across different frequencies $k$, with a concentration of resources in the lower-frequency range. (c) illustrates the scaling of invested magic resources with the number of qubits, with the 'T injection' curve representing MSI with T-counts equal to $n$. This provides a direct comparison between the two methods for generating magic states.}
    \label{fig:qft}
\end{figure*}

The randomness in the outcomes $\{s_i\}$ is crucial for MQC; however, it has no effect from the point of view of magic resources. 
The final projected state of each adaptive measurements is $\frac{1}{\sqrt{2}}(|0\rangle +(-1)^{s_{o}} e^{is_{i}\theta}|1\rangle)$, depending on the values of previous measurement outcomes ($s_{i}$) and this measurement outcome ($s_{o}$).
Indeed, the Pauli spectrum $\Xi_P := d^{-1}\langle \psi|P|\psi\rangle^2$ of these states remains the same regardless of the stochastic measurement outcomes and uses the stabilizer $\alpha$-Renyi entropy (SRE) as magic measure of the output state \cite{leone2022stabilizer}, $\M_\alpha(|\psi\rangle) := (1-\alpha)^{-1}\log  {\sum_P} \Xi_P^{\alpha} -\log d$. Therefore the magic resources required for a single qubit measurement $\M_{\alpha}(\theta ) = \M_{\alpha}((|0\rangle+e^{i\theta}|1\rangle)/\sqrt{2})$, are independent of the signal values. Specifically, for the 2-R\'{e}nyi entropy, we have $\M_2(\theta) = -\log (\frac{1}{2}\cos^4\theta+\frac{1}{2}\sin^4\theta+\frac{1}{2})$. By additivity of the SRE, a $n-$qubit $\otimes J(\theta)$ requires $n\M_2(\theta)$ magic resources. 

The single-qubit state containing maximum magic resources is Kitaev's T-state \cite{bravyi2005universal}, $|T\rangle = \cos\theta_m|0\rangle+\sin\theta_m e^{i\pi/4}|1\rangle$ with $\cos2\theta_m = 1/\sqrt{3}$.  
In the later discussion, we use the one T-state magic resource $1T=\mathbb{M}_2(|T\rangle)$ as a unit to scale magic resources, denoted as \textit{T-count}. 

Since each measurement setting in MQC corresponds to a J gate with angle $\theta$ in the J-decomposition of the target unitary $U$ in Eq.\eqref{eq: CME}, 
we can define the total {\em invested magic resources} for implementing $U$ as:
\begin{equation}\label{eq: invested}
     {\IMR}_{\alpha}(U) := \sum_{J(\theta)} \M_{\alpha}(\theta).
\end{equation}
For example, the arbitrary rotation on the single qubits, $U=J(0)J(\alpha)J(\beta)J(\gamma)$. The corresponding magic resources on MQC is $\IMR_{\alpha}(U) = \M_{\alpha}(\alpha)+\M_{\alpha}(\beta)+\M_{\alpha}(\gamma)$. 
From the Eq.~\eqref{eq: invested}, the magic measure for unitary operations is \textit{gate-order-independent}, thanks to the measurement calculus rule. 
Gate-order independence refers to the fact that the total invested magic resources remain unchanged regardless of the order in which quantum gates are applied. This simplifies the framework for quantifying computational costs in quantum circuits.

This invested magic resource $\IMR_{\alpha}$ possesses the essential properties of a \textit{good} magic resource measure. It exhibits faithfulness, meaning $\IMR_{\alpha}(C)=0$ if and only if $C$ is a Clifford gate. 
Additionally, $\IMR_{\alpha}$ is invariant under Clifford gates, meaning $\IMR_{\alpha}(CU)=\IMR_{\alpha}(U)$ for any Clifford gate $C$, as applying $C$ before or after $U$ does not change the magic resource cost. Finally, $\IMR_{\alpha}$ satisfies additivity, meaning $\IMR_{\alpha}(U\otimes V) = \IMR_{\alpha}(U) + \IMR_{\alpha}(V)$. The proof is left in Sec.~III in \cite{supp}. 

The invested magic resource measure $\IMR_{\alpha}$ offers a unique perspective compared to previous magic resource quantifiers. 
First, $\IMR_{\alpha}$ is naturally defined at the process level. 
Moreover, it is directly derived from the magic content of single-qubit states.
Most importantly, while previous measures typically provide lower bounds on the magic resources required for a given task, $\IMR_{\alpha}$ serves to quantify a sufficient amount of resources for the task and also as a {\em witness} of employed magic resources. This means that if a quantum computation has an invested magic resource cost of $\IMR_{\alpha}$, there exists a concrete MQC implementation using that amount of magic resources.

The invested magic resource framework provides a useful tool for analyzing the complexity of quantum Fourier transformation (QFT), offering an upper bound on the resources required for its implementation. QFT is a circuit which maps position basis $|x\rangle$ into frequency basis $|k\rangle$ with many Hadamard gates and controlled-rotation gates ($ CR_k $) with $ R_k = \text{diag}[1, \exp(i2\pi/2^k)]$. 
Given the J-decomposition of the controlled-rotation gates, we have: 
\begin{equation}
    \begin{aligned}
        \IMR_2(\text{QFT}) &= \sum_{k=2}^n (n+k-1) \IMR_2(CR_k) \\
        &\approx 3.4619n - 5.3388.
    \end{aligned}
\end{equation}
as demonstrated in Fig.~\ref{fig:qft} (c). See Sec.~IV of \cite{supp} for the derivation.

Fig.~\ref{fig:qft} (a) and (b) visualize the contribution of different frequencies to the invested magic resources of QFT. The plots reveal that most of the magic resources are concentrated in the low-frequency range, with a sharp decline in contribution as the frequency increases. This behavior remains consistent across different qubit numbers. 
Previous analyses of classical simulation algorithms for QFT circuits have suggested a complexity of $ O(n\log n) $ T gates \cite{nam2020approximate}. However, the invested magic resource framework provides an upper bound on the required resources, indicating that $ O(n) $ T gates are sufficient. 

\textit{Potential Magic Resources---}
While designing specific graphs for targeted tasks is one approach to MQC, a more general strategy involves preparing versatile graph states capable of performing arbitrary quantum computations. These versatile graphs are often referred to as universal resources for MQC \cite{van2007fundamentals, van2006universal, briegel2009measurement}. However, it is important to recognize that not all invested magic resources effectively contribute to the magic content of the final state. This limitation arises from the inherent maximum potential of magic resources that a particular graph type can hold.

We define reserved magic resources $\R$ and potential magic resources $\mathcal{P}$ in the following way. Given a preparation obtained by a set of single-qubit measurements $[M]$ (obeying MQC rules) on a graph state $|G\rangle$, its reserved and potential magic resource are defined as
\begin{equation}\label{reserved&potential}
\begin{aligned}
    \R ([M]|G\rangle) := \M_2([M]|G\rangle),\ 
    \mathcal{P}(|G\rangle) := \max_{[M]} \mathcal{R}([M]|G\rangle).
\end{aligned}
\end{equation}
The above definition is motivated by the fact that the entanglement structure is fundamental to determine how much magic can be hosted in a certain state, compared to the simple additive structure in factorized states \cite{iannotti2025entanglementstabilizerentropiesrandom}. 
In principle, $\mathcal{P}$ could be calculated by traversing all measurement bases and orders. 

Additionally, MQC allows for a general quantum input. Given an input state $|\psi\rangle \in \mathcal{H}_{in}$ the graph state can be seen as the Choi isomorphic state of the channel  $G: \mathcal{H}_{in}\rightarrow \mathcal{H}_{out}$ as $|G\rangle \in \mathcal{H}_{in}\otimes \mathcal{H}_{out}$.
Then, we have $[E]|\psi\rangle = |G\rangle * |\psi\rangle \in \mathcal{H}_{out}$, with $*$ as link product \cite{chiribella2008quantum, chiribella2009theoretical} between the graph and the input state and $|G\rangle$ as a graph with some links vacant. 
The maximum increment of potential magic resources brought by the entanglement structure with the input $|\psi\rangle$ is thus $\mathcal{P}(G;|\psi\rangle) := \max_{[M]} \M_2([M]|G\rangle*|\psi\rangle) - \M_2(|\psi\rangle)$.

Therefore, if MQC can produce a state $|\phi\rangle$  {or its Clifford equivalents from the input state $|\psi\rangle$ with a graph $|G\rangle$, denoted as $|G\rangle*|\psi\rangle \ge_{\text{MQC}}|\phi\rangle$}, the potential magic resources must satisfy  {$\mathcal{P}(|G\rangle;|\psi\rangle)\ge \M_2(|\phi\rangle)-\M_2(|\psi\rangle)$}. Consequently, a family of graph states can serve as universal resources only if their potential magic resources scale as $n^\alpha$, where $\alpha > 0$ and $n$ is the number of qubits.

A key application of this concept lies in demonstrating the limitations of linear graphs and GHZ states as universal resources, regardless of their size. From the perspective of potential magic resources, we can show that:
\begin{equation}
    \mathcal{P}(|\text{Linear}\rangle) = 1T,\quad \mathcal{P}(|\text{GHZ}\rangle)=1T.
\end{equation}
For a definition of the linear graph states $|\text{Linear}\rangle$ and a proof see Sec.~V in \cite{supp}.

The fact that linear and GHZ states have limited potential magic resources implies that the computational power they offer through magic resources is constant and does not scale with the system size. This finding provides a novel perspective on the concept of universality in MQC, highlighting the importance of potential magic resources as a measure of the capability of different graph structures.

Potential magic resources constrain the computational power of MQC for different graph types. However, detailed calculations can be challenging. Therefore, we provide a general bound for the potential magic resources by considering d-dimensional graphs as   
\begin{equation}
    \mathcal{P}(|G\rangle)_{\min} = O(n^{(d-1)/d})
\end{equation}
For d=1, $P(|G\rangle)_{\min}\sim const$, we observe a constant gap between MQC and classical simulation, as seen in the linear graph. For 2-D graph, $P(|G\rangle)_{\min}\sim \sqrt{n}$, we get a super-linear gap, for $\sqrt{n}\ge \log n$ when $n$ is large. Finally, as a high-dimensional graph, we have $P(|G\rangle)_{\min}\rightarrow n$, which leads to an exponential gap between MQC and classical computation. 
More see Sec.~VI in \cite{supp}. 

The relationship between the invested magic resources ($\IMR$), potential magic resources ($\mathcal{P}$), and reserved magic resources ($\R$) is summarized in Fig.~\ref{fig:Resources_Relationship} using the analogy of pouring water into a glass. 
While invested magic resources represent the total resources required during the computation process, potential magic resources reflect the maximum achievable magic resources that a given graph structure can support. The gap between these two measures provides insights into the efficiency of the quantum computation: any excess water that spills over signifies wasted magic resources, which could be defined as $\mathcal{W} = \mathcal{M}-\mathcal{P}$. Although measures based on the SRE may not be as effective for mixed state cases as they are for pure states \cite{supp}, in practical high-purity experimental settings, SRE-based quantification still provides a reliable estimation of magic resources.

\textit{Experimental Results --- }
The invested magic resources, $\IMR_2$, enable us to quantify the magic resources injected into the network in MQC. The magic resources of the remaining state after each step of measurement, $\R$, represent the reserved magic resources. By introducing these concepts, we can indicate the invested and reserved magic resources step by step in the MQC process.

We focus on two typical processes: single-qubit rotation and QFT. We demonstrate how the magic responds to single-qubit measurements in 1-D and 2-D graph states. The 1D graph state is a linear state, while the 2D graph state is a BOX state. Both originate from the 4-qubits cluster state, $|\text{cluster}\rangle = (|0000\rangle+|0011\rangle+|1100\rangle-|1111\rangle)/2$ \cite{walther2005experimental}. For the specific setup, see Sec.~VII in  \cite{supp}.  
The experimental estimation of reserved magic resources derives from the few-shot randomized measurements \cite{singlezhou,my_p3ppt_paper}, see Sec.~VIII in \cite{supp}.

Our experiment utilizes a linear graph and MQC to generate the $|T\rangle$ state.  
The process can be mathematically expressed with a cluster state in the CME pattern as:
\begin{equation}
    |T\rangle = X^{s_2}Z^{s_1+s_3}[M_{3}^{\pi/4}]^{s_2}[M_2^{\theta_m}]^{s_1}[M_1^0]|\text{cluster}\rangle.
    \label{eq_1d}
\end{equation}
Here, $s_i$ represents the binary measurement outcome on the $i_{\text{th}}$ measurement. Although the measurement outcomes on each qubit are random and influence subsequent measurement settings and corrections, our previous analysis established that they do not affect the overall magic resources within the system. Therefore, we average the final magic resources over all possible single-qubit measurement outcomes to obtain an estimation.

\begin{figure}[t]
    \centering
    \includegraphics[width=0.9\textwidth]{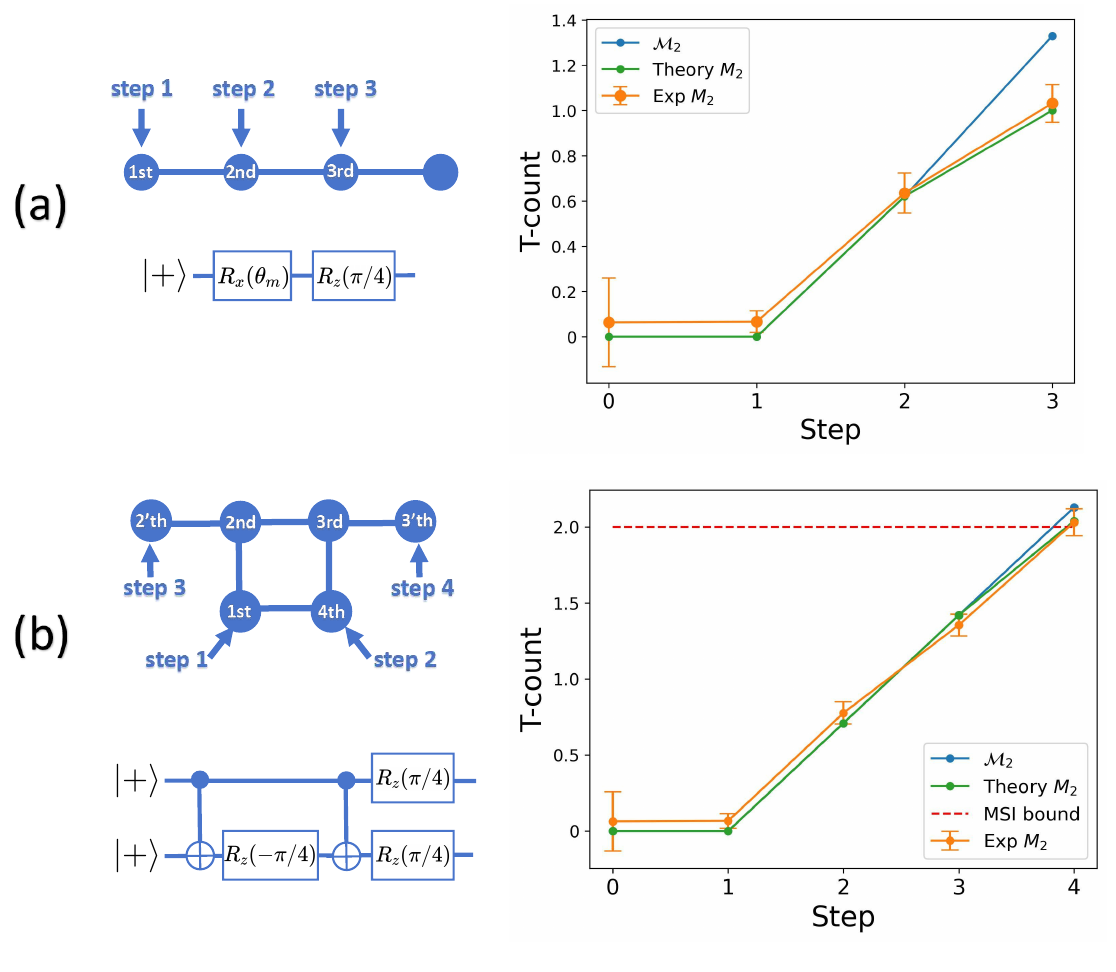}
    \caption{
    \textit{(a)} Experimental generation of the $|T\rangle$ state using a 1D graph in three steps. The angles of $M^\theta$ are ${0,\theta_m,\pi/4}$ for steps 1, 2, and 3, respectively. The plot on the right displays the total invested magic resources $\IMR_2$ and the reserved magic resource $\R$ at each step. 
    \textit{(b) Experimental realization of QFT using a 2D graph.} The left side of the figure depicts the 2D graph structure employed for implementing the QFT circuit of $n=2$ (left bottom) and indicates the order of measurements performed on the qubits (steps 1 to 4). 
    The generation of $|\text{CS}\rangle$ from MQC requires at least 6 qubits, as the figure shows. Step 3 on the 2'th qubit and step 4 on the 3'th qubit are equivalent to the local rotation for the physical qubits 2nd and 3rd qubit.  
    According to Eq.\eqref{eq: Box}, implementing $M_1^0$, $M_4^{(\pi/8,0)}$, $\hat{T}_2$, and $\hat{T}_3$ corresponds to the steps 1 to 4, in which $M_4^{(\pi/8,0)}$ is for non-standard measurement $M^{(\phi,\theta)}$, projection onto $\{\cos\phi|0\rangle\pm e^{-i\theta}\sin\phi|1\rangle\}$, and $\hat{T}$ is T-gate $\hat{T}=\diag[1,e^{i\pi/4}]$. Move experimental results see Sec.~IX and Sec.~X in \cite{supp}. 
    }
    \label{fig: exp}
\end{figure}

The comparison between the invested and reserved magic resources in the process is illustrated in Fig.~\ref{fig: exp}(a). 
The reserved magic resources are 0, 0.62T, and 1T, with accumulated invested magic resources of 0T, 0.62T, and 1.33T, respectively. 
It can be seen that the invested and reserved magic are equal during the first two steps of the process. 
However, for the third step, a waste of magic resources $\mathcal{W}=0.33\T$ is inevitable because the invested magic resources exceed the potential magic resources for a linear graph $\mathcal{P}=1\T$.

We demonstrate the generation of the QFT of $n=2$ in MQC. QFT requires controlled-rotation gates with arbitrary angles. 
Here for $n=2$, it shows $CR_2 = \text{diag}[1,1,1,i]$, also known as $CS$ gate. 
The power of QFT is evident in the resulting QFT state, defined as $|\text{QFT}_n\rangle = \text{QFT}_n |+\rangle|0\rangle^{\otimes(n-1)}$, which possesses a nullity of $\nu(|\text{QFT}_n)\rangle = n-2$ \cite{beverland2020lower}. 
For $n=2$, the QFT state $|\text{QFT}_2\rangle = |\text{CS}\rangle = (|00\rangle+|01\rangle+|10\rangle +i|11\rangle)/2$, representing the state with maximum magic content for a two-qubit system, can be generated from the MQC, see Fig.~\ref{fig: exp}(b). The resulting CME pattern for generating the $|\text{CS}\rangle$ state is given by:
\begin{equation}
    |\text{CS}\rangle = X_3Z_3^{s_4}Z_2^{s_1} \hat{T}_3 \hat{T}_2 [M_4^{(\pi/8, 0)}][M_1^0]|\text{cluster}\rangle.
    \label{eq: Box}
\end{equation}

The invested magic resources increase by approximately $0.71T$ with each step of the process. 
Consequently, the accumulated invested magic resources for steps 1 to 4 are $0\T$, $0.71\T$, $1.42\T$, and $2.13\T$, respectively. The corresponding reserved magic resources are $0T$, $0.71\T$, $1.41\T$, and $2.03\T$. The inevitable wasted magic resources are merely $\mathcal{W}=0.1T$. In the experiments, we obtain $0.066\pm0.024 \T, 0.777\pm 0.036 \T, 1.354\pm 0.036 \T, 2.030\pm 0.045 \T$.
Notably, almost all the invested magic resources are effectively injected into the system until the potential $\mathcal{P}=2.03\T$, bringing the inevitable waste of $\mathcal{W}=0.1\T$.

\textit{Discussion ---}
In conventional MSI, magic resources are injected uniformly at the outset, while a clear synthesis protocol for resource concentration is unattainable. MQC, in contrast, naturally concentrates magic resources into the reserved states, as evidenced by the 2.03T of magic resources injected into a two-qubit cluster state in our experiment.

The key contribution of this work is the introduction of invested and potential magic resources, clarifying the role of non-stabilizerness in MQC, especially from non-Pauli measurements. We employ the SRE for quantifying these resources due to its experimental feasibility; however, the framework is adaptable to alternative measures of magic, such as mana, stabilizer nullity, or the stabilizer norm, among others, depending on the specific scenario and operational constraints. As an explicit example illustrating this point, the QFT circuit is analyzed using the GKP-based magic measure \cite{hahn2022quantifying} (see Sec.~IX of \cite{supp})

Moreover, our work offers a new perspective on the universality of MQC from magic resources, showing that high-dimensional graph states can exhibit superlinear, or even exponential, advantages over classical computation. An intriguing open question is how different graph-state structures influence the potential magic resources they can host.

While calculating potential magic resources for an arbitrary graph state remains challenging due to the unknown optimal measurement, future approaches could explore gradient-based methods, neural-network-assisted optimization, or symmetry-constrained strategies to reduce the computational complexity in the traversal.

In conclusion, our framework establishes MQC as a promising approach for magic state generation and utilization, particularly in resource-efficient implementations of quantum algorithms such as QFT. Future research could focus on optimizing MQC for large-scale quantum networks, paving the way for more robust and fault-tolerant quantum computation.

\ 

\textit{Acknowledgement ---}
This work was supported by the Innovation Program for Quantum Science and Technology (Nos. 2021ZD0301200, 2021ZD0301400), National Natural Science Foundation of China (Grant Nos. 12350006, 11821404),  Anhui Initiative in Quantum Information Technologies (AHY060300), USTC Research Funds of the Double First-Class Initiative (Grant No. YD2030002026). 
YZ acknowledges support from the Innovation Program for Quantum Science and Technology Grant Nos.~2024ZD0301900 and 2021ZD0302000, the National Natural Science Foundation of China (NSFC) Grant No.~12205048 and 12575012, the Shanghai Science and Technology Innovation Action Plan Grant No.~24LZ1400200, Shanghai Pilot Program for Basic Research - Fudan University 21TQ1400100 (25TQ003), and the start-up funding of Fudan University.
AH acknowledges support from PNRR MUR project PE0000023-NQSTI and PNRR MUR project CN $00000013$ -ICSC. 

\textit{Data Availability --- } The data that support the findings of this article are openly available \cite{supp, code_P}. 

\bibliography{reference}

\begin{thebibliography}{53}%
\makeatletter
\providecommand \@ifxundefined [1]{%
 \@ifx{#1\undefined}
}%
\providecommand \@ifnum [1]{%
 \ifnum #1\expandafter \@firstoftwo
 \else \expandafter \@secondoftwo
 \fi
}%
\providecommand \@ifx [1]{%
 \ifx #1\expandafter \@firstoftwo
 \else \expandafter \@secondoftwo
 \fi
}%
\providecommand \natexlab [1]{#1}%
\providecommand \enquote  [1]{``#1''}%
\providecommand \bibnamefont  [1]{#1}%
\providecommand \bibfnamefont [1]{#1}%
\providecommand \citenamefont [1]{#1}%
\providecommand \href@noop [0]{\@secondoftwo}%
\providecommand \href [0]{\begingroup \@sanitize@url \@href}%
\providecommand \@href[1]{\@@startlink{#1}\@@href}%
\providecommand \@@href[1]{\endgroup#1\@@endlink}%
\providecommand \@sanitize@url [0]{\catcode `\\12\catcode `\$12\catcode
  `\&12\catcode `\#12\catcode `\^12\catcode `\_12\catcode `\%12\relax}%
\providecommand \@@startlink[1]{}%
\providecommand \@@endlink[0]{}%
\providecommand \url  [0]{\begingroup\@sanitize@url \@url }%
\providecommand \@url [1]{\endgroup\@href {#1}{\urlprefix }}%
\providecommand \urlprefix  [0]{URL }%
\providecommand \Eprint [0]{\href }%
\providecommand \doibase [0]{https://doi.org/}%
\providecommand \selectlanguage [0]{\@gobble}%
\providecommand \bibinfo  [0]{\@secondoftwo}%
\providecommand \bibfield  [0]{\@secondoftwo}%
\providecommand \translation [1]{[#1]}%
\providecommand \BibitemOpen [0]{}%
\providecommand \bibitemStop [0]{}%
\providecommand \bibitemNoStop [0]{.\EOS\space}%
\providecommand \EOS [0]{\spacefactor3000\relax}%
\providecommand \BibitemShut  [1]{\csname bibitem#1\endcsname}%
\let\auto@bib@innerbib\@empty
\bibitem [{\citenamefont {Campbell}\ \emph {et~al.}(2017)\citenamefont
  {Campbell}, \citenamefont {Terhal},\ and\ \citenamefont
  {Vuillot}}]{campbell2017roads}%
  \BibitemOpen
  \bibfield  {author} {\bibinfo {author} {\bibfnamefont {E.~T.}\ \bibnamefont
  {Campbell}}, \bibinfo {author} {\bibfnamefont {B.~M.}\ \bibnamefont
  {Terhal}},\ and\ \bibinfo {author} {\bibfnamefont {C.}~\bibnamefont
  {Vuillot}},\ }\bibfield  {title} {\bibinfo {title} {Roads towards
  fault-tolerant universal quantum computation},\ }\href
  {https://www.nature.com/articles/nature23460} {\bibfield  {journal} {\bibinfo
   {journal} {Nature}\ }\textbf {\bibinfo {volume} {549}},\ \bibinfo {pages}
  {172} (\bibinfo {year} {2017})}\BibitemShut {NoStop}%
\bibitem [{\citenamefont {Postler}\ \emph {et~al.}(2022)\citenamefont
  {Postler}, \citenamefont {Heu$\beta$en}, \citenamefont {Pogorelov},
  \citenamefont {Rispler}, \citenamefont {Feldker}, \citenamefont {Meth},
  \citenamefont {Marciniak}, \citenamefont {Stricker}, \citenamefont
  {Ringbauer}, \citenamefont {Blatt} \emph
  {et~al.}}]{postler2022demonstration}%
  \BibitemOpen
  \bibfield  {author} {\bibinfo {author} {\bibfnamefont {L.}~\bibnamefont
  {Postler}}, \bibinfo {author} {\bibfnamefont {S.}~\bibnamefont
  {Heu$\beta$en}}, \bibinfo {author} {\bibfnamefont {I.}~\bibnamefont
  {Pogorelov}}, \bibinfo {author} {\bibfnamefont {M.}~\bibnamefont {Rispler}},
  \bibinfo {author} {\bibfnamefont {T.}~\bibnamefont {Feldker}}, \bibinfo
  {author} {\bibfnamefont {M.}~\bibnamefont {Meth}}, \bibinfo {author}
  {\bibfnamefont {C.~D.}\ \bibnamefont {Marciniak}}, \bibinfo {author}
  {\bibfnamefont {R.}~\bibnamefont {Stricker}}, \bibinfo {author}
  {\bibfnamefont {M.}~\bibnamefont {Ringbauer}}, \bibinfo {author}
  {\bibfnamefont {R.}~\bibnamefont {Blatt}}, \emph {et~al.},\ }\bibfield
  {title} {\bibinfo {title} {Demonstration of fault-tolerant universal quantum
  gate operations},\ }\href
  {https://www.nature.com/articles/s41586-022-04721-1} {\bibfield  {journal}
  {\bibinfo  {journal} {Nature}\ }\textbf {\bibinfo {volume} {605}},\ \bibinfo
  {pages} {675} (\bibinfo {year} {2022})}\BibitemShut {NoStop}%
\bibitem [{\citenamefont {AI}(2023)}]{google2023suppressing}%
  \BibitemOpen
  \bibfield  {author} {\bibinfo {author} {\bibfnamefont {G.~Q.}\ \bibnamefont
  {AI}},\ }\bibfield  {title} {\bibinfo {title} {Suppressing quantum errors by
  scaling a surface code logical qubit},\ }\href
  {https://www.nature.com/articles/s41586-022-05434-1} {\bibfield  {journal}
  {\bibinfo  {journal} {Nature}\ }\textbf {\bibinfo {volume} {614}},\ \bibinfo
  {pages} {676} (\bibinfo {year} {2023})}\BibitemShut {NoStop}%
\bibitem [{\citenamefont {Gupta}\ \emph {et~al.}(2024)\citenamefont {Gupta},
  \citenamefont {Sundaresan}, \citenamefont {Alexander}, \citenamefont {Wood},
  \citenamefont {Merkel}, \citenamefont {Healy}, \citenamefont {Hillenbrand},
  \citenamefont {Jochym-O’Connor}, \citenamefont {Wootton}, \citenamefont
  {Yoder} \emph {et~al.}}]{gupta2024encoding}%
  \BibitemOpen
  \bibfield  {author} {\bibinfo {author} {\bibfnamefont {R.~S.}\ \bibnamefont
  {Gupta}}, \bibinfo {author} {\bibfnamefont {N.}~\bibnamefont {Sundaresan}},
  \bibinfo {author} {\bibfnamefont {T.}~\bibnamefont {Alexander}}, \bibinfo
  {author} {\bibfnamefont {C.~J.}\ \bibnamefont {Wood}}, \bibinfo {author}
  {\bibfnamefont {S.~T.}\ \bibnamefont {Merkel}}, \bibinfo {author}
  {\bibfnamefont {M.~B.}\ \bibnamefont {Healy}}, \bibinfo {author}
  {\bibfnamefont {M.}~\bibnamefont {Hillenbrand}}, \bibinfo {author}
  {\bibfnamefont {T.}~\bibnamefont {Jochym-O’Connor}}, \bibinfo {author}
  {\bibfnamefont {J.~R.}\ \bibnamefont {Wootton}}, \bibinfo {author}
  {\bibfnamefont {T.~J.}\ \bibnamefont {Yoder}}, \emph {et~al.},\ }\bibfield
  {title} {\bibinfo {title} {Encoding a magic state with beyond break-even
  fidelity},\ }\href {https://www.nature.com/articles/s41586-023-06846-3}
  {\bibfield  {journal} {\bibinfo  {journal} {Nature}\ }\textbf {\bibinfo
  {volume} {625}},\ \bibinfo {pages} {259} (\bibinfo {year}
  {2024})}\BibitemShut {NoStop}%
\bibitem [{\citenamefont {Bravyi}\ and\ \citenamefont
  {Kitaev}(2005)}]{bravyi2005universal}%
  \BibitemOpen
  \bibfield  {author} {\bibinfo {author} {\bibfnamefont {S.}~\bibnamefont
  {Bravyi}}\ and\ \bibinfo {author} {\bibfnamefont {A.}~\bibnamefont
  {Kitaev}},\ }\bibfield  {title} {\bibinfo {title} {Universal quantum
  computation with ideal clifford gates and noisy ancillas},\ }\href
  {https://journals.aps.org/pra/abstract/10.1103/PhysRevA.71.022316} {\bibfield
   {journal} {\bibinfo  {journal} {Phys. Rev. A}\ }\textbf {\bibinfo {volume}
  {71}},\ \bibinfo {pages} {022316} (\bibinfo {year} {2005})}\BibitemShut
  {NoStop}%
\bibitem [{\citenamefont {Veitch}\ \emph {et~al.}(2014)\citenamefont {Veitch},
  \citenamefont {Mousavian}, \citenamefont {Gottesman},\ and\ \citenamefont
  {Emerson}}]{veitch2014resource}%
  \BibitemOpen
  \bibfield  {author} {\bibinfo {author} {\bibfnamefont {V.}~\bibnamefont
  {Veitch}}, \bibinfo {author} {\bibfnamefont {S.~H.}\ \bibnamefont
  {Mousavian}}, \bibinfo {author} {\bibfnamefont {D.}~\bibnamefont
  {Gottesman}},\ and\ \bibinfo {author} {\bibfnamefont {J.}~\bibnamefont
  {Emerson}},\ }\bibfield  {title} {\bibinfo {title} {The resource theory of
  stabilizer quantum computation},\ }\href
  {https://iopscience.iop.org/article/10.1088/1367-2630/16/1/013009/meta}
  {\bibfield  {journal} {\bibinfo  {journal} {New Journal of Physics}\ }\textbf
  {\bibinfo {volume} {16}},\ \bibinfo {pages} {013009} (\bibinfo {year}
  {2014})}\BibitemShut {NoStop}%
\bibitem [{\citenamefont {Leone}\ \emph {et~al.}(2022)\citenamefont {Leone},
  \citenamefont {Oliviero},\ and\ \citenamefont {Hamma}}]{leone2022stabilizer}%
  \BibitemOpen
  \bibfield  {author} {\bibinfo {author} {\bibfnamefont {L.}~\bibnamefont
  {Leone}}, \bibinfo {author} {\bibfnamefont {S.~F.}\ \bibnamefont
  {Oliviero}},\ and\ \bibinfo {author} {\bibfnamefont {A.}~\bibnamefont
  {Hamma}},\ }\bibfield  {title} {\bibinfo {title} {Stabilizer r{\'e}nyi
  entropy},\ }\href
  {https://journals.aps.org/prl/abstract/10.1103/PhysRevLett.128.050402}
  {\bibfield  {journal} {\bibinfo  {journal} {Phys. Rev. Lett.}\ }\textbf
  {\bibinfo {volume} {128}},\ \bibinfo {pages} {050402} (\bibinfo {year}
  {2022})}\BibitemShut {NoStop}%
\bibitem [{\citenamefont {Leone}\ and\ \citenamefont {Bittel}(2024)}]{bittel}%
  \BibitemOpen
  \bibfield  {author} {\bibinfo {author} {\bibfnamefont {L.}~\bibnamefont
  {Leone}}\ and\ \bibinfo {author} {\bibfnamefont {L.}~\bibnamefont {Bittel}},\
  }\bibfield  {title} {\bibinfo {title} {Stabilizer entropies are monotones for
  magic-state resource theory},\ }\href
  {https://journals.aps.org/pra/abstract/10.1103/PhysRevA.110.L040403}
  {\bibfield  {journal} {\bibinfo  {journal} {Phys. Rev. A}\ }\textbf {\bibinfo
  {volume} {110}},\ \bibinfo {pages} {L040403} (\bibinfo {year}
  {2024})}\BibitemShut {NoStop}%
\bibitem [{\citenamefont {Souza}\ \emph {et~al.}(2011)\citenamefont {Souza},
  \citenamefont {Zhang}, \citenamefont {Ryan},\ and\ \citenamefont
  {Laflamme}}]{souza2011experimental}%
  \BibitemOpen
  \bibfield  {author} {\bibinfo {author} {\bibfnamefont {A.~M.}\ \bibnamefont
  {Souza}}, \bibinfo {author} {\bibfnamefont {J.}~\bibnamefont {Zhang}},
  \bibinfo {author} {\bibfnamefont {C.~A.}\ \bibnamefont {Ryan}},\ and\
  \bibinfo {author} {\bibfnamefont {R.}~\bibnamefont {Laflamme}},\ }\bibfield
  {title} {\bibinfo {title} {Experimental magic state distillation for
  fault-tolerant quantum computing},\ }\href
  {https://www.nature.com/articles/ncomms1166} {\bibfield  {journal} {\bibinfo
  {journal} {Nature communications}\ }\textbf {\bibinfo {volume} {2}},\
  \bibinfo {pages} {169} (\bibinfo {year} {2011})}\BibitemShut {NoStop}%
\bibitem [{\citenamefont {Knill}(2005)}]{knill2005quantum}%
  \BibitemOpen
  \bibfield  {author} {\bibinfo {author} {\bibfnamefont {E.}~\bibnamefont
  {Knill}},\ }\bibfield  {title} {\bibinfo {title} {Quantum computing with
  realistically noisy devices},\ }\href
  {https://www.nature.com/articles/nature03350} {\bibfield  {journal} {\bibinfo
   {journal} {Nature}\ }\textbf {\bibinfo {volume} {434}},\ \bibinfo {pages}
  {39} (\bibinfo {year} {2005})}\BibitemShut {NoStop}%
\bibitem [{\citenamefont {Howard}\ and\ \citenamefont
  {Campbell}(2017)}]{howard2017application}%
  \BibitemOpen
  \bibfield  {author} {\bibinfo {author} {\bibfnamefont {M.}~\bibnamefont
  {Howard}}\ and\ \bibinfo {author} {\bibfnamefont {E.}~\bibnamefont
  {Campbell}},\ }\bibfield  {title} {\bibinfo {title} {Application of a
  resource theory for magic states to fault-tolerant quantum computing},\
  }\href {https://journals.aps.org/prl/abstract/10.1103/PhysRevLett.118.090501}
  {\bibfield  {journal} {\bibinfo  {journal} {Phys. Rev. Lett.}\ }\textbf
  {\bibinfo {volume} {118}},\ \bibinfo {pages} {090501} (\bibinfo {year}
  {2017})}\BibitemShut {NoStop}%
\bibitem [{\citenamefont {Regula}\ and\ \citenamefont
  {Takagi}(2021)}]{regula2021fundamental}%
  \BibitemOpen
  \bibfield  {author} {\bibinfo {author} {\bibfnamefont {B.}~\bibnamefont
  {Regula}}\ and\ \bibinfo {author} {\bibfnamefont {R.}~\bibnamefont
  {Takagi}},\ }\bibfield  {title} {\bibinfo {title} {Fundamental limitations on
  distillation of quantum channel resources},\ }\href
  {https://www.nature.com/articles/s41467-021-24699-0} {\bibfield  {journal}
  {\bibinfo  {journal} {Nature Communications}\ }\textbf {\bibinfo {volume}
  {12}},\ \bibinfo {pages} {4411} (\bibinfo {year} {2021})}\BibitemShut
  {NoStop}%
\bibitem [{\citenamefont {Aaronson}\ and\ \citenamefont
  {Gottesman}(2004)}]{aaronson}%
  \BibitemOpen
  \bibfield  {author} {\bibinfo {author} {\bibfnamefont {S.}~\bibnamefont
  {Aaronson}}\ and\ \bibinfo {author} {\bibfnamefont {D.}~\bibnamefont
  {Gottesman}},\ }\bibfield  {title} {\bibinfo {title} {Improved simulation of
  stabilizer circuits},\ }\href
  {https://journals.aps.org/pra/abstract/10.1103/PhysRevA.70.052328} {\bibfield
   {journal} {\bibinfo  {journal} {Phys. Rev. A}\ }\textbf {\bibinfo {volume}
  {70}},\ \bibinfo {pages} {052328} (\bibinfo {year} {2004})}\BibitemShut
  {NoStop}%
\bibitem [{\citenamefont {Bravyi}\ and\ \citenamefont
  {Gosset}(2016)}]{bravyi2016improved}%
  \BibitemOpen
  \bibfield  {author} {\bibinfo {author} {\bibfnamefont {S.}~\bibnamefont
  {Bravyi}}\ and\ \bibinfo {author} {\bibfnamefont {D.}~\bibnamefont
  {Gosset}},\ }\bibfield  {title} {\bibinfo {title} {Improved classical
  simulation of quantum circuits dominated by clifford gates},\ }\href
  {https://journals.aps.org/prl/abstract/10.1103/PhysRevLett.116.250501}
  {\bibfield  {journal} {\bibinfo  {journal} {Phys. Rev. Lett.}\ }\textbf
  {\bibinfo {volume} {116}},\ \bibinfo {pages} {250501} (\bibinfo {year}
  {2016})}\BibitemShut {NoStop}%
\bibitem [{\citenamefont {Bravyi}\ \emph {et~al.}(2016)\citenamefont {Bravyi},
  \citenamefont {Smith},\ and\ \citenamefont {Smolin}}]{bravyi2016trading}%
  \BibitemOpen
  \bibfield  {author} {\bibinfo {author} {\bibfnamefont {S.}~\bibnamefont
  {Bravyi}}, \bibinfo {author} {\bibfnamefont {G.}~\bibnamefont {Smith}},\ and\
  \bibinfo {author} {\bibfnamefont {J.~A.}\ \bibnamefont {Smolin}},\ }\bibfield
   {title} {\bibinfo {title} {Trading classical and quantum computational
  resources},\ }\href
  {https://journals.aps.org/prx/abstract/10.1103/PhysRevX.6.021043} {\bibfield
  {journal} {\bibinfo  {journal} {Phys. Rev. X}\ }\textbf {\bibinfo {volume}
  {6}},\ \bibinfo {pages} {021043} (\bibinfo {year} {2016})}\BibitemShut
  {NoStop}%
\bibitem [{\citenamefont {Wang}\ \emph {et~al.}(2020)\citenamefont {Wang},
  \citenamefont {Wilde},\ and\ \citenamefont {Su}}]{wang2020efficiently}%
  \BibitemOpen
  \bibfield  {author} {\bibinfo {author} {\bibfnamefont {X.}~\bibnamefont
  {Wang}}, \bibinfo {author} {\bibfnamefont {M.~M.}\ \bibnamefont {Wilde}},\
  and\ \bibinfo {author} {\bibfnamefont {Y.}~\bibnamefont {Su}},\ }\bibfield
  {title} {\bibinfo {title} {Efficiently computable bounds for magic state
  distillation},\ }\href
  {https://journals.aps.org/prl/abstract/10.1103/PhysRevLett.124.090505}
  {\bibfield  {journal} {\bibinfo  {journal} {Phys. Rev. Lett.}\ }\textbf
  {\bibinfo {volume} {124}},\ \bibinfo {pages} {090505} (\bibinfo {year}
  {2020})}\BibitemShut {NoStop}%
\bibitem [{\citenamefont {Beverland}\ \emph {et~al.}(2020)\citenamefont
  {Beverland}, \citenamefont {Campbell}, \citenamefont {Howard},\ and\
  \citenamefont {Kliuchnikov}}]{beverland2020lower}%
  \BibitemOpen
  \bibfield  {author} {\bibinfo {author} {\bibfnamefont {M.}~\bibnamefont
  {Beverland}}, \bibinfo {author} {\bibfnamefont {E.}~\bibnamefont {Campbell}},
  \bibinfo {author} {\bibfnamefont {M.}~\bibnamefont {Howard}},\ and\ \bibinfo
  {author} {\bibfnamefont {V.}~\bibnamefont {Kliuchnikov}},\ }\bibfield
  {title} {\bibinfo {title} {Lower bounds on the non-clifford resources for
  quantum computations},\ }\href
  {https://iopscience.iop.org/article/10.1088/2058-9565/ab8963/meta} {\bibfield
   {journal} {\bibinfo  {journal} {Quantum Science and Technology}\ }\textbf
  {\bibinfo {volume} {5}},\ \bibinfo {pages} {035009} (\bibinfo {year}
  {2020})}\BibitemShut {NoStop}%
\bibitem [{\citenamefont {Hahn}\ \emph {et~al.}(2022)\citenamefont {Hahn},
  \citenamefont {Ferraro}, \citenamefont {Hultquist}, \citenamefont {Ferrini},\
  and\ \citenamefont {Garc{\'\i}a-{\'A}lvarez}}]{hahn2022quantifying}%
  \BibitemOpen
  \bibfield  {author} {\bibinfo {author} {\bibfnamefont {O.}~\bibnamefont
  {Hahn}}, \bibinfo {author} {\bibfnamefont {A.}~\bibnamefont {Ferraro}},
  \bibinfo {author} {\bibfnamefont {L.}~\bibnamefont {Hultquist}}, \bibinfo
  {author} {\bibfnamefont {G.}~\bibnamefont {Ferrini}},\ and\ \bibinfo {author}
  {\bibfnamefont {L.}~\bibnamefont {Garc{\'\i}a-{\'A}lvarez}},\ }\bibfield
  {title} {\bibinfo {title} {Quantifying qubit magic resource with
  gottesman-kitaev-preskill encoding},\ }\href
  {https://journals.aps.org/prl/abstract/10.1103/PhysRevLett.128.210502}
  {\bibfield  {journal} {\bibinfo  {journal} {Phys. Rev. Lett.}\ }\textbf
  {\bibinfo {volume} {128}},\ \bibinfo {pages} {210502} (\bibinfo {year}
  {2022})}\BibitemShut {NoStop}%
\bibitem [{\citenamefont {Oliviero}\ \emph {et~al.}(2021)\citenamefont
  {Oliviero}, \citenamefont {Leone},\ and\ \citenamefont
  {Hamma}}]{oliviero2021transitions}%
  \BibitemOpen
  \bibfield  {author} {\bibinfo {author} {\bibfnamefont {S.~F.}\ \bibnamefont
  {Oliviero}}, \bibinfo {author} {\bibfnamefont {L.}~\bibnamefont {Leone}},\
  and\ \bibinfo {author} {\bibfnamefont {A.}~\bibnamefont {Hamma}},\ }\bibfield
   {title} {\bibinfo {title} {Transitions in entanglement complexity in random
  quantum circuits by measurements},\ }\href
  {https://www.sciencedirect.com/science/article/pii/S0375960121005855}
  {\bibfield  {journal} {\bibinfo  {journal} {Phys. Lett. A}\ }\textbf
  {\bibinfo {volume} {418}},\ \bibinfo {pages} {127721} (\bibinfo {year}
  {2021})}\BibitemShut {NoStop}%
\bibitem [{\citenamefont {Niroula}\ \emph {et~al.}(2024)\citenamefont
  {Niroula}, \citenamefont {White}, \citenamefont {Wang}, \citenamefont
  {Johri}, \citenamefont {Zhu}, \citenamefont {Monroe}, \citenamefont {Noel},\
  and\ \citenamefont {Gullans}}]{niroula2023phase}%
  \BibitemOpen
  \bibfield  {author} {\bibinfo {author} {\bibfnamefont {P.}~\bibnamefont
  {Niroula}}, \bibinfo {author} {\bibfnamefont {C.~D.}\ \bibnamefont {White}},
  \bibinfo {author} {\bibfnamefont {Q.}~\bibnamefont {Wang}}, \bibinfo {author}
  {\bibfnamefont {S.}~\bibnamefont {Johri}}, \bibinfo {author} {\bibfnamefont
  {D.}~\bibnamefont {Zhu}}, \bibinfo {author} {\bibfnamefont {C.}~\bibnamefont
  {Monroe}}, \bibinfo {author} {\bibfnamefont {C.}~\bibnamefont {Noel}},\ and\
  \bibinfo {author} {\bibfnamefont {M.~J.}\ \bibnamefont {Gullans}},\
  }\bibfield  {title} {\bibinfo {title} {Phase transition in magic with random
  quantum circuits},\ }\href
  {https://www.nature.com/articles/s41567-024-02637-3} {\bibfield  {journal}
  {\bibinfo  {journal} {Nature physics}\ }\textbf {\bibinfo {volume} {20}},\
  \bibinfo {pages} {1786} (\bibinfo {year} {2024})}\BibitemShut {NoStop}%
\bibitem [{\citenamefont {Briegel}\ \emph {et~al.}(2009)\citenamefont
  {Briegel}, \citenamefont {Browne}, \citenamefont {D{\"u}r}, \citenamefont
  {Raussendorf},\ and\ \citenamefont {Van~den Nest}}]{briegel2009measurement}%
  \BibitemOpen
  \bibfield  {author} {\bibinfo {author} {\bibfnamefont {H.~J.}\ \bibnamefont
  {Briegel}}, \bibinfo {author} {\bibfnamefont {D.~E.}\ \bibnamefont {Browne}},
  \bibinfo {author} {\bibfnamefont {W.}~\bibnamefont {D{\"u}r}}, \bibinfo
  {author} {\bibfnamefont {R.}~\bibnamefont {Raussendorf}},\ and\ \bibinfo
  {author} {\bibfnamefont {M.}~\bibnamefont {Van~den Nest}},\ }\bibfield
  {title} {\bibinfo {title} {Measurement-based quantum computation},\ }\href
  {https://www.nature.com/articles/nphys1157} {\bibfield  {journal} {\bibinfo
  {journal} {Nature Physics}\ }\textbf {\bibinfo {volume} {5}},\ \bibinfo
  {pages} {19} (\bibinfo {year} {2009})}\BibitemShut {NoStop}%
\bibitem [{\citenamefont {Van~den Nest}\ \emph {et~al.}(2007)\citenamefont
  {Van~den Nest}, \citenamefont {D{\"u}r}, \citenamefont {Miyake},\ and\
  \citenamefont {Briegel}}]{van2007fundamentals}%
  \BibitemOpen
  \bibfield  {author} {\bibinfo {author} {\bibfnamefont {M.}~\bibnamefont
  {Van~den Nest}}, \bibinfo {author} {\bibfnamefont {W.}~\bibnamefont
  {D{\"u}r}}, \bibinfo {author} {\bibfnamefont {A.}~\bibnamefont {Miyake}},\
  and\ \bibinfo {author} {\bibfnamefont {H.~J.}\ \bibnamefont {Briegel}},\
  }\bibfield  {title} {\bibinfo {title} {Fundamentals of universality in
  one-way quantum computation},\ }\href
  {https://iopscience.iop.org/article/10.1088/1367-2630/9/6/204/meta}
  {\bibfield  {journal} {\bibinfo  {journal} {New Journal of Physics}\ }\textbf
  {\bibinfo {volume} {9}},\ \bibinfo {pages} {204} (\bibinfo {year}
  {2007})}\BibitemShut {NoStop}%
\bibitem [{\citenamefont {Van~den Nest}\ \emph {et~al.}(2006)\citenamefont
  {Van~den Nest}, \citenamefont {Miyake}, \citenamefont {D{\"u}r},\ and\
  \citenamefont {Briegel}}]{van2006universal}%
  \BibitemOpen
  \bibfield  {author} {\bibinfo {author} {\bibfnamefont {M.}~\bibnamefont
  {Van~den Nest}}, \bibinfo {author} {\bibfnamefont {A.}~\bibnamefont
  {Miyake}}, \bibinfo {author} {\bibfnamefont {W.}~\bibnamefont {D{\"u}r}},\
  and\ \bibinfo {author} {\bibfnamefont {H.~J.}\ \bibnamefont {Briegel}},\
  }\bibfield  {title} {\bibinfo {title} {Universal resources for
  measurement-based quantum computation},\ }\href
  {https://journals.aps.org/prl/abstract/10.1103/PhysRevLett.97.150504}
  {\bibfield  {journal} {\bibinfo  {journal} {Phys. Rev. Lett.}\ }\textbf
  {\bibinfo {volume} {97}},\ \bibinfo {pages} {150504} (\bibinfo {year}
  {2006})}\BibitemShut {NoStop}%
\bibitem [{\citenamefont {Danos}\ \emph {et~al.}(2005)\citenamefont {Danos},
  \citenamefont {Kashefi},\ and\ \citenamefont
  {Panangaden}}]{danos2005parsimonious}%
  \BibitemOpen
  \bibfield  {author} {\bibinfo {author} {\bibfnamefont {V.}~\bibnamefont
  {Danos}}, \bibinfo {author} {\bibfnamefont {E.}~\bibnamefont {Kashefi}},\
  and\ \bibinfo {author} {\bibfnamefont {P.}~\bibnamefont {Panangaden}},\
  }\bibfield  {title} {\bibinfo {title} {Parsimonious and robust realizations
  of unitary maps in the one-way model},\ }\href
  {https://journals.aps.org/pra/abstract/10.1103/PhysRevA.72.064301} {\bibfield
   {journal} {\bibinfo  {journal} {Phys. Rev. A}\ }\textbf {\bibinfo {volume}
  {72}},\ \bibinfo {pages} {064301} (\bibinfo {year} {2005})}\BibitemShut
  {NoStop}%
\bibitem [{\citenamefont {Danos}\ \emph {et~al.}(2007)\citenamefont {Danos},
  \citenamefont {Kashefi},\ and\ \citenamefont
  {Panangaden}}]{danos2007measurement}%
  \BibitemOpen
  \bibfield  {author} {\bibinfo {author} {\bibfnamefont {V.}~\bibnamefont
  {Danos}}, \bibinfo {author} {\bibfnamefont {E.}~\bibnamefont {Kashefi}},\
  and\ \bibinfo {author} {\bibfnamefont {P.}~\bibnamefont {Panangaden}},\
  }\bibfield  {title} {\bibinfo {title} {The measurement calculus},\ }\href
  {https://dl.acm.org/doi/abs/10.1145/1219092.1219096} {\bibfield  {journal}
  {\bibinfo  {journal} {Journal of the ACM (JACM)}\ }\textbf {\bibinfo {volume}
  {54}},\ \bibinfo {pages} {8} (\bibinfo {year} {2007})}\BibitemShut {NoStop}%
\bibitem [{sup()}]{supp}%
  \BibitemOpen
  \href@noop {} {}\bibinfo {note} {See Supplemental Material at [url] for more
  discussion with invested magic resources and potential magic resources, the
  experimental details, and more experimental results, which include Refs.
  \cite{chitambar2019quantum, leonelearning, campbell2011catalysis,
  oliviero2022measuring, chen2023magic, Haug2023MPS, Lami2023MPS, dalmonte,
  cao, leoneverification, coppersmith2002approximate, zhang2016experimental,
  kurtsiefer2001generation, zhu2017multiqubit, li2023experimental,
  brydges2019probing, elben2019statistical, elben2023randomized,
  elben2020mixed}.}\BibitemShut {Stop}%
\bibitem [{\citenamefont {Nam}\ \emph {et~al.}(2020)\citenamefont {Nam},
  \citenamefont {Su},\ and\ \citenamefont {Maslov}}]{nam2020approximate}%
  \BibitemOpen
  \bibfield  {author} {\bibinfo {author} {\bibfnamefont {Y.}~\bibnamefont
  {Nam}}, \bibinfo {author} {\bibfnamefont {Y.}~\bibnamefont {Su}},\ and\
  \bibinfo {author} {\bibfnamefont {D.}~\bibnamefont {Maslov}},\ }\bibfield
  {title} {\bibinfo {title} {Approximate quantum fourier transform with
  o(nlog(n)) t gates},\ }\href
  {https://www.nature.com/articles/s41534-020-0257-5} {\bibfield  {journal}
  {\bibinfo  {journal} {NPJ Quantum Information}\ }\textbf {\bibinfo {volume}
  {6}},\ \bibinfo {pages} {26} (\bibinfo {year} {2020})}\BibitemShut {NoStop}%
\bibitem [{\citenamefont {Iannotti}\ \emph {et~al.}(2025)\citenamefont
  {Iannotti}, \citenamefont {Esposito}, \citenamefont {Venuti},\ and\
  \citenamefont {Hamma}}]{iannotti2025entanglementstabilizerentropiesrandom}%
  \BibitemOpen
  \bibfield  {author} {\bibinfo {author} {\bibfnamefont {D.}~\bibnamefont
  {Iannotti}}, \bibinfo {author} {\bibfnamefont {G.}~\bibnamefont {Esposito}},
  \bibinfo {author} {\bibfnamefont {L.~C.}\ \bibnamefont {Venuti}},\ and\
  \bibinfo {author} {\bibfnamefont {A.}~\bibnamefont {Hamma}},\ }\href
  {https://arxiv.org/abs/2501.19261} {\bibinfo {title} {Entanglement and
  stabilizer entropies of random bipartite pure quantum states}} (\bibinfo
  {year} {2025}),\ \Eprint {https://arxiv.org/abs/2501.19261} {arXiv:2501.19261
  [quant-ph]} \BibitemShut {NoStop}%
\bibitem [{\citenamefont {Chiribella}\ \emph {et~al.}(2008)\citenamefont
  {Chiribella}, \citenamefont {D’Ariano},\ and\ \citenamefont
  {Perinotti}}]{chiribella2008quantum}%
  \BibitemOpen
  \bibfield  {author} {\bibinfo {author} {\bibfnamefont {G.}~\bibnamefont
  {Chiribella}}, \bibinfo {author} {\bibfnamefont {G.~M.}\ \bibnamefont
  {D’Ariano}},\ and\ \bibinfo {author} {\bibfnamefont {P.}~\bibnamefont
  {Perinotti}},\ }\bibfield  {title} {\bibinfo {title} {Quantum circuit
  architecture},\ }\href
  {https://journals.aps.org/prl/abstract/10.1103/PhysRevLett.101.060401}
  {\bibfield  {journal} {\bibinfo  {journal} {Phys. Rev. Lett.}\ }\textbf
  {\bibinfo {volume} {101}},\ \bibinfo {pages} {060401} (\bibinfo {year}
  {2008})}\BibitemShut {NoStop}%
\bibitem [{\citenamefont {Chiribella}\ \emph {et~al.}(2009)\citenamefont
  {Chiribella}, \citenamefont {D’Ariano},\ and\ \citenamefont
  {Perinotti}}]{chiribella2009theoretical}%
  \BibitemOpen
  \bibfield  {author} {\bibinfo {author} {\bibfnamefont {G.}~\bibnamefont
  {Chiribella}}, \bibinfo {author} {\bibfnamefont {G.~M.}\ \bibnamefont
  {D’Ariano}},\ and\ \bibinfo {author} {\bibfnamefont {P.}~\bibnamefont
  {Perinotti}},\ }\bibfield  {title} {\bibinfo {title} {Theoretical framework
  for quantum networks},\ }\href
  {https://journals.aps.org/pra/abstract/10.1103/PhysRevA.80.022339} {\bibfield
   {journal} {\bibinfo  {journal} {Phys. Rev. A}\ }\textbf {\bibinfo {volume}
  {80}},\ \bibinfo {pages} {022339} (\bibinfo {year} {2009})}\BibitemShut
  {NoStop}%
\bibitem [{\citenamefont {Walther}\ \emph {et~al.}(2005)\citenamefont
  {Walther}, \citenamefont {Resch}, \citenamefont {Rudolph}, \citenamefont
  {Schenck}, \citenamefont {Weinfurter}, \citenamefont {Vedral}, \citenamefont
  {Aspelmeyer},\ and\ \citenamefont {Zeilinger}}]{walther2005experimental}%
  \BibitemOpen
  \bibfield  {author} {\bibinfo {author} {\bibfnamefont {P.}~\bibnamefont
  {Walther}}, \bibinfo {author} {\bibfnamefont {K.~J.}\ \bibnamefont {Resch}},
  \bibinfo {author} {\bibfnamefont {T.}~\bibnamefont {Rudolph}}, \bibinfo
  {author} {\bibfnamefont {E.}~\bibnamefont {Schenck}}, \bibinfo {author}
  {\bibfnamefont {H.}~\bibnamefont {Weinfurter}}, \bibinfo {author}
  {\bibfnamefont {V.}~\bibnamefont {Vedral}}, \bibinfo {author} {\bibfnamefont
  {M.}~\bibnamefont {Aspelmeyer}},\ and\ \bibinfo {author} {\bibfnamefont
  {A.}~\bibnamefont {Zeilinger}},\ }\bibfield  {title} {\bibinfo {title}
  {Experimental one-way quantum computing},\ }\href
  {https://www.nature.com/articles/nature03347} {\bibfield  {journal} {\bibinfo
   {journal} {Nature}\ }\textbf {\bibinfo {volume} {434}},\ \bibinfo {pages}
  {169} (\bibinfo {year} {2005})}\BibitemShut {NoStop}%
\bibitem [{\citenamefont {Zhou}\ \emph {et~al.}(2020)\citenamefont {Zhou},
  \citenamefont {Zeng},\ and\ \citenamefont {Liu}}]{singlezhou}%
  \BibitemOpen
  \bibfield  {author} {\bibinfo {author} {\bibfnamefont {Y.}~\bibnamefont
  {Zhou}}, \bibinfo {author} {\bibfnamefont {P.}~\bibnamefont {Zeng}},\ and\
  \bibinfo {author} {\bibfnamefont {Z.}~\bibnamefont {Liu}},\ }\bibfield
  {title} {\bibinfo {title} {Single-copies estimation of entanglement
  negativity},\ }\href {https://doi.org/10.1103/PhysRevLett.125.200502}
  {\bibfield  {journal} {\bibinfo  {journal} {Phys. Rev. Lett.}\ }\textbf
  {\bibinfo {volume} {125}},\ \bibinfo {pages} {200502} (\bibinfo {year}
  {2020})}\BibitemShut {NoStop}%
\bibitem [{\citenamefont {Li}\ \emph {et~al.}(2024)\citenamefont {Li},
  \citenamefont {Chen}, \citenamefont {Zhang}, \citenamefont {Hong},
  \citenamefont {Zhou}, \citenamefont {Chen}, \citenamefont {Li},\ and\
  \citenamefont {Guo}}]{my_p3ppt_paper}%
  \BibitemOpen
  \bibfield  {author} {\bibinfo {author} {\bibfnamefont {G.-C.}\ \bibnamefont
  {Li}}, \bibinfo {author} {\bibfnamefont {L.}~\bibnamefont {Chen}}, \bibinfo
  {author} {\bibfnamefont {S.-Q.}\ \bibnamefont {Zhang}}, \bibinfo {author}
  {\bibfnamefont {X.-S.}\ \bibnamefont {Hong}}, \bibinfo {author}
  {\bibfnamefont {Y.}~\bibnamefont {Zhou}}, \bibinfo {author} {\bibfnamefont
  {G.}~\bibnamefont {Chen}}, \bibinfo {author} {\bibfnamefont {C.-F.}\
  \bibnamefont {Li}},\ and\ \bibinfo {author} {\bibfnamefont {G.-C.}\
  \bibnamefont {Guo}},\ }\bibfield  {title} {\bibinfo {title} {Directly
  estimating mixed-state entanglement with bell measurement assistance},\
  }\href {https://arxiv.org/abs/2405.20696} {\bibfield  {journal} {\bibinfo
  {journal} {arXiv preprint arXiv:2405.20696}\ } (\bibinfo {year}
  {2024})}\BibitemShut {NoStop}%
\bibitem [{cod()}]{code_P}%
  \BibitemOpen
  \href@noop {} {}\bibinfo {note} {Codes for calculating potential magic
  resources for many graph states are at Github Repository
  \url{https://github.com/SuccessorOfPredecessor/Potential-Magic-in-Quantum-Graph-States}.
  Detailed Explanations are in the Sec.~XI of the Supplementary
  Material.}\BibitemShut {Stop}%
\bibitem [{\citenamefont {Chitambar}\ and\ \citenamefont
  {Gour}(2019)}]{chitambar2019quantum}%
  \BibitemOpen
  \bibfield  {author} {\bibinfo {author} {\bibfnamefont {E.}~\bibnamefont
  {Chitambar}}\ and\ \bibinfo {author} {\bibfnamefont {G.}~\bibnamefont
  {Gour}},\ }\bibfield  {title} {\bibinfo {title} {Quantum resource theories},\
  }\href {https://journals.aps.org/rmp/abstract/10.1103/RevModPhys.91.025001}
  {\bibfield  {journal} {\bibinfo  {journal} {Reviews of modern physics}\
  }\textbf {\bibinfo {volume} {91}},\ \bibinfo {pages} {025001} (\bibinfo
  {year} {2019})}\BibitemShut {NoStop}%
\bibitem [{\citenamefont {Leone}\ \emph {et~al.}(2024)\citenamefont {Leone},
  \citenamefont {Oliviero},\ and\ \citenamefont {Hamma}}]{leonelearning}%
  \BibitemOpen
  \bibfield  {author} {\bibinfo {author} {\bibfnamefont {L.}~\bibnamefont
  {Leone}}, \bibinfo {author} {\bibfnamefont {S.~F.}\ \bibnamefont
  {Oliviero}},\ and\ \bibinfo {author} {\bibfnamefont {A.}~\bibnamefont
  {Hamma}},\ }\bibfield  {title} {\bibinfo {title} {Learning t-doped stabilizer
  states},\ }\href {https://quantum-journal.org/papers/q-2024-05-27-1361}
  {\bibfield  {journal} {\bibinfo  {journal} {Quantum}\ }\textbf {\bibinfo
  {volume} {8}},\ \bibinfo {pages} {1361} (\bibinfo {year} {2024})}\BibitemShut
  {NoStop}%
\bibitem [{\citenamefont {Campbell}(2011)}]{campbell2011catalysis}%
  \BibitemOpen
  \bibfield  {author} {\bibinfo {author} {\bibfnamefont {E.~T.}\ \bibnamefont
  {Campbell}},\ }\bibfield  {title} {\bibinfo {title} {Catalysis and activation
  of magic states in fault-tolerant architectures},\ }\href
  {https://arxiv.org/pdf/1010.0104.pdf} {\bibfield  {journal} {\bibinfo
  {journal} {Physical Review A}\ }\textbf {\bibinfo {volume} {83}},\ \bibinfo
  {pages} {032317} (\bibinfo {year} {2011})}\BibitemShut {NoStop}%
\bibitem [{\citenamefont {Oliviero}\ \emph {et~al.}(2022)\citenamefont
  {Oliviero}, \citenamefont {Leone}, \citenamefont {Hamma},\ and\ \citenamefont
  {Lloyd}}]{oliviero2022measuring}%
  \BibitemOpen
  \bibfield  {author} {\bibinfo {author} {\bibfnamefont {S.~F.}\ \bibnamefont
  {Oliviero}}, \bibinfo {author} {\bibfnamefont {L.}~\bibnamefont {Leone}},
  \bibinfo {author} {\bibfnamefont {A.}~\bibnamefont {Hamma}},\ and\ \bibinfo
  {author} {\bibfnamefont {S.}~\bibnamefont {Lloyd}},\ }\bibfield  {title}
  {\bibinfo {title} {Measuring magic on a quantum processor},\ }\href
  {https://www.nature.com/articles/s41534-022-00666-5} {\bibfield  {journal}
  {\bibinfo  {journal} {npj Quantum Information}\ }\textbf {\bibinfo {volume}
  {8}},\ \bibinfo {pages} {148} (\bibinfo {year} {2022})}\BibitemShut {NoStop}%
\bibitem [{\citenamefont {Chen}\ \emph {et~al.}(2023)\citenamefont {Chen},
  \citenamefont {Yan},\ and\ \citenamefont {Zhou}}]{chen2023magic}%
  \BibitemOpen
  \bibfield  {author} {\bibinfo {author} {\bibfnamefont {J.}~\bibnamefont
  {Chen}}, \bibinfo {author} {\bibfnamefont {Y.}~\bibnamefont {Yan}},\ and\
  \bibinfo {author} {\bibfnamefont {Y.}~\bibnamefont {Zhou}},\ }\bibfield
  {title} {\bibinfo {title} {Magic of quantum hypergraph states},\ }\bibfield
  {journal} {\bibinfo  {journal} {arXiv preprint arXiv:2308.01886}\ }\href
  {https://doi.org/10.48550/arXiv.2308.01886} {10.48550/arXiv.2308.01886}
  (\bibinfo {year} {2023})\BibitemShut {NoStop}%
\bibitem [{\citenamefont {Haug}\ and\ \citenamefont
  {Piroli}(2023)}]{Haug2023MPS}%
  \BibitemOpen
  \bibfield  {author} {\bibinfo {author} {\bibfnamefont {T.}~\bibnamefont
  {Haug}}\ and\ \bibinfo {author} {\bibfnamefont {L.}~\bibnamefont {Piroli}},\
  }\bibfield  {title} {\bibinfo {title} {Quantifying nonstabilizerness of
  matrix product states},\ }\href {https://doi.org/10.1103/PhysRevB.107.035148}
  {\bibfield  {journal} {\bibinfo  {journal} {Phys. Rev. B}\ }\textbf {\bibinfo
  {volume} {107}},\ \bibinfo {pages} {035148} (\bibinfo {year}
  {2023})}\BibitemShut {NoStop}%
\bibitem [{\citenamefont {Lami}\ and\ \citenamefont
  {Collura}(2023)}]{Lami2023MPS}%
  \BibitemOpen
  \bibfield  {author} {\bibinfo {author} {\bibfnamefont {G.}~\bibnamefont
  {Lami}}\ and\ \bibinfo {author} {\bibfnamefont {M.}~\bibnamefont {Collura}},\
  }\bibfield  {title} {\bibinfo {title} {Nonstabilizerness via perfect pauli
  sampling of matrix product states},\ }\href
  {https://doi.org/10.1103/PhysRevLett.131.180401} {\bibfield  {journal}
  {\bibinfo  {journal} {Phys. Rev. Lett.}\ }\textbf {\bibinfo {volume} {131}},\
  \bibinfo {pages} {180401} (\bibinfo {year} {2023})}\BibitemShut {NoStop}%
\bibitem [{\citenamefont {Tirrito}\ \emph {et~al.}(2024)\citenamefont
  {Tirrito}, \citenamefont {Tarabunga}, \citenamefont {Lami}, \citenamefont
  {Chanda}, \citenamefont {Leone}, \citenamefont {Oliviero}, \citenamefont
  {Dalmonte}, \citenamefont {Collura},\ and\ \citenamefont {Hamma}}]{dalmonte}%
  \BibitemOpen
  \bibfield  {author} {\bibinfo {author} {\bibfnamefont {E.}~\bibnamefont
  {Tirrito}}, \bibinfo {author} {\bibfnamefont {P.~S.}\ \bibnamefont
  {Tarabunga}}, \bibinfo {author} {\bibfnamefont {G.}~\bibnamefont {Lami}},
  \bibinfo {author} {\bibfnamefont {T.}~\bibnamefont {Chanda}}, \bibinfo
  {author} {\bibfnamefont {L.}~\bibnamefont {Leone}}, \bibinfo {author}
  {\bibfnamefont {S.~F.}\ \bibnamefont {Oliviero}}, \bibinfo {author}
  {\bibfnamefont {M.}~\bibnamefont {Dalmonte}}, \bibinfo {author}
  {\bibfnamefont {M.}~\bibnamefont {Collura}},\ and\ \bibinfo {author}
  {\bibfnamefont {A.}~\bibnamefont {Hamma}},\ }\bibfield  {title} {\bibinfo
  {title} {Quantifying nonstabilizerness through entanglement spectrum
  flatness},\ }\href
  {https://journals.aps.org/pra/abstract/10.1103/PhysRevA.109.L040401}
  {\bibfield  {journal} {\bibinfo  {journal} {Physical Review A}\ }\textbf
  {\bibinfo {volume} {109}},\ \bibinfo {pages} {L040401} (\bibinfo {year}
  {2024})}\BibitemShut {NoStop}%
\bibitem [{\citenamefont {Cao}\ \emph {et~al.}(2024)\citenamefont {Cao},
  \citenamefont {Cheng}, \citenamefont {Hamma}, \citenamefont {Leone},
  \citenamefont {Munizzi},\ and\ \citenamefont {Oliviero}}]{cao}%
  \BibitemOpen
  \bibfield  {author} {\bibinfo {author} {\bibfnamefont {C.}~\bibnamefont
  {Cao}}, \bibinfo {author} {\bibfnamefont {G.}~\bibnamefont {Cheng}}, \bibinfo
  {author} {\bibfnamefont {A.}~\bibnamefont {Hamma}}, \bibinfo {author}
  {\bibfnamefont {L.}~\bibnamefont {Leone}}, \bibinfo {author} {\bibfnamefont
  {W.}~\bibnamefont {Munizzi}},\ and\ \bibinfo {author} {\bibfnamefont {S.~F.}\
  \bibnamefont {Oliviero}},\ }\bibfield  {title} {\bibinfo {title}
  {Gravitational back-reaction is the holographic dual of magic},\ }\href
  {https://arxiv.org/abs/2403.07056} {\bibfield  {journal} {\bibinfo  {journal}
  {arXiv preprint arXiv:2403.07056}\ } (\bibinfo {year} {2024})}\BibitemShut
  {NoStop}%
\bibitem [{\citenamefont {Leone}\ \emph {et~al.}(2023)\citenamefont {Leone},
  \citenamefont {Oliviero},\ and\ \citenamefont {Hamma}}]{leoneverification}%
  \BibitemOpen
  \bibfield  {author} {\bibinfo {author} {\bibfnamefont {L.}~\bibnamefont
  {Leone}}, \bibinfo {author} {\bibfnamefont {S.~F.}\ \bibnamefont
  {Oliviero}},\ and\ \bibinfo {author} {\bibfnamefont {A.}~\bibnamefont
  {Hamma}},\ }\bibfield  {title} {\bibinfo {title} {Nonstabilizerness
  determining the hardness of direct fidelity estimation},\ }\href
  {https://journals.aps.org/pra/abstract/10.1103/PhysRevA.107.022429}
  {\bibfield  {journal} {\bibinfo  {journal} {Physical Review A}\ }\textbf
  {\bibinfo {volume} {107}},\ \bibinfo {pages} {022429} (\bibinfo {year}
  {2023})}\BibitemShut {NoStop}%
\bibitem [{\citenamefont {Coppersmith}(2002)}]{coppersmith2002approximate}%
  \BibitemOpen
  \bibfield  {author} {\bibinfo {author} {\bibfnamefont {D.}~\bibnamefont
  {Coppersmith}},\ }\bibfield  {title} {\bibinfo {title} {An approximate
  fourier transform useful in quantum factoring},\ }\href
  {https://arxiv.org/abs/quant-ph/0201067} {\bibfield  {journal} {\bibinfo
  {journal} {arXiv preprint quant-ph/0201067}\ } (\bibinfo {year}
  {2002})}\BibitemShut {NoStop}%
\bibitem [{\citenamefont {Zhang}\ \emph {et~al.}(2016)\citenamefont {Zhang},
  \citenamefont {Huang}, \citenamefont {Liu}, \citenamefont {Li},\ and\
  \citenamefont {Guo}}]{zhang2016experimental}%
  \BibitemOpen
  \bibfield  {author} {\bibinfo {author} {\bibfnamefont {C.}~\bibnamefont
  {Zhang}}, \bibinfo {author} {\bibfnamefont {Y.-F.}\ \bibnamefont {Huang}},
  \bibinfo {author} {\bibfnamefont {B.-H.}\ \bibnamefont {Liu}}, \bibinfo
  {author} {\bibfnamefont {C.-F.}\ \bibnamefont {Li}},\ and\ \bibinfo {author}
  {\bibfnamefont {G.-C.}\ \bibnamefont {Guo}},\ }\bibfield  {title} {\bibinfo
  {title} {Experimental generation of a high-fidelity four-photon linear
  cluster state},\ }\href
  {https://journals.aps.org/pra/abstract/10.1103/PhysRevA.93.062329} {\bibfield
   {journal} {\bibinfo  {journal} {Physical Review A}\ }\textbf {\bibinfo
  {volume} {93}},\ \bibinfo {pages} {062329} (\bibinfo {year}
  {2016})}\BibitemShut {NoStop}%
\bibitem [{\citenamefont {Kurtsiefer}\ \emph {et~al.}(2001)\citenamefont
  {Kurtsiefer}, \citenamefont {Oberparleiter},\ and\ \citenamefont
  {Weinfurter}}]{kurtsiefer2001generation}%
  \BibitemOpen
  \bibfield  {author} {\bibinfo {author} {\bibfnamefont {C.}~\bibnamefont
  {Kurtsiefer}}, \bibinfo {author} {\bibfnamefont {M.}~\bibnamefont
  {Oberparleiter}},\ and\ \bibinfo {author} {\bibfnamefont {H.}~\bibnamefont
  {Weinfurter}},\ }\bibfield  {title} {\bibinfo {title} {Generation of
  correlated photon pairs in type-ii parametric down conversion—revisited},\
  }\href {https://www.tandfonline.com/doi/abs/10.1080/09500340108240902}
  {\bibfield  {journal} {\bibinfo  {journal} {Journal of Modern Optics}\
  }\textbf {\bibinfo {volume} {48}},\ \bibinfo {pages} {1997} (\bibinfo {year}
  {2001})}\BibitemShut {NoStop}%
\bibitem [{\citenamefont {Zhu}(2017)}]{zhu2017multiqubit}%
  \BibitemOpen
  \bibfield  {author} {\bibinfo {author} {\bibfnamefont {H.}~\bibnamefont
  {Zhu}},\ }\bibfield  {title} {\bibinfo {title} {Multiqubit clifford groups
  are unitary 3-designs},\ }\href
  {https://journals.aps.org/pra/abstract/10.1103/PhysRevA.96.062336} {\bibfield
   {journal} {\bibinfo  {journal} {Physical Review A}\ }\textbf {\bibinfo
  {volume} {96}},\ \bibinfo {pages} {062336} (\bibinfo {year}
  {2017})}\BibitemShut {NoStop}%
\bibitem [{\citenamefont {Li}\ \emph {et~al.}(2023)\citenamefont {Li},
  \citenamefont {Yin}, \citenamefont {Zhang}, \citenamefont {Chen},
  \citenamefont {Yin}, \citenamefont {Peng}, \citenamefont {Hong},
  \citenamefont {Chen}, \citenamefont {Li},\ and\ \citenamefont
  {Guo}}]{li2023experimental}%
  \BibitemOpen
  \bibfield  {author} {\bibinfo {author} {\bibfnamefont {G.-C.}\ \bibnamefont
  {Li}}, \bibinfo {author} {\bibfnamefont {Z.-Q.}\ \bibnamefont {Yin}},
  \bibinfo {author} {\bibfnamefont {W.-H.}\ \bibnamefont {Zhang}}, \bibinfo
  {author} {\bibfnamefont {L.}~\bibnamefont {Chen}}, \bibinfo {author}
  {\bibfnamefont {P.}~\bibnamefont {Yin}}, \bibinfo {author} {\bibfnamefont
  {X.-X.}\ \bibnamefont {Peng}}, \bibinfo {author} {\bibfnamefont {X.-S.}\
  \bibnamefont {Hong}}, \bibinfo {author} {\bibfnamefont {G.}~\bibnamefont
  {Chen}}, \bibinfo {author} {\bibfnamefont {C.-F.}\ \bibnamefont {Li}},\ and\
  \bibinfo {author} {\bibfnamefont {G.-C.}\ \bibnamefont {Guo}},\ }\bibfield
  {title} {\bibinfo {title} {Experimental full calibration of quantum devices
  in a semi-device-independent way},\ }\href
  {https://opg.optica.org/abstract.cfm?uri=optica-10-12-1723} {\bibfield
  {journal} {\bibinfo  {journal} {Optica}\ }\textbf {\bibinfo {volume} {10}},\
  \bibinfo {pages} {1723} (\bibinfo {year} {2023})}\BibitemShut {NoStop}%
\bibitem [{\citenamefont {Brydges}\ \emph {et~al.}(2019)\citenamefont
  {Brydges}, \citenamefont {Elben}, \citenamefont {Jurcevic}, \citenamefont
  {Vermersch}, \citenamefont {Maier}, \citenamefont {Lanyon}, \citenamefont
  {Zoller}, \citenamefont {Blatt},\ and\ \citenamefont
  {Roos}}]{brydges2019probing}%
  \BibitemOpen
  \bibfield  {author} {\bibinfo {author} {\bibfnamefont {T.}~\bibnamefont
  {Brydges}}, \bibinfo {author} {\bibfnamefont {A.}~\bibnamefont {Elben}},
  \bibinfo {author} {\bibfnamefont {P.}~\bibnamefont {Jurcevic}}, \bibinfo
  {author} {\bibfnamefont {B.}~\bibnamefont {Vermersch}}, \bibinfo {author}
  {\bibfnamefont {C.}~\bibnamefont {Maier}}, \bibinfo {author} {\bibfnamefont
  {B.~P.}\ \bibnamefont {Lanyon}}, \bibinfo {author} {\bibfnamefont
  {P.}~\bibnamefont {Zoller}}, \bibinfo {author} {\bibfnamefont
  {R.}~\bibnamefont {Blatt}},\ and\ \bibinfo {author} {\bibfnamefont {C.~F.}\
  \bibnamefont {Roos}},\ }\bibfield  {title} {\bibinfo {title} {Probing
  r{\'e}nyi entanglement entropy via randomized measurements},\ }\href
  {https://www.science.org/doi/abs/10.1126/science.aau4963} {\bibfield
  {journal} {\bibinfo  {journal} {Science}\ }\textbf {\bibinfo {volume}
  {364}},\ \bibinfo {pages} {260} (\bibinfo {year} {2019})}\BibitemShut
  {NoStop}%
\bibitem [{\citenamefont {Elben}\ \emph {et~al.}(2019)\citenamefont {Elben},
  \citenamefont {Vermersch}, \citenamefont {Roos},\ and\ \citenamefont
  {Zoller}}]{elben2019statistical}%
  \BibitemOpen
  \bibfield  {author} {\bibinfo {author} {\bibfnamefont {A.}~\bibnamefont
  {Elben}}, \bibinfo {author} {\bibfnamefont {B.}~\bibnamefont {Vermersch}},
  \bibinfo {author} {\bibfnamefont {C.~F.}\ \bibnamefont {Roos}},\ and\
  \bibinfo {author} {\bibfnamefont {P.}~\bibnamefont {Zoller}},\ }\bibfield
  {title} {\bibinfo {title} {Statistical correlations between locally
  randomized measurements: A toolbox for probing entanglement in many-body
  quantum states},\ }\href
  {https://journals.aps.org/pra/abstract/10.1103/PhysRevA.99.052323} {\bibfield
   {journal} {\bibinfo  {journal} {Physical Review A}\ }\textbf {\bibinfo
  {volume} {99}},\ \bibinfo {pages} {052323} (\bibinfo {year}
  {2019})}\BibitemShut {NoStop}%
\bibitem [{\citenamefont {Elben}\ \emph {et~al.}(2023)\citenamefont {Elben},
  \citenamefont {Flammia}, \citenamefont {Huang}, \citenamefont {Kueng},
  \citenamefont {Preskill}, \citenamefont {Vermersch},\ and\ \citenamefont
  {Zoller}}]{elben2023randomized}%
  \BibitemOpen
  \bibfield  {author} {\bibinfo {author} {\bibfnamefont {A.}~\bibnamefont
  {Elben}}, \bibinfo {author} {\bibfnamefont {S.~T.}\ \bibnamefont {Flammia}},
  \bibinfo {author} {\bibfnamefont {H.-Y.}\ \bibnamefont {Huang}}, \bibinfo
  {author} {\bibfnamefont {R.}~\bibnamefont {Kueng}}, \bibinfo {author}
  {\bibfnamefont {J.}~\bibnamefont {Preskill}}, \bibinfo {author}
  {\bibfnamefont {B.}~\bibnamefont {Vermersch}},\ and\ \bibinfo {author}
  {\bibfnamefont {P.}~\bibnamefont {Zoller}},\ }\bibfield  {title} {\bibinfo
  {title} {The randomized measurement toolbox},\ }\href
  {https://www.nature.com/articles/s42254-022-00535-2} {\bibfield  {journal}
  {\bibinfo  {journal} {Nature Reviews Physics}\ }\textbf {\bibinfo {volume}
  {5}},\ \bibinfo {pages} {9} (\bibinfo {year} {2023})}\BibitemShut {NoStop}%
\bibitem [{\citenamefont {Elben}\ \emph {et~al.}(2020)\citenamefont {Elben},
  \citenamefont {Kueng}, \citenamefont {Huang}, \citenamefont {van Bijnen},
  \citenamefont {Kokail}, \citenamefont {Dalmonte}, \citenamefont {Calabrese},
  \citenamefont {Kraus}, \citenamefont {Preskill}, \citenamefont {Zoller} \emph
  {et~al.}}]{elben2020mixed}%
  \BibitemOpen
  \bibfield  {author} {\bibinfo {author} {\bibfnamefont {A.}~\bibnamefont
  {Elben}}, \bibinfo {author} {\bibfnamefont {R.}~\bibnamefont {Kueng}},
  \bibinfo {author} {\bibfnamefont {H.-Y.~R.}\ \bibnamefont {Huang}}, \bibinfo
  {author} {\bibfnamefont {R.}~\bibnamefont {van Bijnen}}, \bibinfo {author}
  {\bibfnamefont {C.}~\bibnamefont {Kokail}}, \bibinfo {author} {\bibfnamefont
  {M.}~\bibnamefont {Dalmonte}}, \bibinfo {author} {\bibfnamefont
  {P.}~\bibnamefont {Calabrese}}, \bibinfo {author} {\bibfnamefont
  {B.}~\bibnamefont {Kraus}}, \bibinfo {author} {\bibfnamefont
  {J.}~\bibnamefont {Preskill}}, \bibinfo {author} {\bibfnamefont
  {P.}~\bibnamefont {Zoller}}, \emph {et~al.},\ }\bibfield  {title} {\bibinfo
  {title} {Mixed-state entanglement from local randomized measurements},\
  }\href {https://journals.aps.org/prl/abstract/10.1103/PhysRevLett.125.200501}
  {\bibfield  {journal} {\bibinfo  {journal} {Physical Review Letters}\
  }\textbf {\bibinfo {volume} {125}},\ \bibinfo {pages} {200501} (\bibinfo
  {year} {2020})}\BibitemShut {NoStop}%
\end{thebibliography}%


\begin{thebibliography}{44}%
\makeatletter
\providecommand \@ifxundefined [1]{%
 \@ifx{#1\undefined}
}%
\providecommand \@ifnum [1]{%
 \ifnum #1\expandafter \@firstoftwo
 \else \expandafter \@secondoftwo
 \fi
}%
\providecommand \@ifx [1]{%
 \ifx #1\expandafter \@firstoftwo
 \else \expandafter \@secondoftwo
 \fi
}%
\providecommand \natexlab [1]{#1}%
\providecommand \enquote  [1]{``#1''}%
\providecommand \bibnamefont  [1]{#1}%
\providecommand \bibfnamefont [1]{#1}%
\providecommand \citenamefont [1]{#1}%
\providecommand \href@noop [0]{\@secondoftwo}%
\providecommand \href [0]{\begingroup \@sanitize@url \@href}%
\providecommand \@href[1]{\@@startlink{#1}\@@href}%
\providecommand \@@href[1]{\endgroup#1\@@endlink}%
\providecommand \@sanitize@url [0]{\catcode `\\12\catcode `\$12\catcode
  `\&12\catcode `\#12\catcode `\^12\catcode `\_12\catcode `\%12\relax}%
\providecommand \@@startlink[1]{}%
\providecommand \@@endlink[0]{}%
\providecommand \url  [0]{\begingroup\@sanitize@url \@url }%
\providecommand \@url [1]{\endgroup\@href {#1}{\urlprefix }}%
\providecommand \urlprefix  [0]{URL }%
\providecommand \Eprint [0]{\href }%
\providecommand \doibase [0]{https://doi.org/}%
\providecommand \selectlanguage [0]{\@gobble}%
\providecommand \bibinfo  [0]{\@secondoftwo}%
\providecommand \bibfield  [0]{\@secondoftwo}%
\providecommand \translation [1]{[#1]}%
\providecommand \BibitemOpen [0]{}%
\providecommand \bibitemStop [0]{}%
\providecommand \bibitemNoStop [0]{.\EOS\space}%
\providecommand \EOS [0]{\spacefactor3000\relax}%
\providecommand \BibitemShut  [1]{\csname bibitem#1\endcsname}%
\let\auto@bib@innerbib\@empty
\bibitem [{\citenamefont {Leone}\ \emph {et~al.}(2022)\citenamefont {Leone},
  \citenamefont {Oliviero},\ and\ \citenamefont {Hamma}}]{leone2022stabilizer}%
  \BibitemOpen
  \bibfield  {author} {\bibinfo {author} {\bibfnamefont {L.}~\bibnamefont
  {Leone}}, \bibinfo {author} {\bibfnamefont {S.~F.}\ \bibnamefont
  {Oliviero}},\ and\ \bibinfo {author} {\bibfnamefont {A.}~\bibnamefont
  {Hamma}},\ }\bibfield  {title} {\bibinfo {title} {Stabilizer r{\'e}nyi
  entropy},\ }\href
  {https://journals.aps.org/prl/abstract/10.1103/PhysRevLett.128.050402}
  {\bibfield  {journal} {\bibinfo  {journal} {Phys. Rev. Lett.}\ }\textbf
  {\bibinfo {volume} {128}},\ \bibinfo {pages} {050402} (\bibinfo {year}
  {2022})}\BibitemShut {NoStop}%
\bibitem [{\citenamefont {Bravyi}\ and\ \citenamefont
  {Kitaev}(2005)}]{bravyi2005universal}%
  \BibitemOpen
  \bibfield  {author} {\bibinfo {author} {\bibfnamefont {S.}~\bibnamefont
  {Bravyi}}\ and\ \bibinfo {author} {\bibfnamefont {A.}~\bibnamefont
  {Kitaev}},\ }\bibfield  {title} {\bibinfo {title} {Universal quantum
  computation with ideal clifford gates and noisy ancillas},\ }\href
  {https://journals.aps.org/pra/abstract/10.1103/PhysRevA.71.022316} {\bibfield
   {journal} {\bibinfo  {journal} {Phys. Rev. A}\ }\textbf {\bibinfo {volume}
  {71}},\ \bibinfo {pages} {022316} (\bibinfo {year} {2005})}\BibitemShut
  {NoStop}%
\bibitem [{\citenamefont {Postler}\ \emph {et~al.}(2022)\citenamefont
  {Postler}, \citenamefont {Heu$\beta$en}, \citenamefont {Pogorelov},
  \citenamefont {Rispler}, \citenamefont {Feldker}, \citenamefont {Meth},
  \citenamefont {Marciniak}, \citenamefont {Stricker}, \citenamefont
  {Ringbauer}, \citenamefont {Blatt} \emph
  {et~al.}}]{postler2022demonstration}%
  \BibitemOpen
  \bibfield  {author} {\bibinfo {author} {\bibfnamefont {L.}~\bibnamefont
  {Postler}}, \bibinfo {author} {\bibfnamefont {S.}~\bibnamefont
  {Heu$\beta$en}}, \bibinfo {author} {\bibfnamefont {I.}~\bibnamefont
  {Pogorelov}}, \bibinfo {author} {\bibfnamefont {M.}~\bibnamefont {Rispler}},
  \bibinfo {author} {\bibfnamefont {T.}~\bibnamefont {Feldker}}, \bibinfo
  {author} {\bibfnamefont {M.}~\bibnamefont {Meth}}, \bibinfo {author}
  {\bibfnamefont {C.~D.}\ \bibnamefont {Marciniak}}, \bibinfo {author}
  {\bibfnamefont {R.}~\bibnamefont {Stricker}}, \bibinfo {author}
  {\bibfnamefont {M.}~\bibnamefont {Ringbauer}}, \bibinfo {author}
  {\bibfnamefont {R.}~\bibnamefont {Blatt}}, \emph {et~al.},\ }\bibfield
  {title} {\bibinfo {title} {Demonstration of fault-tolerant universal quantum
  gate operations},\ }\href
  {https://www.nature.com/articles/s41586-022-04721-1} {\bibfield  {journal}
  {\bibinfo  {journal} {Nature}\ }\textbf {\bibinfo {volume} {605}},\ \bibinfo
  {pages} {675} (\bibinfo {year} {2022})}\BibitemShut {NoStop}%
\bibitem [{\citenamefont {AI}(2023)}]{google2023suppressing}%
  \BibitemOpen
  \bibfield  {author} {\bibinfo {author} {\bibfnamefont {G.~Q.}\ \bibnamefont
  {AI}},\ }\bibfield  {title} {\bibinfo {title} {Suppressing quantum errors by
  scaling a surface code logical qubit},\ }\href
  {https://www.nature.com/articles/s41586-022-05434-1} {\bibfield  {journal}
  {\bibinfo  {journal} {Nature}\ }\textbf {\bibinfo {volume} {614}},\ \bibinfo
  {pages} {676} (\bibinfo {year} {2023})}\BibitemShut {NoStop}%
\bibitem [{\citenamefont {Gupta}\ \emph {et~al.}(2024)\citenamefont {Gupta},
  \citenamefont {Sundaresan}, \citenamefont {Alexander}, \citenamefont {Wood},
  \citenamefont {Merkel}, \citenamefont {Healy}, \citenamefont {Hillenbrand},
  \citenamefont {Jochym-O’Connor}, \citenamefont {Wootton}, \citenamefont
  {Yoder} \emph {et~al.}}]{gupta2024encoding}%
  \BibitemOpen
  \bibfield  {author} {\bibinfo {author} {\bibfnamefont {R.~S.}\ \bibnamefont
  {Gupta}}, \bibinfo {author} {\bibfnamefont {N.}~\bibnamefont {Sundaresan}},
  \bibinfo {author} {\bibfnamefont {T.}~\bibnamefont {Alexander}}, \bibinfo
  {author} {\bibfnamefont {C.~J.}\ \bibnamefont {Wood}}, \bibinfo {author}
  {\bibfnamefont {S.~T.}\ \bibnamefont {Merkel}}, \bibinfo {author}
  {\bibfnamefont {M.~B.}\ \bibnamefont {Healy}}, \bibinfo {author}
  {\bibfnamefont {M.}~\bibnamefont {Hillenbrand}}, \bibinfo {author}
  {\bibfnamefont {T.}~\bibnamefont {Jochym-O’Connor}}, \bibinfo {author}
  {\bibfnamefont {J.~R.}\ \bibnamefont {Wootton}}, \bibinfo {author}
  {\bibfnamefont {T.~J.}\ \bibnamefont {Yoder}}, \emph {et~al.},\ }\bibfield
  {title} {\bibinfo {title} {Encoding a magic state with beyond break-even
  fidelity},\ }\href {https://www.nature.com/articles/s41586-023-06846-3}
  {\bibfield  {journal} {\bibinfo  {journal} {Nature}\ }\textbf {\bibinfo
  {volume} {625}},\ \bibinfo {pages} {259} (\bibinfo {year}
  {2024})}\BibitemShut {NoStop}%
\bibitem [{\citenamefont {Veitch}\ \emph {et~al.}(2014)\citenamefont {Veitch},
  \citenamefont {Mousavian}, \citenamefont {Gottesman},\ and\ \citenamefont
  {Emerson}}]{veitch2014resource}%
  \BibitemOpen
  \bibfield  {author} {\bibinfo {author} {\bibfnamefont {V.}~\bibnamefont
  {Veitch}}, \bibinfo {author} {\bibfnamefont {S.~H.}\ \bibnamefont
  {Mousavian}}, \bibinfo {author} {\bibfnamefont {D.}~\bibnamefont
  {Gottesman}},\ and\ \bibinfo {author} {\bibfnamefont {J.}~\bibnamefont
  {Emerson}},\ }\bibfield  {title} {\bibinfo {title} {The resource theory of
  stabilizer quantum computation},\ }\href
  {https://iopscience.iop.org/article/10.1088/1367-2630/16/1/013009/meta}
  {\bibfield  {journal} {\bibinfo  {journal} {New Journal of Physics}\ }\textbf
  {\bibinfo {volume} {16}},\ \bibinfo {pages} {013009} (\bibinfo {year}
  {2014})}\BibitemShut {NoStop}%
\bibitem [{\citenamefont {Leone}\ and\ \citenamefont {Bittel}(2024)}]{bittel}%
  \BibitemOpen
  \bibfield  {author} {\bibinfo {author} {\bibfnamefont {L.}~\bibnamefont
  {Leone}}\ and\ \bibinfo {author} {\bibfnamefont {L.}~\bibnamefont {Bittel}},\
  }\bibfield  {title} {\bibinfo {title} {Stabilizer entropies are monotones for
  magic-state resource theory},\ }\href
  {https://journals.aps.org/pra/abstract/10.1103/PhysRevA.110.L040403}
  {\bibfield  {journal} {\bibinfo  {journal} {Phys. Rev. A}\ }\textbf {\bibinfo
  {volume} {110}},\ \bibinfo {pages} {L040403} (\bibinfo {year}
  {2024})}\BibitemShut {NoStop}%
\bibitem [{\citenamefont {Souza}\ \emph {et~al.}(2011)\citenamefont {Souza},
  \citenamefont {Zhang}, \citenamefont {Ryan},\ and\ \citenamefont
  {Laflamme}}]{souza2011experimental}%
  \BibitemOpen
  \bibfield  {author} {\bibinfo {author} {\bibfnamefont {A.~M.}\ \bibnamefont
  {Souza}}, \bibinfo {author} {\bibfnamefont {J.}~\bibnamefont {Zhang}},
  \bibinfo {author} {\bibfnamefont {C.~A.}\ \bibnamefont {Ryan}},\ and\
  \bibinfo {author} {\bibfnamefont {R.}~\bibnamefont {Laflamme}},\ }\bibfield
  {title} {\bibinfo {title} {Experimental magic state distillation for
  fault-tolerant quantum computing},\ }\href
  {https://www.nature.com/articles/ncomms1166} {\bibfield  {journal} {\bibinfo
  {journal} {Nature communications}\ }\textbf {\bibinfo {volume} {2}},\
  \bibinfo {pages} {169} (\bibinfo {year} {2011})}\BibitemShut {NoStop}%
\bibitem [{\citenamefont {Knill}(2005)}]{knill2005quantum}%
  \BibitemOpen
  \bibfield  {author} {\bibinfo {author} {\bibfnamefont {E.}~\bibnamefont
  {Knill}},\ }\bibfield  {title} {\bibinfo {title} {Quantum computing with
  realistically noisy devices},\ }\href
  {https://www.nature.com/articles/nature03350} {\bibfield  {journal} {\bibinfo
   {journal} {Nature}\ }\textbf {\bibinfo {volume} {434}},\ \bibinfo {pages}
  {39} (\bibinfo {year} {2005})}\BibitemShut {NoStop}%
\bibitem [{\citenamefont {Chitambar}\ and\ \citenamefont
  {Gour}(2019)}]{chitambar2019quantum}%
  \BibitemOpen
  \bibfield  {author} {\bibinfo {author} {\bibfnamefont {E.}~\bibnamefont
  {Chitambar}}\ and\ \bibinfo {author} {\bibfnamefont {G.}~\bibnamefont
  {Gour}},\ }\bibfield  {title} {\bibinfo {title} {Quantum resource theories},\
  }\href {https://journals.aps.org/rmp/abstract/10.1103/RevModPhys.91.025001}
  {\bibfield  {journal} {\bibinfo  {journal} {Reviews of modern physics}\
  }\textbf {\bibinfo {volume} {91}},\ \bibinfo {pages} {025001} (\bibinfo
  {year} {2019})}\BibitemShut {NoStop}%
\bibitem [{\citenamefont {Regula}\ and\ \citenamefont
  {Takagi}(2021)}]{regula2021fundamental}%
  \BibitemOpen
  \bibfield  {author} {\bibinfo {author} {\bibfnamefont {B.}~\bibnamefont
  {Regula}}\ and\ \bibinfo {author} {\bibfnamefont {R.}~\bibnamefont
  {Takagi}},\ }\bibfield  {title} {\bibinfo {title} {Fundamental limitations on
  distillation of quantum channel resources},\ }\href
  {https://www.nature.com/articles/s41467-021-24699-0} {\bibfield  {journal}
  {\bibinfo  {journal} {Nature Communications}\ }\textbf {\bibinfo {volume}
  {12}},\ \bibinfo {pages} {4411} (\bibinfo {year} {2021})}\BibitemShut
  {NoStop}%
\bibitem [{\citenamefont {Aaronson}\ and\ \citenamefont
  {Gottesman}(2004)}]{aaronson}%
  \BibitemOpen
  \bibfield  {author} {\bibinfo {author} {\bibfnamefont {S.}~\bibnamefont
  {Aaronson}}\ and\ \bibinfo {author} {\bibfnamefont {D.}~\bibnamefont
  {Gottesman}},\ }\bibfield  {title} {\bibinfo {title} {Improved simulation of
  stabilizer circuits},\ }\href
  {https://journals.aps.org/pra/abstract/10.1103/PhysRevA.70.052328} {\bibfield
   {journal} {\bibinfo  {journal} {Phys. Rev. A}\ }\textbf {\bibinfo {volume}
  {70}},\ \bibinfo {pages} {052328} (\bibinfo {year} {2004})}\BibitemShut
  {NoStop}%
\bibitem [{\citenamefont {Bravyi}\ and\ \citenamefont
  {Gosset}(2016)}]{bravyi2016improved}%
  \BibitemOpen
  \bibfield  {author} {\bibinfo {author} {\bibfnamefont {S.}~\bibnamefont
  {Bravyi}}\ and\ \bibinfo {author} {\bibfnamefont {D.}~\bibnamefont
  {Gosset}},\ }\bibfield  {title} {\bibinfo {title} {Improved classical
  simulation of quantum circuits dominated by clifford gates},\ }\href
  {https://journals.aps.org/prl/abstract/10.1103/PhysRevLett.116.250501}
  {\bibfield  {journal} {\bibinfo  {journal} {Phys. Rev. Lett.}\ }\textbf
  {\bibinfo {volume} {116}},\ \bibinfo {pages} {250501} (\bibinfo {year}
  {2016})}\BibitemShut {NoStop}%
\bibitem [{\citenamefont {Bravyi}\ \emph {et~al.}(2016)\citenamefont {Bravyi},
  \citenamefont {Smith},\ and\ \citenamefont {Smolin}}]{bravyi2016trading}%
  \BibitemOpen
  \bibfield  {author} {\bibinfo {author} {\bibfnamefont {S.}~\bibnamefont
  {Bravyi}}, \bibinfo {author} {\bibfnamefont {G.}~\bibnamefont {Smith}},\ and\
  \bibinfo {author} {\bibfnamefont {J.~A.}\ \bibnamefont {Smolin}},\ }\bibfield
   {title} {\bibinfo {title} {Trading classical and quantum computational
  resources},\ }\href
  {https://journals.aps.org/prx/abstract/10.1103/PhysRevX.6.021043} {\bibfield
  {journal} {\bibinfo  {journal} {Phys. Rev. X}\ }\textbf {\bibinfo {volume}
  {6}},\ \bibinfo {pages} {021043} (\bibinfo {year} {2016})}\BibitemShut
  {NoStop}%
\bibitem [{\citenamefont {Howard}\ and\ \citenamefont
  {Campbell}(2017)}]{howard2017application}%
  \BibitemOpen
  \bibfield  {author} {\bibinfo {author} {\bibfnamefont {M.}~\bibnamefont
  {Howard}}\ and\ \bibinfo {author} {\bibfnamefont {E.}~\bibnamefont
  {Campbell}},\ }\bibfield  {title} {\bibinfo {title} {Application of a
  resource theory for magic states to fault-tolerant quantum computing},\
  }\href {https://journals.aps.org/prl/abstract/10.1103/PhysRevLett.118.090501}
  {\bibfield  {journal} {\bibinfo  {journal} {Phys. Rev. Lett.}\ }\textbf
  {\bibinfo {volume} {118}},\ \bibinfo {pages} {090501} (\bibinfo {year}
  {2017})}\BibitemShut {NoStop}%
\bibitem [{\citenamefont {Wang}\ \emph {et~al.}(2020)\citenamefont {Wang},
  \citenamefont {Wilde},\ and\ \citenamefont {Su}}]{wang2020efficiently}%
  \BibitemOpen
  \bibfield  {author} {\bibinfo {author} {\bibfnamefont {X.}~\bibnamefont
  {Wang}}, \bibinfo {author} {\bibfnamefont {M.~M.}\ \bibnamefont {Wilde}},\
  and\ \bibinfo {author} {\bibfnamefont {Y.}~\bibnamefont {Su}},\ }\bibfield
  {title} {\bibinfo {title} {Efficiently computable bounds for magic state
  distillation},\ }\href
  {https://journals.aps.org/prl/abstract/10.1103/PhysRevLett.124.090505}
  {\bibfield  {journal} {\bibinfo  {journal} {Phys. Rev. Lett.}\ }\textbf
  {\bibinfo {volume} {124}},\ \bibinfo {pages} {090505} (\bibinfo {year}
  {2020})}\BibitemShut {NoStop}%
\bibitem [{\citenamefont {Beverland}\ \emph {et~al.}(2020)\citenamefont
  {Beverland}, \citenamefont {Campbell}, \citenamefont {Howard},\ and\
  \citenamefont {Kliuchnikov}}]{beverland2020lower}%
  \BibitemOpen
  \bibfield  {author} {\bibinfo {author} {\bibfnamefont {M.}~\bibnamefont
  {Beverland}}, \bibinfo {author} {\bibfnamefont {E.}~\bibnamefont {Campbell}},
  \bibinfo {author} {\bibfnamefont {M.}~\bibnamefont {Howard}},\ and\ \bibinfo
  {author} {\bibfnamefont {V.}~\bibnamefont {Kliuchnikov}},\ }\bibfield
  {title} {\bibinfo {title} {Lower bounds on the non-clifford resources for
  quantum computations},\ }\href
  {https://iopscience.iop.org/article/10.1088/2058-9565/ab8963/meta} {\bibfield
   {journal} {\bibinfo  {journal} {Quantum Science and Technology}\ }\textbf
  {\bibinfo {volume} {5}},\ \bibinfo {pages} {035009} (\bibinfo {year}
  {2020})}\BibitemShut {NoStop}%
\bibitem [{\citenamefont {Hahn}\ \emph {et~al.}(2022)\citenamefont {Hahn},
  \citenamefont {Ferraro}, \citenamefont {Hultquist}, \citenamefont {Ferrini},\
  and\ \citenamefont {Garc{\'\i}a-{\'A}lvarez}}]{hahn2022quantifying}%
  \BibitemOpen
  \bibfield  {author} {\bibinfo {author} {\bibfnamefont {O.}~\bibnamefont
  {Hahn}}, \bibinfo {author} {\bibfnamefont {A.}~\bibnamefont {Ferraro}},
  \bibinfo {author} {\bibfnamefont {L.}~\bibnamefont {Hultquist}}, \bibinfo
  {author} {\bibfnamefont {G.}~\bibnamefont {Ferrini}},\ and\ \bibinfo {author}
  {\bibfnamefont {L.}~\bibnamefont {Garc{\'\i}a-{\'A}lvarez}},\ }\bibfield
  {title} {\bibinfo {title} {Quantifying qubit magic resource with
  gottesman-kitaev-preskill encoding},\ }\href
  {https://journals.aps.org/prl/abstract/10.1103/PhysRevLett.128.210502}
  {\bibfield  {journal} {\bibinfo  {journal} {Phys. Rev. Lett.}\ }\textbf
  {\bibinfo {volume} {128}},\ \bibinfo {pages} {210502} (\bibinfo {year}
  {2022})}\BibitemShut {NoStop}%
\bibitem [{\citenamefont {Leone}\ \emph {et~al.}(2024)\citenamefont {Leone},
  \citenamefont {Oliviero},\ and\ \citenamefont {Hamma}}]{leonelearning}%
  \BibitemOpen
  \bibfield  {author} {\bibinfo {author} {\bibfnamefont {L.}~\bibnamefont
  {Leone}}, \bibinfo {author} {\bibfnamefont {S.~F.}\ \bibnamefont
  {Oliviero}},\ and\ \bibinfo {author} {\bibfnamefont {A.}~\bibnamefont
  {Hamma}},\ }\bibfield  {title} {\bibinfo {title} {Learning t-doped stabilizer
  states},\ }\href {https://quantum-journal.org/papers/q-2024-05-27-1361}
  {\bibfield  {journal} {\bibinfo  {journal} {Quantum}\ }\textbf {\bibinfo
  {volume} {8}},\ \bibinfo {pages} {1361} (\bibinfo {year} {2024})}\BibitemShut
  {NoStop}%
\bibitem [{\citenamefont {Campbell}(2011)}]{campbell2011catalysis}%
  \BibitemOpen
  \bibfield  {author} {\bibinfo {author} {\bibfnamefont {E.~T.}\ \bibnamefont
  {Campbell}},\ }\bibfield  {title} {\bibinfo {title} {Catalysis and activation
  of magic states in fault-tolerant architectures},\ }\href
  {https://arxiv.org/pdf/1010.0104.pdf} {\bibfield  {journal} {\bibinfo
  {journal} {Physical Review A}\ }\textbf {\bibinfo {volume} {83}},\ \bibinfo
  {pages} {032317} (\bibinfo {year} {2011})}\BibitemShut {NoStop}%
\bibitem [{\citenamefont {Oliviero}\ \emph {et~al.}(2022)\citenamefont
  {Oliviero}, \citenamefont {Leone}, \citenamefont {Hamma},\ and\ \citenamefont
  {Lloyd}}]{oliviero2022measuring}%
  \BibitemOpen
  \bibfield  {author} {\bibinfo {author} {\bibfnamefont {S.~F.}\ \bibnamefont
  {Oliviero}}, \bibinfo {author} {\bibfnamefont {L.}~\bibnamefont {Leone}},
  \bibinfo {author} {\bibfnamefont {A.}~\bibnamefont {Hamma}},\ and\ \bibinfo
  {author} {\bibfnamefont {S.}~\bibnamefont {Lloyd}},\ }\bibfield  {title}
  {\bibinfo {title} {Measuring magic on a quantum processor},\ }\href
  {https://www.nature.com/articles/s41534-022-00666-5} {\bibfield  {journal}
  {\bibinfo  {journal} {npj Quantum Information}\ }\textbf {\bibinfo {volume}
  {8}},\ \bibinfo {pages} {148} (\bibinfo {year} {2022})}\BibitemShut {NoStop}%
\bibitem [{\citenamefont {Chen}\ \emph {et~al.}(2023)\citenamefont {Chen},
  \citenamefont {Yan},\ and\ \citenamefont {Zhou}}]{chen2023magic}%
  \BibitemOpen
  \bibfield  {author} {\bibinfo {author} {\bibfnamefont {J.}~\bibnamefont
  {Chen}}, \bibinfo {author} {\bibfnamefont {Y.}~\bibnamefont {Yan}},\ and\
  \bibinfo {author} {\bibfnamefont {Y.}~\bibnamefont {Zhou}},\ }\bibfield
  {title} {\bibinfo {title} {Magic of quantum hypergraph states},\ }\bibfield
  {journal} {\bibinfo  {journal} {arXiv preprint arXiv:2308.01886}\ }\href
  {https://doi.org/10.48550/arXiv.2308.01886} {10.48550/arXiv.2308.01886}
  (\bibinfo {year} {2023})\BibitemShut {NoStop}%
\bibitem [{\citenamefont {Haug}\ and\ \citenamefont
  {Piroli}(2023)}]{Haug2023MPS}%
  \BibitemOpen
  \bibfield  {author} {\bibinfo {author} {\bibfnamefont {T.}~\bibnamefont
  {Haug}}\ and\ \bibinfo {author} {\bibfnamefont {L.}~\bibnamefont {Piroli}},\
  }\bibfield  {title} {\bibinfo {title} {Quantifying nonstabilizerness of
  matrix product states},\ }\href {https://doi.org/10.1103/PhysRevB.107.035148}
  {\bibfield  {journal} {\bibinfo  {journal} {Phys. Rev. B}\ }\textbf {\bibinfo
  {volume} {107}},\ \bibinfo {pages} {035148} (\bibinfo {year}
  {2023})}\BibitemShut {NoStop}%
\bibitem [{\citenamefont {Lami}\ and\ \citenamefont
  {Collura}(2023)}]{Lami2023MPS}%
  \BibitemOpen
  \bibfield  {author} {\bibinfo {author} {\bibfnamefont {G.}~\bibnamefont
  {Lami}}\ and\ \bibinfo {author} {\bibfnamefont {M.}~\bibnamefont {Collura}},\
  }\bibfield  {title} {\bibinfo {title} {Nonstabilizerness via perfect pauli
  sampling of matrix product states},\ }\href
  {https://doi.org/10.1103/PhysRevLett.131.180401} {\bibfield  {journal}
  {\bibinfo  {journal} {Phys. Rev. Lett.}\ }\textbf {\bibinfo {volume} {131}},\
  \bibinfo {pages} {180401} (\bibinfo {year} {2023})}\BibitemShut {NoStop}%
\bibitem [{\citenamefont {Tirrito}\ \emph {et~al.}(2024)\citenamefont
  {Tirrito}, \citenamefont {Tarabunga}, \citenamefont {Lami}, \citenamefont
  {Chanda}, \citenamefont {Leone}, \citenamefont {Oliviero}, \citenamefont
  {Dalmonte}, \citenamefont {Collura},\ and\ \citenamefont {Hamma}}]{dalmonte}%
  \BibitemOpen
  \bibfield  {author} {\bibinfo {author} {\bibfnamefont {E.}~\bibnamefont
  {Tirrito}}, \bibinfo {author} {\bibfnamefont {P.~S.}\ \bibnamefont
  {Tarabunga}}, \bibinfo {author} {\bibfnamefont {G.}~\bibnamefont {Lami}},
  \bibinfo {author} {\bibfnamefont {T.}~\bibnamefont {Chanda}}, \bibinfo
  {author} {\bibfnamefont {L.}~\bibnamefont {Leone}}, \bibinfo {author}
  {\bibfnamefont {S.~F.}\ \bibnamefont {Oliviero}}, \bibinfo {author}
  {\bibfnamefont {M.}~\bibnamefont {Dalmonte}}, \bibinfo {author}
  {\bibfnamefont {M.}~\bibnamefont {Collura}},\ and\ \bibinfo {author}
  {\bibfnamefont {A.}~\bibnamefont {Hamma}},\ }\bibfield  {title} {\bibinfo
  {title} {Quantifying nonstabilizerness through entanglement spectrum
  flatness},\ }\href
  {https://journals.aps.org/pra/abstract/10.1103/PhysRevA.109.L040401}
  {\bibfield  {journal} {\bibinfo  {journal} {Physical Review A}\ }\textbf
  {\bibinfo {volume} {109}},\ \bibinfo {pages} {L040401} (\bibinfo {year}
  {2024})}\BibitemShut {NoStop}%
\bibitem [{\citenamefont {Cao}\ \emph {et~al.}(2024)\citenamefont {Cao},
  \citenamefont {Cheng}, \citenamefont {Hamma}, \citenamefont {Leone},
  \citenamefont {Munizzi},\ and\ \citenamefont {Oliviero}}]{cao}%
  \BibitemOpen
  \bibfield  {author} {\bibinfo {author} {\bibfnamefont {C.}~\bibnamefont
  {Cao}}, \bibinfo {author} {\bibfnamefont {G.}~\bibnamefont {Cheng}}, \bibinfo
  {author} {\bibfnamefont {A.}~\bibnamefont {Hamma}}, \bibinfo {author}
  {\bibfnamefont {L.}~\bibnamefont {Leone}}, \bibinfo {author} {\bibfnamefont
  {W.}~\bibnamefont {Munizzi}},\ and\ \bibinfo {author} {\bibfnamefont {S.~F.}\
  \bibnamefont {Oliviero}},\ }\bibfield  {title} {\bibinfo {title}
  {Gravitational back-reaction is the holographic dual of magic},\ }\href
  {https://arxiv.org/abs/2403.07056} {\bibfield  {journal} {\bibinfo  {journal}
  {arXiv preprint arXiv:2403.07056}\ } (\bibinfo {year} {2024})}\BibitemShut
  {NoStop}%
\bibitem [{\citenamefont {Leone}\ \emph {et~al.}(2023)\citenamefont {Leone},
  \citenamefont {Oliviero},\ and\ \citenamefont {Hamma}}]{leoneverification}%
  \BibitemOpen
  \bibfield  {author} {\bibinfo {author} {\bibfnamefont {L.}~\bibnamefont
  {Leone}}, \bibinfo {author} {\bibfnamefont {S.~F.}\ \bibnamefont
  {Oliviero}},\ and\ \bibinfo {author} {\bibfnamefont {A.}~\bibnamefont
  {Hamma}},\ }\bibfield  {title} {\bibinfo {title} {Nonstabilizerness
  determining the hardness of direct fidelity estimation},\ }\href
  {https://journals.aps.org/pra/abstract/10.1103/PhysRevA.107.022429}
  {\bibfield  {journal} {\bibinfo  {journal} {Physical Review A}\ }\textbf
  {\bibinfo {volume} {107}},\ \bibinfo {pages} {022429} (\bibinfo {year}
  {2023})}\BibitemShut {NoStop}%
\bibitem [{inp()}]{inpreparation}%
  \BibitemOpen
  \href@noop {} {}\bibinfo {note} {A. Hamma et al., in
  preparation.}\BibitemShut {Stop}%
\bibitem [{\citenamefont {Danos}\ \emph {et~al.}(2005)\citenamefont {Danos},
  \citenamefont {Kashefi},\ and\ \citenamefont
  {Panangaden}}]{danos2005parsimonious}%
  \BibitemOpen
  \bibfield  {author} {\bibinfo {author} {\bibfnamefont {V.}~\bibnamefont
  {Danos}}, \bibinfo {author} {\bibfnamefont {E.}~\bibnamefont {Kashefi}},\
  and\ \bibinfo {author} {\bibfnamefont {P.}~\bibnamefont {Panangaden}},\
  }\bibfield  {title} {\bibinfo {title} {Parsimonious and robust realizations
  of unitary maps in the one-way model},\ }\href
  {https://journals.aps.org/pra/abstract/10.1103/PhysRevA.72.064301} {\bibfield
   {journal} {\bibinfo  {journal} {Phys. Rev. A}\ }\textbf {\bibinfo {volume}
  {72}},\ \bibinfo {pages} {064301} (\bibinfo {year} {2005})}\BibitemShut
  {NoStop}%
\bibitem [{\citenamefont {Danos}\ \emph {et~al.}(2007)\citenamefont {Danos},
  \citenamefont {Kashefi},\ and\ \citenamefont
  {Panangaden}}]{danos2007measurement}%
  \BibitemOpen
  \bibfield  {author} {\bibinfo {author} {\bibfnamefont {V.}~\bibnamefont
  {Danos}}, \bibinfo {author} {\bibfnamefont {E.}~\bibnamefont {Kashefi}},\
  and\ \bibinfo {author} {\bibfnamefont {P.}~\bibnamefont {Panangaden}},\
  }\bibfield  {title} {\bibinfo {title} {The measurement calculus},\ }\href
  {https://dl.acm.org/doi/abs/10.1145/1219092.1219096} {\bibfield  {journal}
  {\bibinfo  {journal} {Journal of the ACM (JACM)}\ }\textbf {\bibinfo {volume}
  {54}},\ \bibinfo {pages} {8} (\bibinfo {year} {2007})}\BibitemShut {NoStop}%
\bibitem [{\citenamefont {Nam}\ \emph {et~al.}(2020)\citenamefont {Nam},
  \citenamefont {Su},\ and\ \citenamefont {Maslov}}]{nam2020approximate}%
  \BibitemOpen
  \bibfield  {author} {\bibinfo {author} {\bibfnamefont {Y.}~\bibnamefont
  {Nam}}, \bibinfo {author} {\bibfnamefont {Y.}~\bibnamefont {Su}},\ and\
  \bibinfo {author} {\bibfnamefont {D.}~\bibnamefont {Maslov}},\ }\bibfield
  {title} {\bibinfo {title} {Approximate quantum fourier transform with
  o(nlog(n)) t gates},\ }\href
  {https://www.nature.com/articles/s41534-020-0257-5} {\bibfield  {journal}
  {\bibinfo  {journal} {NPJ Quantum Information}\ }\textbf {\bibinfo {volume}
  {6}},\ \bibinfo {pages} {26} (\bibinfo {year} {2020})}\BibitemShut {NoStop}%
\bibitem [{\citenamefont {Coppersmith}(2002)}]{coppersmith2002approximate}%
  \BibitemOpen
  \bibfield  {author} {\bibinfo {author} {\bibfnamefont {D.}~\bibnamefont
  {Coppersmith}},\ }\bibfield  {title} {\bibinfo {title} {An approximate
  fourier transform useful in quantum factoring},\ }\href
  {https://arxiv.org/abs/quant-ph/0201067} {\bibfield  {journal} {\bibinfo
  {journal} {arXiv preprint quant-ph/0201067}\ } (\bibinfo {year}
  {2002})}\BibitemShut {NoStop}%
\bibitem [{cod()}]{code_P}%
  \BibitemOpen
  \href@noop {} {}\bibinfo {note} {Codes for calculating potential magic
  resources for many graph states are at Github Repository
  \url{https://github.com/SuccessorOfPredecessor/Potential-Magic-in-Quantum-Graph-States}.
  Detailed Explanations are in the Sec.~XI of the Supplementary
  Material.}\BibitemShut {Stop}%
\bibitem [{\citenamefont {Zhang}\ \emph {et~al.}(2016)\citenamefont {Zhang},
  \citenamefont {Huang}, \citenamefont {Liu}, \citenamefont {Li},\ and\
  \citenamefont {Guo}}]{zhang2016experimental}%
  \BibitemOpen
  \bibfield  {author} {\bibinfo {author} {\bibfnamefont {C.}~\bibnamefont
  {Zhang}}, \bibinfo {author} {\bibfnamefont {Y.-F.}\ \bibnamefont {Huang}},
  \bibinfo {author} {\bibfnamefont {B.-H.}\ \bibnamefont {Liu}}, \bibinfo
  {author} {\bibfnamefont {C.-F.}\ \bibnamefont {Li}},\ and\ \bibinfo {author}
  {\bibfnamefont {G.-C.}\ \bibnamefont {Guo}},\ }\bibfield  {title} {\bibinfo
  {title} {Experimental generation of a high-fidelity four-photon linear
  cluster state},\ }\href
  {https://journals.aps.org/pra/abstract/10.1103/PhysRevA.93.062329} {\bibfield
   {journal} {\bibinfo  {journal} {Physical Review A}\ }\textbf {\bibinfo
  {volume} {93}},\ \bibinfo {pages} {062329} (\bibinfo {year}
  {2016})}\BibitemShut {NoStop}%
\bibitem [{\citenamefont {Kurtsiefer}\ \emph {et~al.}(2001)\citenamefont
  {Kurtsiefer}, \citenamefont {Oberparleiter},\ and\ \citenamefont
  {Weinfurter}}]{kurtsiefer2001generation}%
  \BibitemOpen
  \bibfield  {author} {\bibinfo {author} {\bibfnamefont {C.}~\bibnamefont
  {Kurtsiefer}}, \bibinfo {author} {\bibfnamefont {M.}~\bibnamefont
  {Oberparleiter}},\ and\ \bibinfo {author} {\bibfnamefont {H.}~\bibnamefont
  {Weinfurter}},\ }\bibfield  {title} {\bibinfo {title} {Generation of
  correlated photon pairs in type-ii parametric down conversion—revisited},\
  }\href {https://www.tandfonline.com/doi/abs/10.1080/09500340108240902}
  {\bibfield  {journal} {\bibinfo  {journal} {Journal of Modern Optics}\
  }\textbf {\bibinfo {volume} {48}},\ \bibinfo {pages} {1997} (\bibinfo {year}
  {2001})}\BibitemShut {NoStop}%
\bibitem [{\citenamefont {Zhu}(2017)}]{zhu2017multiqubit}%
  \BibitemOpen
  \bibfield  {author} {\bibinfo {author} {\bibfnamefont {H.}~\bibnamefont
  {Zhu}},\ }\bibfield  {title} {\bibinfo {title} {Multiqubit clifford groups
  are unitary 3-designs},\ }\href
  {https://journals.aps.org/pra/abstract/10.1103/PhysRevA.96.062336} {\bibfield
   {journal} {\bibinfo  {journal} {Physical Review A}\ }\textbf {\bibinfo
  {volume} {96}},\ \bibinfo {pages} {062336} (\bibinfo {year}
  {2017})}\BibitemShut {NoStop}%
\bibitem [{\citenamefont {Li}\ \emph {et~al.}(2023)\citenamefont {Li},
  \citenamefont {Yin}, \citenamefont {Zhang}, \citenamefont {Chen},
  \citenamefont {Yin}, \citenamefont {Peng}, \citenamefont {Hong},
  \citenamefont {Chen}, \citenamefont {Li},\ and\ \citenamefont
  {Guo}}]{li2023experimental}%
  \BibitemOpen
  \bibfield  {author} {\bibinfo {author} {\bibfnamefont {G.-C.}\ \bibnamefont
  {Li}}, \bibinfo {author} {\bibfnamefont {Z.-Q.}\ \bibnamefont {Yin}},
  \bibinfo {author} {\bibfnamefont {W.-H.}\ \bibnamefont {Zhang}}, \bibinfo
  {author} {\bibfnamefont {L.}~\bibnamefont {Chen}}, \bibinfo {author}
  {\bibfnamefont {P.}~\bibnamefont {Yin}}, \bibinfo {author} {\bibfnamefont
  {X.-X.}\ \bibnamefont {Peng}}, \bibinfo {author} {\bibfnamefont {X.-S.}\
  \bibnamefont {Hong}}, \bibinfo {author} {\bibfnamefont {G.}~\bibnamefont
  {Chen}}, \bibinfo {author} {\bibfnamefont {C.-F.}\ \bibnamefont {Li}},\ and\
  \bibinfo {author} {\bibfnamefont {G.-C.}\ \bibnamefont {Guo}},\ }\bibfield
  {title} {\bibinfo {title} {Experimental full calibration of quantum devices
  in a semi-device-independent way},\ }\href
  {https://opg.optica.org/abstract.cfm?uri=optica-10-12-1723} {\bibfield
  {journal} {\bibinfo  {journal} {Optica}\ }\textbf {\bibinfo {volume} {10}},\
  \bibinfo {pages} {1723} (\bibinfo {year} {2023})}\BibitemShut {NoStop}%
\bibitem [{\citenamefont {Brydges}\ \emph {et~al.}(2019)\citenamefont
  {Brydges}, \citenamefont {Elben}, \citenamefont {Jurcevic}, \citenamefont
  {Vermersch}, \citenamefont {Maier}, \citenamefont {Lanyon}, \citenamefont
  {Zoller}, \citenamefont {Blatt},\ and\ \citenamefont
  {Roos}}]{brydges2019probing}%
  \BibitemOpen
  \bibfield  {author} {\bibinfo {author} {\bibfnamefont {T.}~\bibnamefont
  {Brydges}}, \bibinfo {author} {\bibfnamefont {A.}~\bibnamefont {Elben}},
  \bibinfo {author} {\bibfnamefont {P.}~\bibnamefont {Jurcevic}}, \bibinfo
  {author} {\bibfnamefont {B.}~\bibnamefont {Vermersch}}, \bibinfo {author}
  {\bibfnamefont {C.}~\bibnamefont {Maier}}, \bibinfo {author} {\bibfnamefont
  {B.~P.}\ \bibnamefont {Lanyon}}, \bibinfo {author} {\bibfnamefont
  {P.}~\bibnamefont {Zoller}}, \bibinfo {author} {\bibfnamefont
  {R.}~\bibnamefont {Blatt}},\ and\ \bibinfo {author} {\bibfnamefont {C.~F.}\
  \bibnamefont {Roos}},\ }\bibfield  {title} {\bibinfo {title} {Probing
  r{\'e}nyi entanglement entropy via randomized measurements},\ }\href
  {https://www.science.org/doi/abs/10.1126/science.aau4963} {\bibfield
  {journal} {\bibinfo  {journal} {Science}\ }\textbf {\bibinfo {volume}
  {364}},\ \bibinfo {pages} {260} (\bibinfo {year} {2019})}\BibitemShut
  {NoStop}%
\bibitem [{\citenamefont {Elben}\ \emph {et~al.}(2019)\citenamefont {Elben},
  \citenamefont {Vermersch}, \citenamefont {Roos},\ and\ \citenamefont
  {Zoller}}]{elben2019statistical}%
  \BibitemOpen
  \bibfield  {author} {\bibinfo {author} {\bibfnamefont {A.}~\bibnamefont
  {Elben}}, \bibinfo {author} {\bibfnamefont {B.}~\bibnamefont {Vermersch}},
  \bibinfo {author} {\bibfnamefont {C.~F.}\ \bibnamefont {Roos}},\ and\
  \bibinfo {author} {\bibfnamefont {P.}~\bibnamefont {Zoller}},\ }\bibfield
  {title} {\bibinfo {title} {Statistical correlations between locally
  randomized measurements: A toolbox for probing entanglement in many-body
  quantum states},\ }\href
  {https://journals.aps.org/pra/abstract/10.1103/PhysRevA.99.052323} {\bibfield
   {journal} {\bibinfo  {journal} {Physical Review A}\ }\textbf {\bibinfo
  {volume} {99}},\ \bibinfo {pages} {052323} (\bibinfo {year}
  {2019})}\BibitemShut {NoStop}%
\bibitem [{\citenamefont {Elben}\ \emph {et~al.}(2023)\citenamefont {Elben},
  \citenamefont {Flammia}, \citenamefont {Huang}, \citenamefont {Kueng},
  \citenamefont {Preskill}, \citenamefont {Vermersch},\ and\ \citenamefont
  {Zoller}}]{elben2023randomized}%
  \BibitemOpen
  \bibfield  {author} {\bibinfo {author} {\bibfnamefont {A.}~\bibnamefont
  {Elben}}, \bibinfo {author} {\bibfnamefont {S.~T.}\ \bibnamefont {Flammia}},
  \bibinfo {author} {\bibfnamefont {H.-Y.}\ \bibnamefont {Huang}}, \bibinfo
  {author} {\bibfnamefont {R.}~\bibnamefont {Kueng}}, \bibinfo {author}
  {\bibfnamefont {J.}~\bibnamefont {Preskill}}, \bibinfo {author}
  {\bibfnamefont {B.}~\bibnamefont {Vermersch}},\ and\ \bibinfo {author}
  {\bibfnamefont {P.}~\bibnamefont {Zoller}},\ }\bibfield  {title} {\bibinfo
  {title} {The randomized measurement toolbox},\ }\href
  {https://www.nature.com/articles/s42254-022-00535-2} {\bibfield  {journal}
  {\bibinfo  {journal} {Nature Reviews Physics}\ }\textbf {\bibinfo {volume}
  {5}},\ \bibinfo {pages} {9} (\bibinfo {year} {2023})}\BibitemShut {NoStop}%
\bibitem [{\citenamefont {Zhou}\ \emph {et~al.}(2020)\citenamefont {Zhou},
  \citenamefont {Zeng},\ and\ \citenamefont {Liu}}]{singlezhou}%
  \BibitemOpen
  \bibfield  {author} {\bibinfo {author} {\bibfnamefont {Y.}~\bibnamefont
  {Zhou}}, \bibinfo {author} {\bibfnamefont {P.}~\bibnamefont {Zeng}},\ and\
  \bibinfo {author} {\bibfnamefont {Z.}~\bibnamefont {Liu}},\ }\bibfield
  {title} {\bibinfo {title} {Single-copies estimation of entanglement
  negativity},\ }\href {https://doi.org/10.1103/PhysRevLett.125.200502}
  {\bibfield  {journal} {\bibinfo  {journal} {Phys. Rev. Lett.}\ }\textbf
  {\bibinfo {volume} {125}},\ \bibinfo {pages} {200502} (\bibinfo {year}
  {2020})}\BibitemShut {NoStop}%
\bibitem [{\citenamefont {Li}\ \emph {et~al.}(2024)\citenamefont {Li},
  \citenamefont {Chen}, \citenamefont {Zhang}, \citenamefont {Hong},
  \citenamefont {Zhou}, \citenamefont {Chen}, \citenamefont {Li},\ and\
  \citenamefont {Guo}}]{my_p3ppt_paper}%
  \BibitemOpen
  \bibfield  {author} {\bibinfo {author} {\bibfnamefont {G.-C.}\ \bibnamefont
  {Li}}, \bibinfo {author} {\bibfnamefont {L.}~\bibnamefont {Chen}}, \bibinfo
  {author} {\bibfnamefont {S.-Q.}\ \bibnamefont {Zhang}}, \bibinfo {author}
  {\bibfnamefont {X.-S.}\ \bibnamefont {Hong}}, \bibinfo {author}
  {\bibfnamefont {Y.}~\bibnamefont {Zhou}}, \bibinfo {author} {\bibfnamefont
  {G.}~\bibnamefont {Chen}}, \bibinfo {author} {\bibfnamefont {C.-F.}\
  \bibnamefont {Li}},\ and\ \bibinfo {author} {\bibfnamefont {G.-C.}\
  \bibnamefont {Guo}},\ }\bibfield  {title} {\bibinfo {title} {Directly
  estimating mixed-state entanglement with bell measurement assistance},\
  }\href {https://arxiv.org/abs/2405.20696} {\bibfield  {journal} {\bibinfo
  {journal} {arXiv preprint arXiv:2405.20696}\ } (\bibinfo {year}
  {2024})}\BibitemShut {NoStop}%
\bibitem [{\citenamefont {Elben}\ \emph {et~al.}(2020)\citenamefont {Elben},
  \citenamefont {Kueng}, \citenamefont {Huang}, \citenamefont {van Bijnen},
  \citenamefont {Kokail}, \citenamefont {Dalmonte}, \citenamefont {Calabrese},
  \citenamefont {Kraus}, \citenamefont {Preskill}, \citenamefont {Zoller} \emph
  {et~al.}}]{elben2020mixed}%
  \BibitemOpen
  \bibfield  {author} {\bibinfo {author} {\bibfnamefont {A.}~\bibnamefont
  {Elben}}, \bibinfo {author} {\bibfnamefont {R.}~\bibnamefont {Kueng}},
  \bibinfo {author} {\bibfnamefont {H.-Y.~R.}\ \bibnamefont {Huang}}, \bibinfo
  {author} {\bibfnamefont {R.}~\bibnamefont {van Bijnen}}, \bibinfo {author}
  {\bibfnamefont {C.}~\bibnamefont {Kokail}}, \bibinfo {author} {\bibfnamefont
  {M.}~\bibnamefont {Dalmonte}}, \bibinfo {author} {\bibfnamefont
  {P.}~\bibnamefont {Calabrese}}, \bibinfo {author} {\bibfnamefont
  {B.}~\bibnamefont {Kraus}}, \bibinfo {author} {\bibfnamefont
  {J.}~\bibnamefont {Preskill}}, \bibinfo {author} {\bibfnamefont
  {P.}~\bibnamefont {Zoller}}, \emph {et~al.},\ }\bibfield  {title} {\bibinfo
  {title} {Mixed-state entanglement from local randomized measurements},\
  }\href {https://journals.aps.org/prl/abstract/10.1103/PhysRevLett.125.200501}
  {\bibfield  {journal} {\bibinfo  {journal} {Physical Review Letters}\
  }\textbf {\bibinfo {volume} {125}},\ \bibinfo {pages} {200501} (\bibinfo
  {year} {2020})}\BibitemShut {NoStop}%
\bibitem [{\citenamefont {Walther}\ \emph {et~al.}(2005)\citenamefont
  {Walther}, \citenamefont {Resch}, \citenamefont {Rudolph}, \citenamefont
  {Schenck}, \citenamefont {Weinfurter}, \citenamefont {Vedral}, \citenamefont
  {Aspelmeyer},\ and\ \citenamefont {Zeilinger}}]{walther2005experimental}%
  \BibitemOpen
  \bibfield  {author} {\bibinfo {author} {\bibfnamefont {P.}~\bibnamefont
  {Walther}}, \bibinfo {author} {\bibfnamefont {K.~J.}\ \bibnamefont {Resch}},
  \bibinfo {author} {\bibfnamefont {T.}~\bibnamefont {Rudolph}}, \bibinfo
  {author} {\bibfnamefont {E.}~\bibnamefont {Schenck}}, \bibinfo {author}
  {\bibfnamefont {H.}~\bibnamefont {Weinfurter}}, \bibinfo {author}
  {\bibfnamefont {V.}~\bibnamefont {Vedral}}, \bibinfo {author} {\bibfnamefont
  {M.}~\bibnamefont {Aspelmeyer}},\ and\ \bibinfo {author} {\bibfnamefont
  {A.}~\bibnamefont {Zeilinger}},\ }\bibfield  {title} {\bibinfo {title}
  {Experimental one-way quantum computing},\ }\href
  {https://www.nature.com/articles/nature03347} {\bibfield  {journal} {\bibinfo
   {journal} {Nature}\ }\textbf {\bibinfo {volume} {434}},\ \bibinfo {pages}
  {169} (\bibinfo {year} {2005})}\BibitemShut {NoStop}%
\end{thebibliography}%

\end{document}


\title{Supplementary Material for ``Invested and Potential Magic Resources in Measurement-Based Quantum Computation"}

\maketitle

\tableofcontents

\newpage

\section*{Symbol List}

\SetTblrInner{rowsep=1pt}
\begin{center}
\renewcommand\arraystretch{1.5}
\begin{tabular}{cp{9cm}}
\toprule
\textbf{Symbol} & \textbf{Description} \\
\midrule
$M^{\theta}$ & a single-qubit measurement project onto state $\frac{1}{\sqrt{2}}(|0\rangle+e^{i\theta}|1\rangle)$.\\
$M^{(\theta,\phi)}$ & a single-qubit measurement project onto state $\cos\theta|0\rangle+\sin\theta e^{i\phi}|1\rangle)$.\\
$[C]$ & A series of Pauli gates given the measurement outcomes to correct the final state into the desired one. \\
$[M]$ & A series adaptive single-qubit measurement obeying the MQC rule, and generating a series of measurement outcomes, also known as signals.\\
$[E]$ & A series of CZ gates to generate a given graph state $|G\rangle$.\\
$\Xi_P$ & Elements of Pauli spectrum with $P\in\text{Pauli}(n)$ and $\Xi_P=d^{-1}\langle \psi|P|\psi\rangle^2$. Pauli spectrum of pure state satisfies $\sum_{P}\Xi_P=1$ and $\Xi_P\ge0$, thus it could also be regarded as a special list of possibilities. \\
$\mathbb{S}_{\alpha}(\rho)$ & $\alpha$--R\'{e}nyi entropy of quantum state $\rho$, defined as $\mathbb{S}_{\alpha}(\rho)=(1-\alpha)^{-1}\log tr(\rho^{\alpha})$\\
$\mathbb{M}_{\alpha}(\psi)$ & $\alpha$-R\'{e}nyi entropy of Pauli spectrum $\{\Xi_P\}$ of the state $\psi$, with $\mathbb{M}_\alpha=(1-\alpha)^{-1}\log\sum_{P}\Xi_P^{\alpha}-\log d$. It is also a good measure for magic resources; for more discussion, see \cite{leone2022stabilizer}. In this article, we mainly focus on the $\mathbb{M}_2$. \\
$\mathcal{M}(U)$ & Invested Magic Resources of a unitary $U$, defined by Eq.(2) in the main text. \\
$\mathcal{R}(|\psi\rangle)$ & Reserved Magic Resources, just as $\mathbb{M}_2$ of the reserved state. \\
$\mathcal{P}(|G\rangle)$ & Potential Magic Resources of the graph state $|G\rangle$, defined by Eq.(4) in the main text. \\
$\mathcal{W}$ & Waste of magic resources, defined as wastes in the injection magic resources by non-Pauli measurements in the MQC, $\mathcal{W}=\mathcal{M}-\mathcal{P}$. See Fig.~2 in the main text.\\ 
$|T\rangle$ & The magic T state, firstly defined in the \cite{bravyi2005universal} as $|T\rangle=\cos\theta_m|0\rangle+\sin\theta_me^{i\pi/4}|1\rangle$ with $\cos2\theta_m = 1/\sqrt{3}$. \\
$\hat{T}$ & T-gate, defined as $\hat{T}=\text{diag}(1, e^{i\pi/4})$.\\
$1T$ & Unit of magic resources in this article, defined as $1T=\mathbb{M}_2(|T\rangle)$. It serves as the maximum single-qubit magic resource. \\
\bottomrule
\end{tabular}
\end{center}

\newpage

\section{Review of Magic Resources}

To achieve large-scale, fault-tolerant universal quantum computation, one utilizes error correction codes to encode logical qubits and introduce sets of fault-tolerant universal gates \cite{postler2022demonstration, google2023suppressing, gupta2024encoding}. Within this framework, T states/gates have been recognized as crucial resources, often referred to as magic states/gates, for achieving universal quantum computation. 
While Clifford gates alone are not universal, the combination of Clifford gates and T-states/gates provides the necessary ingredients for universal quantum computation, allowing for the generation of arbitrary quantum states with arbitrary precision. Additionally, T-states can be distilled from noisy encoded states, enhancing their practicality. These magic states/gates are fundamental resources towards fault-tolerant universal computation and can be 
quantified by the appropriate resource theory \cite{veitch2014resource, leone2022stabilizer, bittel}. 

\begin{figure}[hbp]
    \centering
    \includegraphics[width=\textwidth]{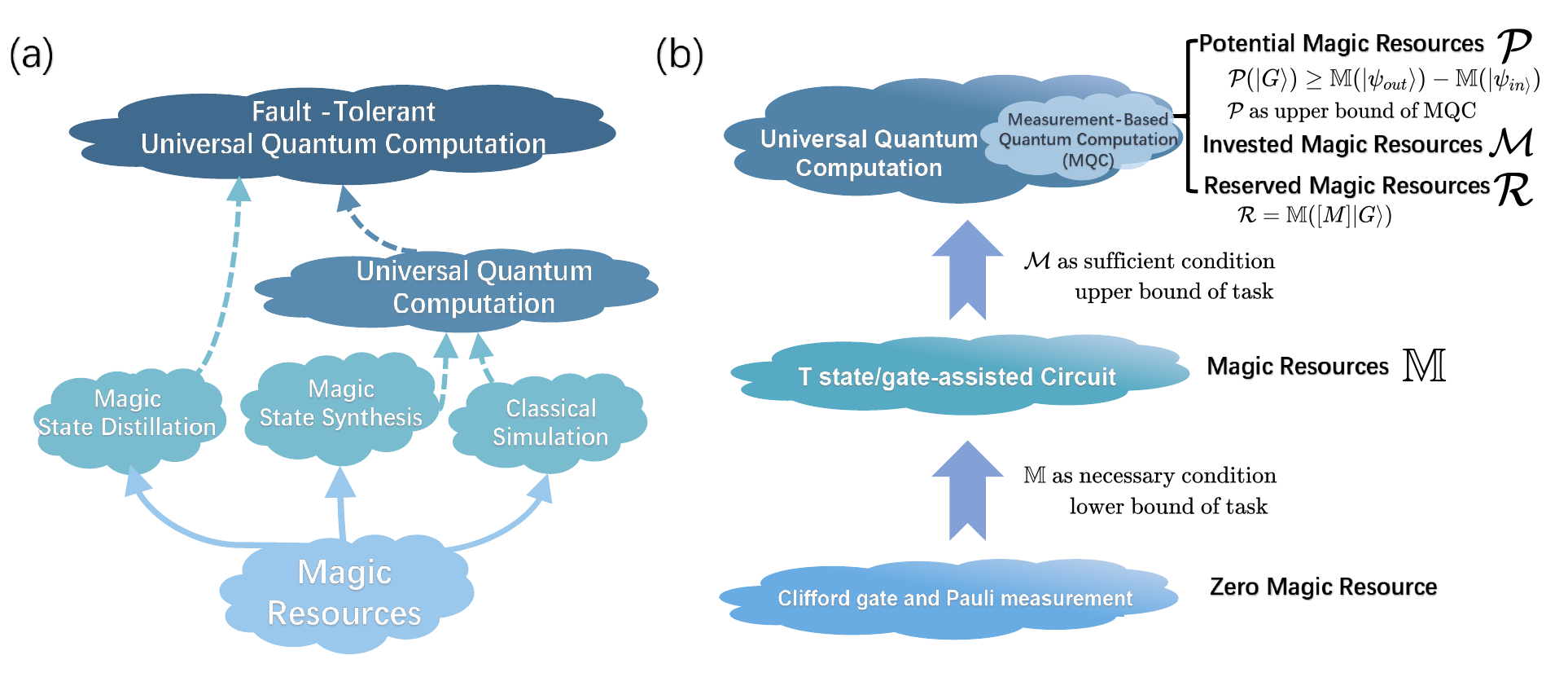}
    \caption{ (a) Cartoon describing the relationship between magic resources and quantum computation tasks; (b) description of the three layers of quantum computation and their relationship with invested and potential magic resources. 
    }
    \label{fig:demo1}
\end{figure}

As depicted in Fig.\ref{fig:demo1}(a), three key research areas are closely connected to magic resources and play a vital role in progressing towards fault-tolerant universal quantum computation: magic state distillation, magic state synthesis, and classical simulation complexity.

Magic state distillation (MSD) focuses on transforming noisy T-state encodings into higher-quality T-states suitable for practical applications \cite{bravyi2005universal}. Experimental progress has been made in this area, including demonstrations with 5-qubit \cite{souza2011experimental} and 4-qubit codes \cite{gupta2024encoding}.  Another application of the magic state is magic state synthesis (MSS), which leverages many T states/gates with Clifford gates and Pauli measurement to generate an arbitrary unitary or state \cite{knill2005quantum}. Suppose a magic resource contained in an arbitrary state $|\psi\rangle$ is quantified by an appropriate measure $m(|\psi\rangle)$, then at least $n = \lceil m(|\psi\rangle)/m(|T\rangle)\rceil$ T-states are needed to synthesize such a state \cite{chitambar2019quantum}. Therefore, magic resources quantify the difficulties of the implementation of universal gates. 
Within the framework of quantum resource theory, the concept of magic resources provides fundamental lower bounds on the number of encodes required for distillation and synthesis\cite{regula2021fundamental}.

Understanding universal quantum computation also involves considering the complexity of simulating the same tasks on classical computers. The Gottesman-Knill theorem \cite{aaronson} establishes that quantum computation on a system of $n$ qubits involving only computational basis states, Clifford gates, and Pauli measurements can be efficiently simulated classically, that is, by a polynomial (in $n$) overhead.   Moving beyond this first layer of computation, see Fig.\ref{fig:demo1}(b)), it has been shown  \cite{bravyi2016improved, bravyi2016trading} that incorporating magic state injection (MSI) exponentially increases the difficulty of classical simulation. MSI involves preparing $n$ qubits in computational basis states and $k$ qubits in the T states, resulting in an $n+k$ qubit system. Simulating such a system with Clifford gates and Pauli measurements on a classical computer requires a runtime scaling as $2^{O(k)}\text{poly}(n)$.
From the perspective of classical simulation complexity, universal quantum computation is characterized by an exponential increase in resource requirements, which can be quantified by the magic resources present in the system. 

Quantifying magic resources has been a challenging task in the field of quantum information. A fundamental approach involves measuring the extent to which a quantum state deviates from being a stabilizer state, which are states efficiently simulable on classical computers according to the Gottesman-Knill theorem. A reliable magic resource measure should satisfy three key properties: faithfulness, non-increasingness under the free operations of the theory (that is, Clifford operations), and (sub-)additivity.
Faithfulness ensures that a state has zero magic if and only if it is a stabilizer state. Invariance under Clifford operations implies that applying Clifford gates, which are considered free operations in magic resource theories, does not alter the magic content of a state. Additivity dictates that the magic of a combined system is equal to 
the sum of the magic within its individual components. While measures like mana\cite{veitch2014resource}, robustness\cite{howard2017application}, thauma\cite{wang2020efficiently}, and nullity\cite{beverland2020lower} satisfy faithfulness and Clifford invariance, they lack the additivity property. Recent approaches based on stabilizer norm \cite{howard2017application}, stabilizer R\'enyi entropy\cite{leone2022stabilizer, bittel}, and GKP codes\cite{hahn2022quantifying} have successfully addressed this limitation, providing measures that fulfill all three desired criteria. 

All these methods are deeply connected to the Pauli spectrum of a quantum state. For an n-qubit system, the Pauli group  {with quotient the overall phase}, $\mathcal{P}_n$, consists of $\{I,X,Y,Z\}^{\otimes n}$. For any pure state $|\psi\rangle$, we can define the Pauli spectrum as the set of values ${\Xi_P}$, where  {$\Xi_P = d^{-1}\langle \psi|P|\psi\rangle^2$} for each Pauli operator $P\in\mathcal{P}_n$ and $d=2^n$ is the dimension of the Hilbert space. The Pauli spectrum satisfies the properties $\sum_P\Xi_P = 1$ and $\Xi_P\ge 0$  for pure states, making it a probability distribution. Operationally, 
This is the probability of obtaining the state $|P\rangle:= I\otimes P|R\rangle$ when preparing $|\psi\rangle\otimes |\psi^*\rangle$, where $|R\rangle:=d^{-1/2}\sum_i|ii\rangle$ is the maximally entangled state on two copies of the Hilbert space\cite{leonelearning}.
Using the Pauli spectrum, we can define the $\alpha$-Rényi entropy of the state for $\alpha\ge 0$:
\begin{equation}
    \M_\alpha(|\psi\rangle) = (1-\alpha)^{-1}\log  {\sum_P} \Xi_P^{\alpha} -\log d.
\end{equation}
It has been proven that $\M_{\alpha}$ with $\alpha\ge 0$ serves as a reliable magic resource quantifier\cite{leone2022stabilizer, bittel}, satisfying faithfulness ($M_\alpha(|\psi\rangle)=0$ if and only if $|\psi\rangle$ is a stabilizer state), Clifford invariance ($\M_{\alpha}(C|\psi\rangle) = \M_{\alpha}(|\psi\rangle)$ for any Clifford operation $C$), and additivity ($\M_{\alpha}(|\psi\rangle\otimes |\phi\rangle) = \M_{\alpha}(|\psi\rangle)+\M_{\alpha}(|\phi\rangle)$). Furthermore, $M_\alpha$ provides a lower bound for both the mana and nullity measures of magic and is also bounded by twice the robustness of magic. Notably, the case of $\alpha=1/2$ corresponds to the stabilizer-norm as $\log(\mathcal{D}(\psi))=1/2\mathbb{M}_{1/2}(|\psi\rangle)$ \cite{campbell2011catalysis} and also the magic resource measure derived from GKP codes. 
Among all possible choices of $\alpha$, the 2-Rényi entropy ($\M_2$) stands out due to its generalizability to mixed states, which are crucial for describing realistic quantum systems, and its feasibility on the experiments \cite{oliviero2022measuring}. Also, $\M_2$ is analytically tractable for a few important cases in many-qubit systems \cite{chen2023magic}, and also can be numerically calculated by tensor network algorithms to large-scale \cite{Haug2023MPS,Lami2023MPS, dalmonte}.

For mixed states $\rho$, the Pauli spectrum no longer sums to 1 but instead reflects the purity of the state, with $\sum_P\Xi_P=\tr(\rho^2)$. To address this, we consider a generalized Pauli spectrum given by $\Xi_P/\tr(\rho^2)$. Consequently, the 2-Rényi entropy of the generalized Pauli spectrum takes the form:
\begin{equation}
    \M_2(\rho) = -\log\sum_P\Xi_P^2-\log d - \mathbb{S}_2(\rho),
\end{equation}
where $\mathbb{S}_{2}(\rho)=-\log \tr(\rho^2)$ is the 2-R\'enyi entropy of the quantum state. In the resource theory for mixed states, there is also the partial trace, and although  $\M_2(\rho)$ can sometimes increase under partial trace, numerical evidence shows that this is overwhelmingly rare, and therefore $\M_2(\rho)$ is a good {\em proxy} for this resource theory\cite{leone2022stabilizer}. 

Another important comment is that the meaning of $\M_2(\rho)$ for mixed states is somewhat different. In fact, there are states with $\M_2(\rho)=0$ from which no magic states can be extracted. However, a key significance is that such states cannot be purified into magic-free states.\cite{aaronson, cao}. Moreover, this quantity is related to the cost of other protocols, such as learning and quantum verification \cite{leonelearning, leoneverification}. 
And SRE has the operational quantities, the hardness of fidelity estimation, which is a very relevant issue in experimental quantum computation. Moreover, the mixed state SRE is related to the non-stabilizerness of their purifications. \cite{inpreparation}.

\section{Review of Measurement-Based Quantum Computation}

This section delves into the role of non-Pauli measurements in MQC, also known as one-way computation due to the destructive nature of measurements on the quantum state, and their connection to the investment of magic resources.

The setup of MQC starts with a graph state $|G\rangle$, which is an entangled state of n qubits defined by a graph $G = (V, E)$ with n vertices and edges representing controlled-Z (CZ) gates. The graph state is obtained by applying CZ gates between qubits connected by edges in the graph, starting from an initial state of all qubits in $|+\rangle=(|0\rangle+|1\rangle)/\sqrt{2}$. Graph states are a special type of stabilizer state, meaning they possess a set of stabilizing operators that leave the state unchanged. For each qubit j in a graph state, there exists a stabilizer operator $K_j = X^{(j)}\prod_{i\in  {N}(j)}Z^{(i)}$ involving the Pauli X operator on qubit j and Pauli Z operators on its neighboring qubits ($ {N}(j)$). Unlike the quantum circuit model, where quantum computations are achieved through a sequence of unitary gates and measurements, MQC relies on preparing a suitable graph state and performing a series of adaptive single-qubit measurements in the X-Y plane of the Bloch sphere. The final state of the computation is locally Pauli equivalent to the desired target state. Based on the stochastic outcomes of the measurements, additional Pauli corrections are applied to obtain the final result. This process can be summarized as the CME pattern, consisting of Entanglement (preparing the graph state), Measurement (adaptive single-qubit measurements), and Correction (Pauli corrections).

\begin{figure}[t]
    \centering
    \includegraphics[width=\textwidth]{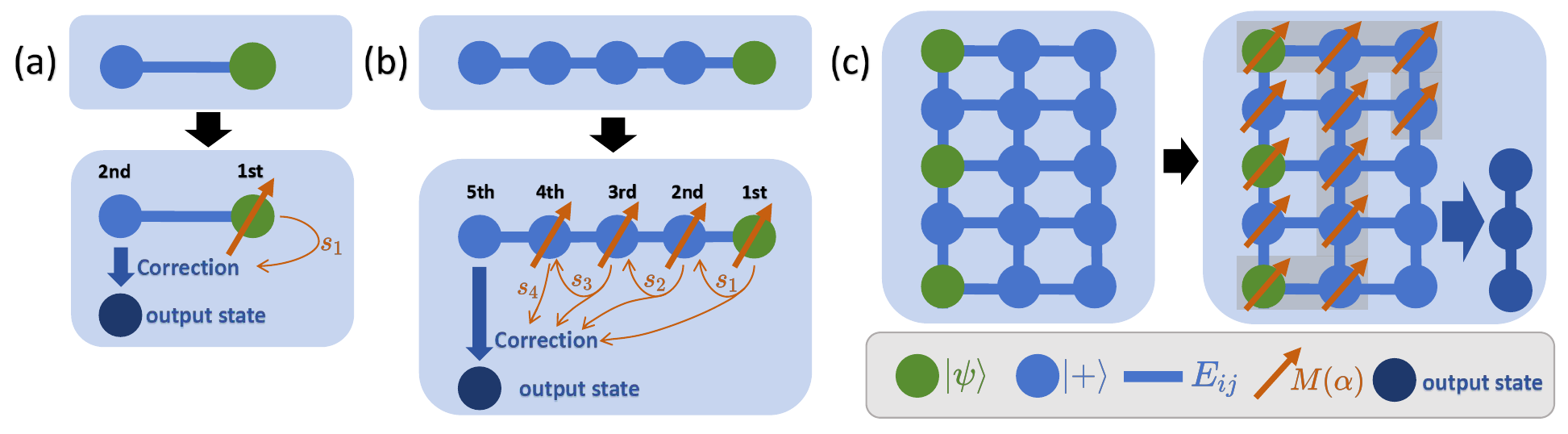}
    \caption{
    \textit{Demonstration of Measurement-based quantum computation.} (a) shows how one rotation on one qubit is satisfied by MQC. The first is to build up the entanglement state $E_{12}|+\rangle|\psi\rangle$, then after the measurement on the input state and correction on the reserved qubit, we get $J(\alpha)|\psi\rangle$. (b) shows an arbitrary rotation for a single qubit. (c) shows how the 2D graph is used for MQC. The grey-shaped area is for the transfer of the signals. 
    }
    \label{fig: MQC_demo}
\end{figure}

Such measurement calculus provides a framework for translating quantum circuits into equivalent MQC procedures\cite{danos2005parsimonious, danos2007measurement}. It relies on the fact that any unitary operation can be decomposed into a sequence of CZ gates and J gates, where the J gate is represented by the matrix $J(\alpha)=\frac{1}{\sqrt{2}}\begin{pmatrix}1& e^{i\alpha}\\ 1&-e^{i\alpha}\end{pmatrix}$. This decomposition process, known as J-decomposition, allows the expression of fundamental quantum gates like Hadamard($H=J(0)$), Pauli X($X=J(\pi)J(0)$), Pauli Z ($Z=J(0)J(\pi)$), Phase gate $S=J(0)J(\frac{\pi}{2})$ and rotations around the X and Z axes($R_x(\alpha)=J(\alpha)J(0) \ \&\ R_z(\alpha)=J(0)J(\alpha)$) using CZ and J gates. Each J gate in the decomposition corresponds to a small CME structure. For instance, the J gate with angle $\alpha$ can be implemented as:
\begin{equation}
    J(\alpha) := X_2^{s_1}M_1^{-\alpha}E_{12},
\end{equation}
where $E_{12}$ represents a CZ gate between qubits 1 and 2, $M_1^{-\alpha}$ denotes a measurement on the first qubit in the basis $(|0\rangle\pm e^{-i\alpha}|1\rangle)/\sqrt{2}$, $s_1$ is the measurement outcome, and $X_2^{s_1}$ is a Pauli X correction applied to the second qubit based on the measurement outcome. This equation illustrates how the action of $J(\alpha)$ on a state $|\psi\rangle$ can be achieved using an auxiliary qubit  {as the 2nd qubit} initialized in the $|+\rangle$ state and an appropriate MQC procedure, as illustrated in Fig.\ref{fig: MQC_demo}(a).

For more complex unitary operations, the measurement calculus utilizes swap relationships between operators to decompose them into sequences of J gates\cite{danos2007measurement}. For example, an arbitrary single-qubit rotation $U$ can be expressed as $U = J(0)J(-\alpha)J(-\beta)J(-\gamma)$,  {which is equivalent to the Z-X-Z decomposition for SU(2),} 
leading to an MQC implementation involving a linear graph of five qubits and four auxiliary qubits:
\begin{equation}
\begin{aligned}
    U|\psi\rangle &= \underbrace{X_5^{s_4}M^0_4E_{45}}_{J(0)}\underbrace{X_4^{s_3}M^{\alpha}_3E_{43}}_{J(\alpha)}\underbrace{X_3^{s_2}M^\beta_2E_{32}}_{J(\beta)}\underbrace{X_2^{s_1}M^\gamma_1E_{21}}_{J(\gamma)}|\psi\rangle|+\rangle^{\otimes 4} \\
    & = \underbrace{X_5^{s_2+s_4}Z_5^{s_1+s_3}}_{ {\text{correction}}}
         \underbrace{M_4^{0}M_3^{(-1)^{s_2}\alpha}M^{(-1)^{s_1}\beta}_2{ M}_1^{\gamma}}_{ {\text{measurement}}}
         \underbrace{E_{12345}|\psi\rangle|+\rangle^{\otimes 4}}_{ {\text{entanglement}}}.
\end{aligned}
\end{equation}
 Fig.\ref{fig: MQC_demo}(b) demonstrates this process and the transfer of signals of the measurement results. For the implementation of more complex unitary operations on multi-qubit states, we could use a 2-D graph to realize them. See Fig.~\ref{fig: MQC_demo}(c). 

By reordering the operators and grouping the entanglement operations and correction operations, we obtain a practical MQC plan where qubits are connected in a linear chain, and measurements are performed sequentially, adapting the measurement basis based on previous outcomes. This highlights the adaptive nature of measurements in MQC. Overall, the CME pattern for the MQC could be summarized as:
\begin{equation}
    U := [C][M][E].
\end{equation}

\section{Proof of Invested Magic Resources as a Good Measure}\label{Mth:IMR}

For an arbitrary unitary $U$, there exists a possible MQC process to realize it. We calculated the magic resources needed in the MQC to guarantee the necessary conditions to realize $U$. This depends on the J-decomposition of the unitary with $\IMR(U) = \sum_{J(\theta)} \M_\alpha[M(\theta)]$. We will prove that it satisfies faithfulness, invariance with Clifford gates, and additivity.

First, it exhibits faithfulness, meaning $\IMR_{\alpha}(C)=0$ if and only if $C$ is a Clifford gate. This follows from the fact that Clifford gates can be decomposed into CZ gates, Hadamard gates (equivalent to $ J(0) $), and phase gates (equivalent to $ J(0)J(\frac{\pi}{2}) $), all of which have zero magic cost according to $\IMR_{\alpha}$. Conversely, if $\IMR_{\alpha}(U)=0$, then the unitary $ U $ can only involve gates generated by $\{CZ, J(0), J(\frac{\pi}{2}), J(\pi)\}$, which correspond to Clifford operations.

Additionally, invariance under Clifford gates means $\IMR_{\alpha}(CU)=\IMR_{\alpha}(U)$ for any Clifford gate $ C $, as applying $ C $ before or after $ U $ does not change the magic resource cost. Similar to faithfulness, Clifford gates can be decomposed by the Hadamard gate, the phase gate, and the CZ gate, and they can all be implemented with no invested magic resources.

Finally, additivity means $\IMR_{\alpha}(U \otimes V) = \IMR_{\alpha}(U) + \IMR_{\alpha}(V)$. $ U $ and $ V $ represent two independent MQC procedures. Suppose a greater graph with $ |G\rangle = |G_U\rangle \otimes |G_V\rangle $. The half graph $ |G_U\rangle $ realizes $ U $, and the other half $ |G_V\rangle $ realizes $ V $. Thus, the additivity is always satisfied. 

\section{Invested Magic Resources of Quantum Fourier Transformation}

\begin{figure}
    \centering
    \includegraphics[width=\linewidth]{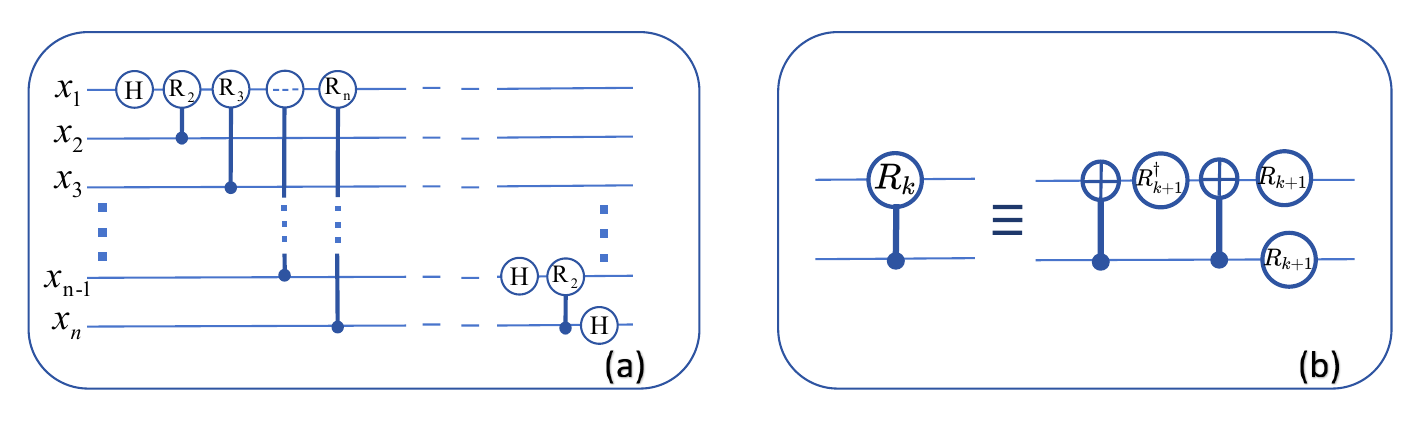}
    \caption{\textit{(a)Quantum Fourier Transformation Circuit. (b) Decomposition of control rotation gates.}}
    \label{fig:qft}
\end{figure}

QFT is a linear transformation that maps computational basis states $ |x\rangle $ to the Fourier basis, $ |x\rangle \xrightarrow[]{QFT} \sum_{k}\omega^{xk}|k\rangle $ with $ \omega = \exp(i2\pi/2^n) $ and $ |k\rangle $ for Fourier bases. Fig.~\ref{fig:qft}(a) shows the circuit implementation of QFT using Hadamard gates and controlled-rotation gates ($ CR_k $) with $ R_k = \text{diag}[1, \exp(i2\pi/2^k)] $. The QFT circuit from the $ j $-th qubit to the $ n $-th qubit can be expressed iteratively as
$$\text{QFT}^{(j:n)} = \big(H^{(j)} \bigotimes_{i=j+1}^{n} C^{(i)}R_{i-j+1}^{(j)}\big) \text{QFT}^{(j+1:n)},$$
where the superscripts indicate the qubit indices involved in each operation.

To analyze the magic resource cost of QFT, we utilize the J-decomposition technique, as shown in Fig.~\ref{fig:qft}(b). Each $ CR_k $ gate can be decomposed into two CNOT gates and three $ R_{k+1} $ gates, with $ R_{k+1} $ further decomposed as $ J(0)J(2\pi/2^{k+1}) $. Consequently, the invested magic resource cost of a $ CR_k $ gate is given by
$$\IMR_2(CR_k) = 3\IMR_2(R_{k+1}) = -3 \log\left(\frac{1}{2} + \frac{1}{2}\cos^4\left(\frac{\pi}{2^k}\right) + \frac{1}{2}\sin^4\left(\frac{\pi}{2^k}\right)\right).$$
For large values of $ k $, this expression can be approximated as
$$\IMR_2(CR_k) \approx -3 \log\left(1 + \left(\frac{\pi}{2^k}\right)^4\right) \approx \frac{3\pi^4}{2^{4k}},$$
indicating that high-frequency components contribute minimally to the overall magic resource cost. Fig.~\ref{fig:qft}(c) and (d) visualize the contribution of different frequencies to the invested magic resources of QFT. The plots reveal that most of the magic resources are concentrated in the low-frequency range, with a sharp decline in contribution as the frequency increases. This behavior remains consistent across different qubit numbers. Based on this observation, it is natural to consider a truncated QFT where only low-frequency components are included, potentially reducing the magic resource cost without significant loss of accuracy.

The full QFT for $ n $ qubits requires $ (n+k-1) $ instances of $ CR_k $ gates for each frequency $ k $. By summing the magic resource contributions from all frequencies up to a cut-off value $ m $ and approximating the contributions from higher frequencies, we obtain:
\begin{equation}
    \begin{aligned}
        \IMR_2(\text{QFT}) &= \sum_{k=2}^n (n+k-1) \IMR_2(CR_k) \\
        &\approx \sum_{k=2}^m (n+k-1) \IMR_2(CR_k) + \sum_{k=m}^n (n+k-1) \frac{3\pi^4}{2^{4k}} \\
        &\approx 3.4619n - 5.3388.
    \end{aligned}
\end{equation}
There are a total of $3(m^2+2mn-2n-m)/2$ J gates for the truncated QFT and $9(n^2-n)/2$ J gates for the complete QFT.  
For a sufficiently large cut-off frequency $ m $, the second term becomes negligible, and the invested magic resource cost of QFT scales linearly with the number of qubits: $ \IMR_2(\text{QFT}) \approx 3.46n - 5.34 $ (in units of T-count). This result, visualized in Fig.~\ref{fig:qft}(e), suggests that QFT can be implemented with a relatively modest amount of magic resources compared to the full complexity of the circuit.

Previous analyses of classical simulation algorithms for QFT circuits have suggested a complexity of $ O(n\log n) $ T gates \cite{nam2020approximate}. However, the invested magic resource framework provides an upper bound on the required resources, indicating that $ O(n) $ T gates might be sufficient. This aligns with the observation that truncated QFT, which primarily involves low-frequency components, can provide a good approximation of the full QFT with reduced complexity \cite{coppersmith2002approximate}. Our analysis further supports this notion by demonstrating that the majority of magic resources are concentrated in the lower frequencies, and the cut-off frequency for an accurate approximation is independent of the qubit number. Since magic resources are directly related to classical simulation complexity, we can conclude that truncated QFT exhibits similar complexity to the full QFT while requiring significantly fewer resources.

\section{Analysis on Potential Magic Resources of Linear and GHZ state}\label{Mth:1T}

\begin{figure} [htbp]
    \centering
    \includegraphics[width=0.5\linewidth]{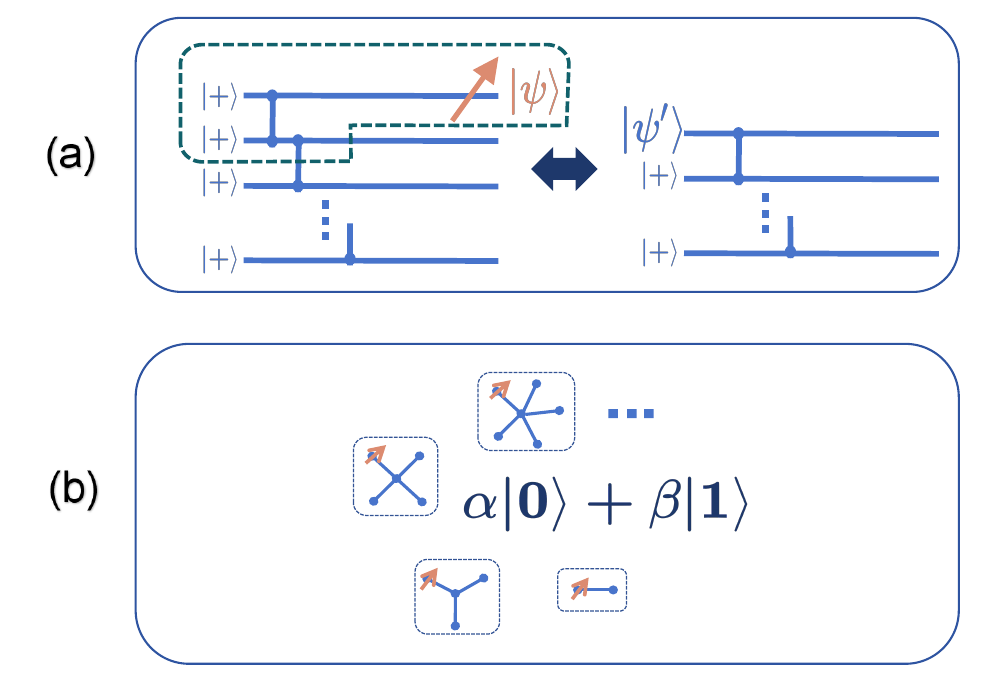}
    \caption{\textit{Computation process of the Potential Magic Resources for Linear and GHZ state}
    }
    \label{fig:potential}
\end{figure}
In this section, we prove that the potential magic resources of linear graph states and GHZ states are both $ 1\T $. 

The proof for the linear graph states is illustrated in Fig.~\ref{fig:potential}(a). A linear graph with $ n $ qubits, denoted as $ |L_n\rangle $, is defined by
$|L_n\rangle = \bigotimes_{j=1}^{n-1} E_{j+1,j}|+\rangle^{\otimes n}$,
where $ E_{j+1,j} $ is the CZ gate between the $(j+1)_{th}$ and $ j_{th} $ qubits. Suppose we perform a projection measurement with the state $ |\psi\rangle = \alpha |0\rangle + \beta |1\rangle $, the resulting state after the projection is given by $\tr_{1}\big[|\psi\rangle\langle \psi| \otimes |L_n\rangle\big] = E_{12}|\psi'\rangle|L_{n-2}\rangle$ 
with normalization, where $ |\psi'\rangle = \alpha^* |+\rangle + \beta^* |-\rangle $. As shown in Fig.~\ref{fig:potential}(a), we can move the remaining CZ gates that do not involve the measured qubit to a position after the measurement. Thus, the measurement on the first qubit is equivalent to state preparation on $ |\psi'\rangle $. This process can be repeated for each subsequent qubit measurement. Suppose we have $ m $ qubits to be measured with measurements $ M_1 $ to $ M_m $. Then, we have:
\begin{equation}
    \begin{aligned}
        &\M_2 \Big[M_m\dots M_2M_1|L_n\rangle|\Big]\\
        = &\M_2 \Big[M_m\dots M_2 E_{12}|\psi'\rangle|L_{n-2}\rangle\Big] \\
        = &\M_2 \Big [E_{12} |\psi''\rangle|L_{n-m-1}\rangle]\\ 
        = &\M_2[|\psi''\rangle] \le 1\T,
    \end{aligned}
\end{equation}
where $ |\psi''\rangle $ is some 1-qubit state related to the measurement settings and results. According to the definition of potential magic resources, we have $ \mathcal{P}(|L_n\rangle) = 1\T $.

For GHZ states, as shown in Fig.~\ref{fig:potential}(b), an $ n $-qubit GHZ state can be written as $|\text{GHZ}_n\rangle = \frac{1}{\sqrt{2}} (|\mathbf{0}_n\rangle + |\mathbf{1}_n\rangle)$,
which is locally Clifford equivalent to "star-like" graphs with a central qubit connected to $n-1$ outer qubits. Consider a projection state $|\psi\rangle = \alpha |0\rangle + \beta |1\rangle $. After projection, the remaining state becomes $ \alpha^* |\mathbf{0}_{n-1}\rangle + \beta^* |\mathbf{1}_{n-1}\rangle $. This process can be repeated multiple times. After consecutive single-qubit measurements, the GHZ state is always equivalent to $ \alpha |\mathbf{0}_{n-1}\rangle + \beta |\mathbf{1}_{n-1}\rangle $ with some $ \alpha $ and $ \beta $. Thus, we have:
\begin{equation}
    \begin{aligned}
        &\M_2 \Big[M_m\dots M_2M_1|\text{GHZ}_n\rangle\Big]\\ = & \M_2\Big[\alpha|\mathbf{0}_{n-m}\rangle+\beta|\mathbf{1}_{n-m}\rangle\Big] \\
        = & \M_2\Big[CX|0\rangle\otimes (\alpha|\mathbf{0}_{n-m-1}\rangle+\beta|\mathbf{1}_{n-m-1}\rangle) \Big]\\
        = & \M_2\Big[\alpha|0\rangle + \beta|1\rangle\Big] \le 1\T,  
    \end{aligned}
\end{equation}
where $ CX $ is the CNOT gate between $ |0\rangle $ and the remaining $ \alpha |\mathbf{0}_{n-m-1}\rangle + \beta |\mathbf{1}_{n-m-1}\rangle $. This process can be repeated until the final single qubit. According to the definition of potential magic resources, we have $ \mathcal{P}(|\text{GHZ}_n\rangle) = 1\T $.

Thus, we have proven that the potential magic resources for both linear graph states and GHZ states are $ 1\T $. 

\section{Discussion of Potential Magic Resources and its bounds}\label{Methods_P_Bounds}

The potential magic resources are defined as $\mathcal{P}(|G\rangle)=\max_{[M]}\mathbb{M}_2([M]|G\rangle)$. It is important to note that the measurement set $[M]$ is constrained by the laws of MQC, meaning that single-qubit measurements must follow the order of signal propagation. Calculating the exact potential magic resources for a specific graph state can be highly complex. 

The difficulty in computing $\mathcal{P}$ arises from its definition as the maximum achievable magic over all allowed measurement configurations. Our numerical investigations indicate that this maximum is often far from the average and typically lies in a highly non-convex region of the parameter space. This renders gradient-based optimization techniques, such as gradient descent, often ineffective. To date, the most reliable method we have employed is brute-force enumeration over a wide range of candidate measurement patterns, using different combinations of Pauli and magic bases—an approach that is computationally demanding. 

As further examples, we calculated the potential magic for several graph states using the brute-force method. The results are presented in Tab.~\ref{tab: potential-graph-state}. The code is publicly available at \cite{code_P}.

\begin{table}[ht]
    \centering
    \begin{tabular}{c|c|c}
         \textbf{Graph State} & \textbf{ $\mathcal{P}$ (X-Y measurement) } &  \textbf{ $\mathcal{P}$ (arbitrary measurement) }  \\
         \hline\hline
         \includegraphics[width=0.05\textwidth]{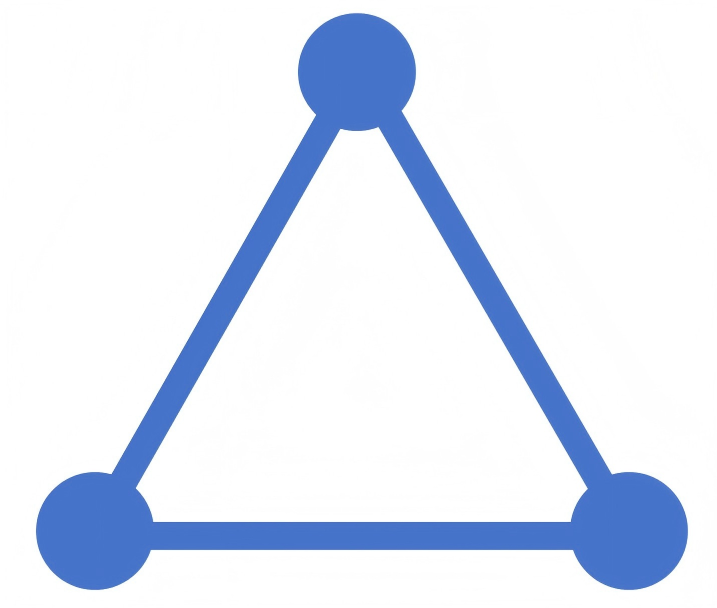} & 0.71 T & 1 T\\
         \hline
         \includegraphics[width=0.05\textwidth]{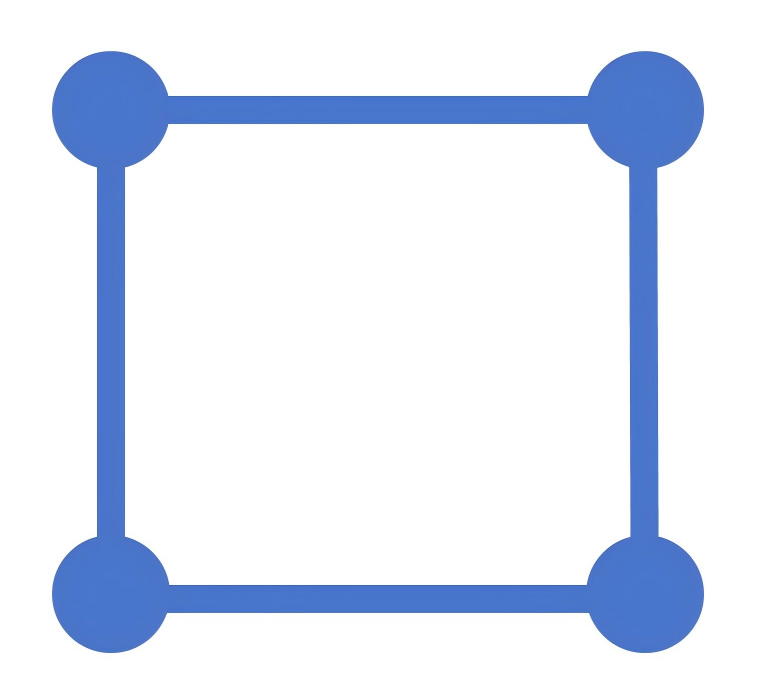} & 1.41 T & 2 T\\
         \hline
         \includegraphics[width=0.05\textwidth]{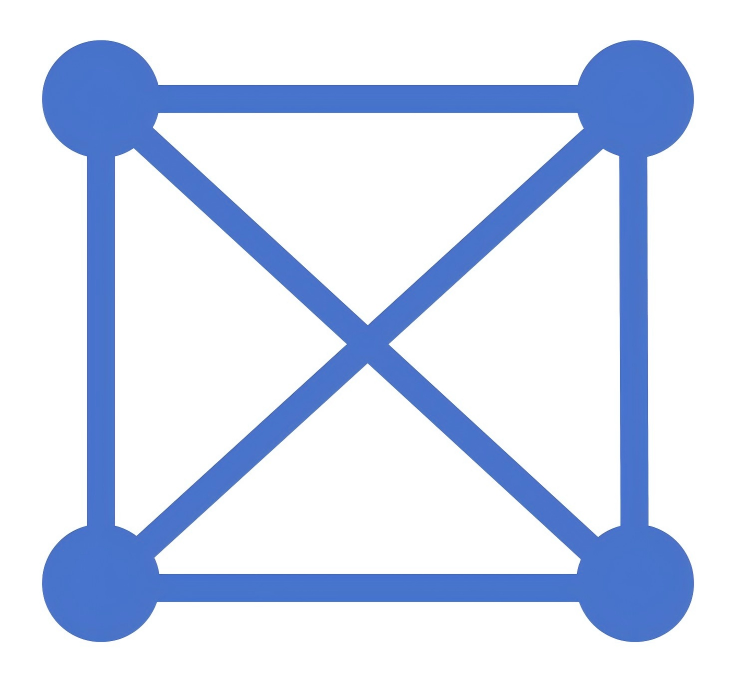} & 0.71 T & 1 T\\
         \hline
         \includegraphics[width=0.05\textwidth]{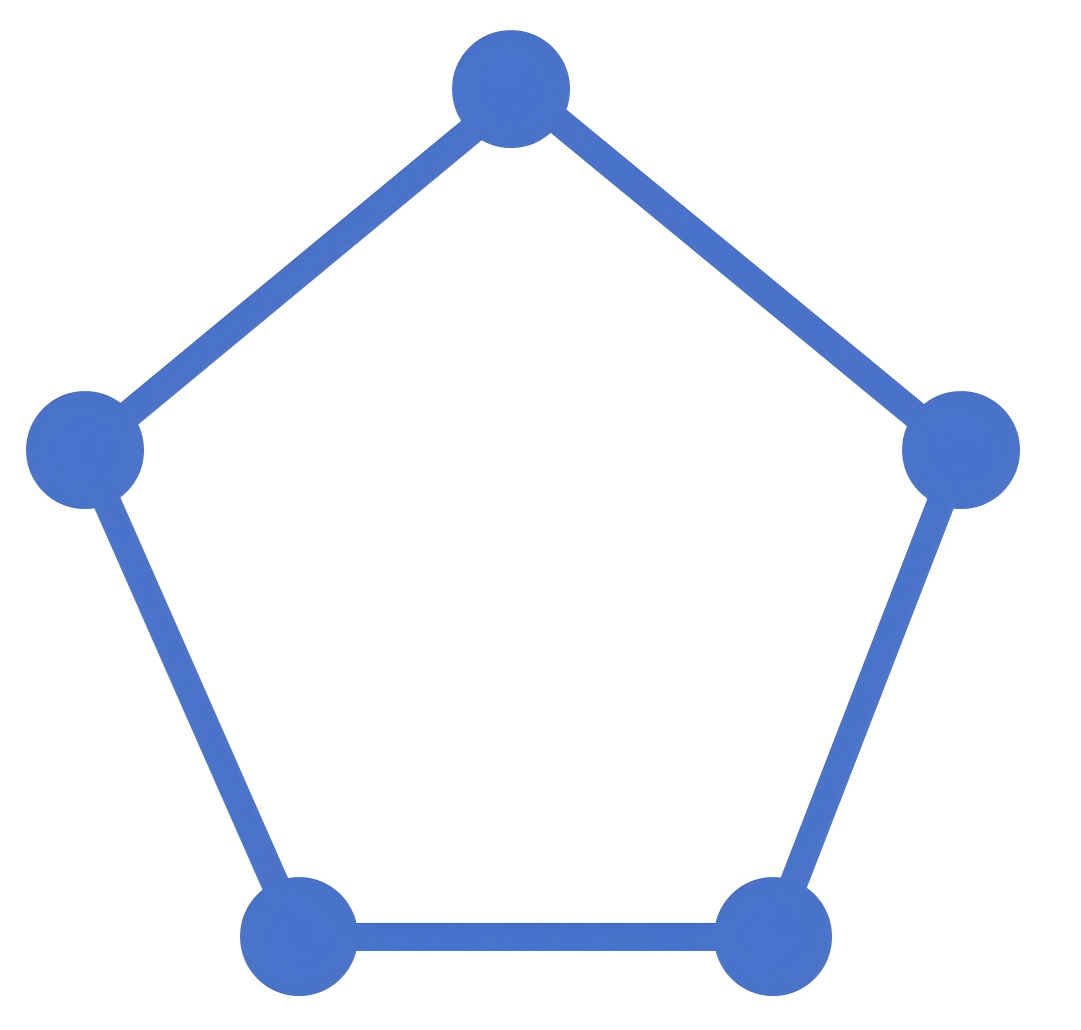} & 1.44 T & 2 T\\
         \hline
         \includegraphics[width=0.05\textwidth]{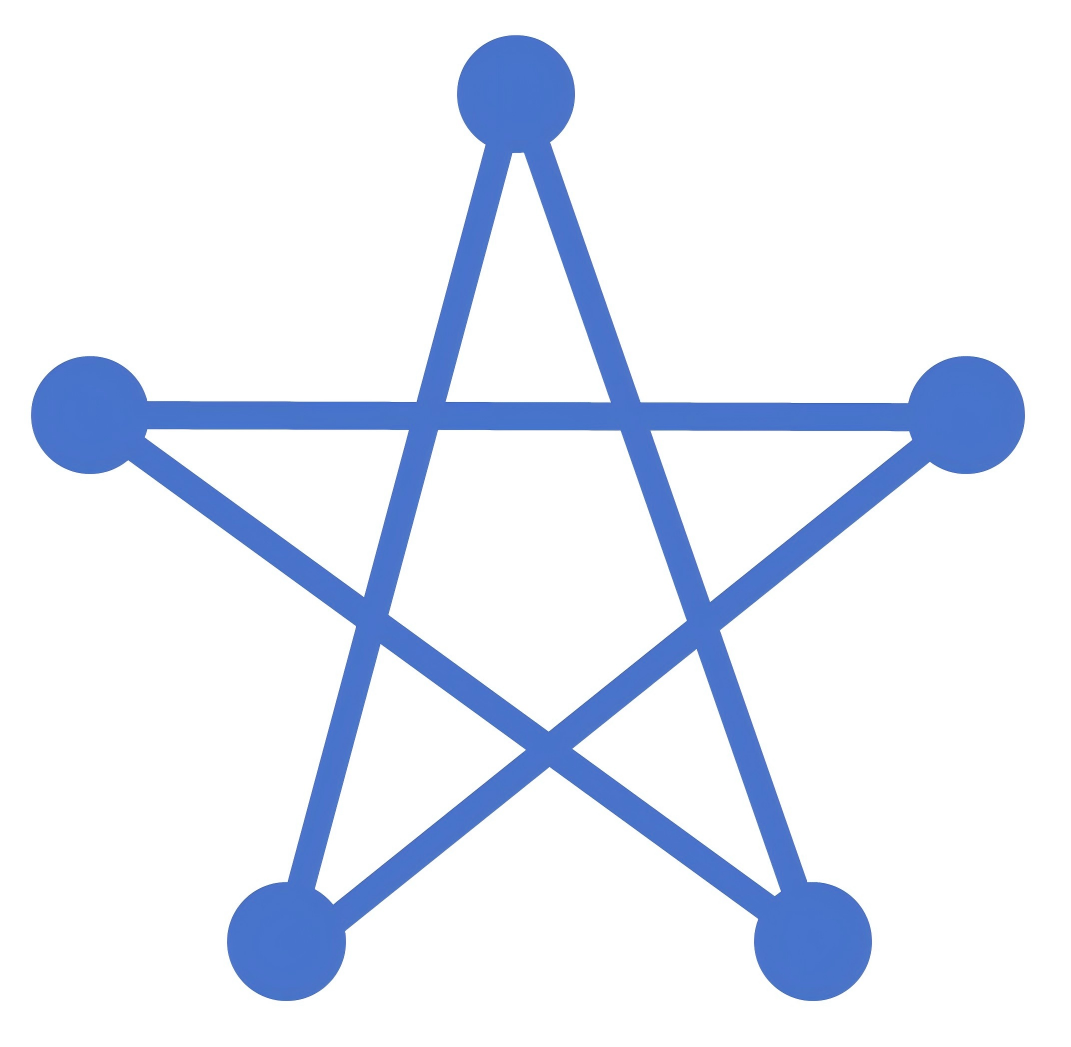} & 1.45 T & 2 T\\
         \hline
         \includegraphics[width=0.05\textwidth]{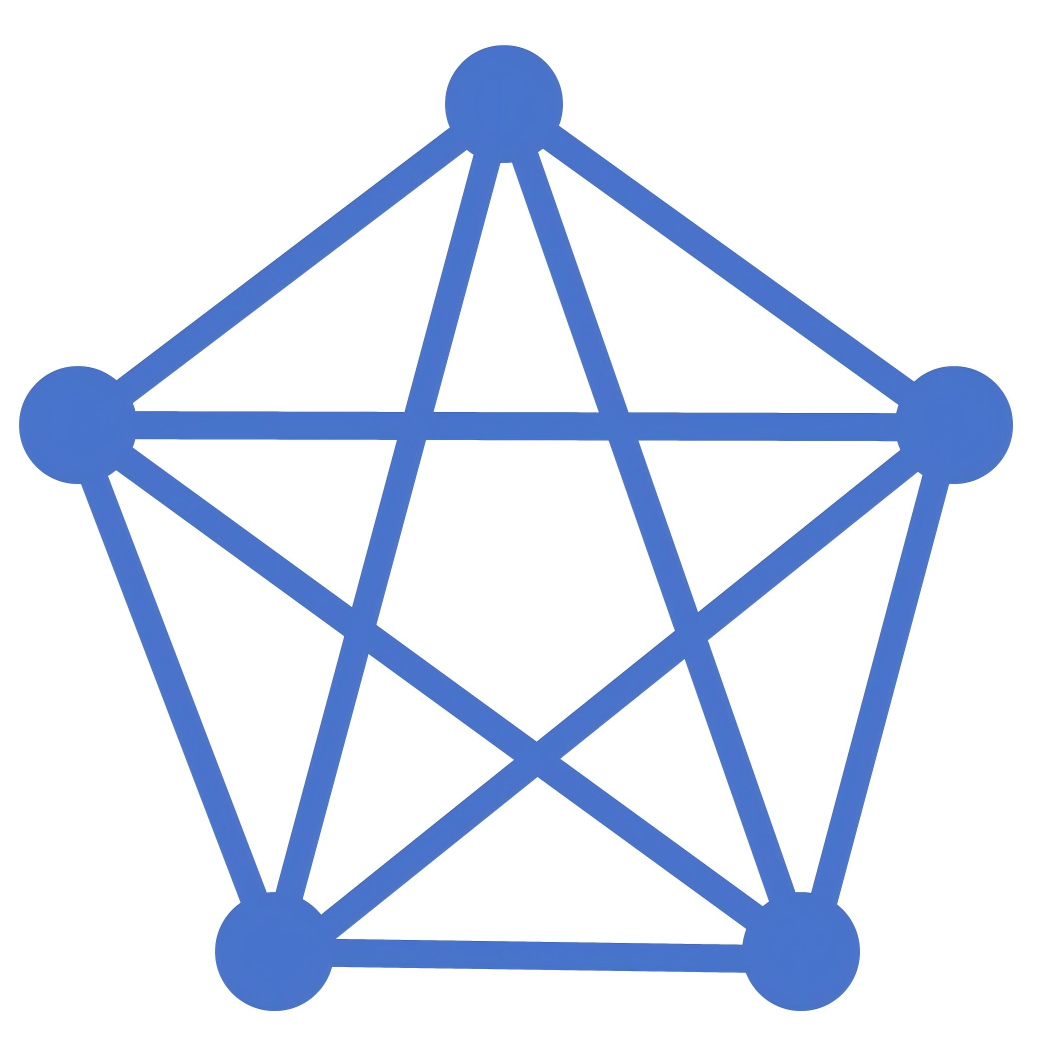} & 0.71 T & 1 T\\
    \end{tabular}
    \caption{
    \textit{Potential magic of some graph states, calculated via the brute-force method.} The columns specify the type of single-qubit measurement allowed. ``X-Y measurement" refers to projective measurements onto the equatorial plane of the Bloch sphere (basis states $(|0\rangle\pm e^{i\phi}|1\rangle)/\sqrt{2}$). ``Arbitrary measurement" refers to projective measurements onto any basis, defined by the general state $\cos\frac{\theta}{2}|0\rangle\pm \sin\frac{\theta}{2} e^{i\phi}|1\rangle$. Details of the calculation codes in \cite{code_P}. }
    \label{tab: potential-graph-state}
\end{table}

To make the problem more tractable, two complementary strategies can be employed:
(i) Restricting the problem to experimentally realistic scenarios, where measurement operations are subject to limited quantum control. In this case, one can model control imperfections as unitary noise, where each ideal measurement $M$ is replaced by an ensemble $M_V := V^\dagger M V$ with $V$ drawn from a suitable distribution—e.g., the Clifford group. Since such unitaries are isospectral, one may instead compute the average potential magic resource as $\mathcal{P}(|G\rangle) := \mathbb E_V \mathcal{R}([M]|G\rangle)$, and assess the typicality of the outcome.
(ii) Imposing structural constraints on the measurements, such as requiring symmetry (e.g., $[T,M]=0$), which reflects experimental accessibility while significantly reducing the size of the optimization landscape.

Future work could also focus on establishing upper and lower bounds for the potential magic resources, which would provide useful insight even without exact optimization. We hope this work will motivate further research into scalable estimation and optimization techniques for potential magic resources.

Additionally, we will attempt to establish the scaling behavior of potential magic resources with respect to graph dimensions.

To begin, we analyze how measurements affect the state. Suppose $[m]$ represents a set of 
$m$ single-qubit measurements. For each single-qubit measurement, we define the state as $|\psi\rangle = J(\alpha)J(\beta)J(\gamma)J(0)|0\rangle$. The graph state $|G\rangle$ is defined as $|G\rangle = E_H|+\rangle^{\otimes n}$. Using the swap law between J-gates and CZ-gates, we have the relation:
\begin{equation}
    J(\alpha)\cdot CZ = CX \cdot J(\alpha).
\end{equation}
From this, we derive the following expressions for the measurement process:
\begin{equation}
\begin{aligned}\relax
    [M]|G\rangle &= \ (\langle 0|^{\otimes m})[J]E_G|+\rangle^{\otimes n}\\
    &= \ (\langle 0|^{\otimes m})|\tilde{E}_G[J]|+\rangle^{\otimes n}\\
    &= \ (\langle 0|^{\otimes m}) \tilde{E}_G(|\psi_1\rangle|\psi_2\rangle...|\psi_m\rangle)\otimes|+\rangle^{n-m}
\end{aligned}    
\end{equation}
where $[J]$ denotes a series of J-gates and $\tilde{E}_G$ represents $E_G$ with some CZ gates replaced by CX gates, depending on the number of J-gates applied. The process then effectively becomes quantum teleportation, where each $|\psi_i\rangle$ is teleported to a connected qubit in the graph. Thus, we arrive at the following relationship for the magic resources:
\begin{equation}
    \mathbb{M}_2\big([M]|G\rangle\big) \le \sum_{i}\mathbb{M}_2(|\psi_i\rangle)
    \label{Eq: M2_bound}
\end{equation}
This indicates that the potential magic resources are directly related to the teleportation power within the graph. With the selective choice of measurement, we could make the teleportation process will not affect each other and make the equation in Eq.~\eqref{Eq: M2_bound} satisfied. 

Next, we establish bounds for the potential magic resources. The natural upper bound for
$\mathcal{P}(|G\rangle)\le nT$, corresponding to the scenario where all qubits are in the maximum single-qubit T state.

\begin{figure}[ht]
    \centering
    \includegraphics[width=0.6\linewidth]{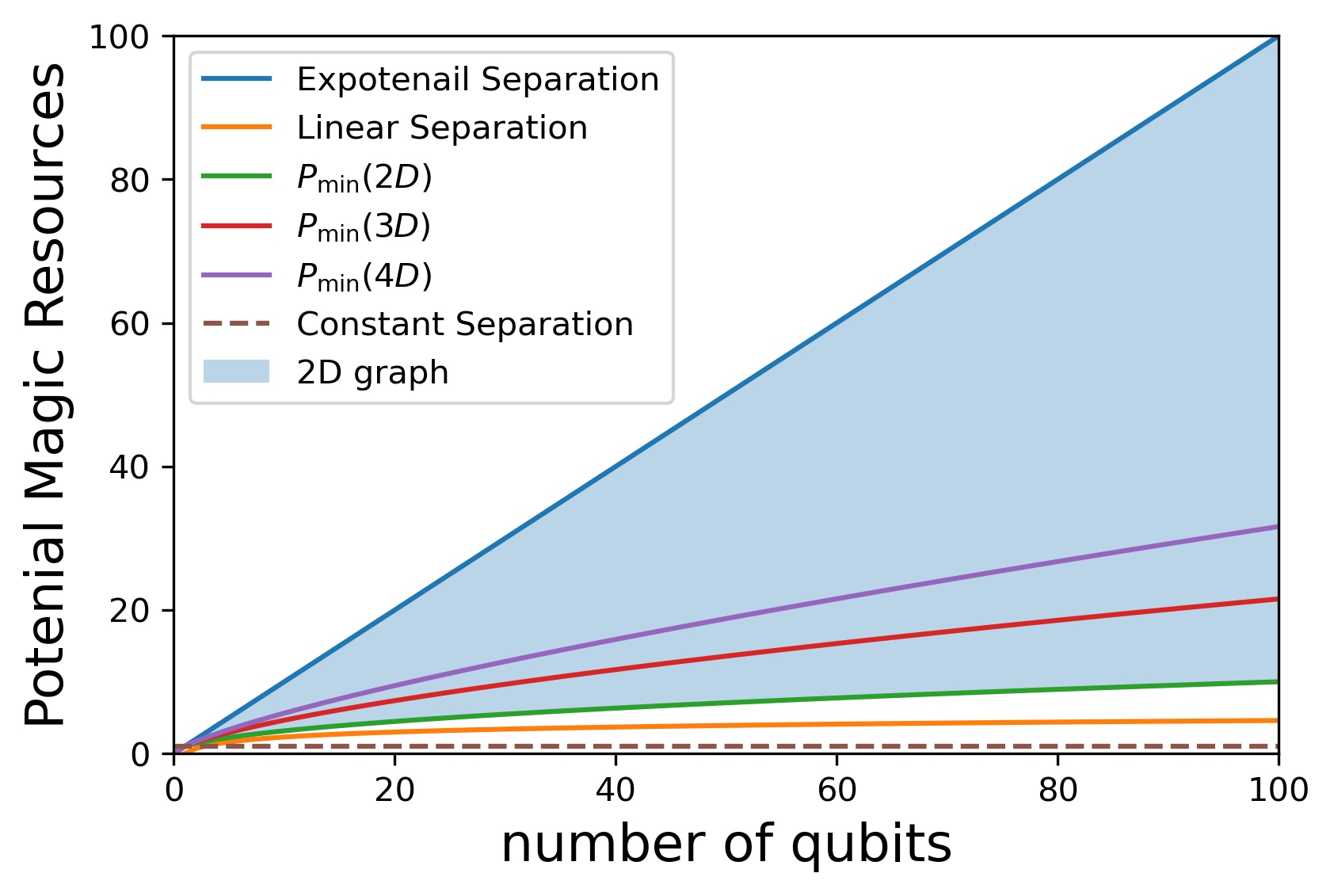}
    \caption{\textit{Demonstration of Bounds of Potential Magic Resources.} 
    The graph lines represent the potential magic resources as a function of qubit number for different graph dimensionalities: 2D, 3D, and 4D. Based on the Gottesman-Knill theorem, we define the following scaling behaviors for the potential magic resources:
    $\mathcal{P}\sim 1$ as constant separation between MQC and CC, $\mathcal{P}\sim \log{n}$ as linear gap, $\mathcal{P}\sim \sqrt{n}$ as the minimum potential magic resources for 2-D graph and $\mathcal{P}\sim n^{2/3}$ for 3-D graph, and $\mathcal{P}\sim n^{3/4}$ for 4D graph, and $\mathcal{P}\sim n$ for exponential separation. The shaded area represents the possible range of potential magic resources for the most commonly used 2D graph.}
    \label{fig: potential_bound}
\end{figure}

For the lower bound, the goal is to teleport as many qubits as possible. A natural approach is to exploit the entanglement between the outermost layer and the second outermost layer, ensuring that all qubits in the second outermost layer achieve the maximum possible magic resources. With proper measurement order and basis, we could always satisfy this goal without affecting each other. Let $N_{2nd}$ represent the number of qubits in the second outermost layer of the graph. The lower bound is then given by:
\begin{equation}
    \mathcal{P}(|G\rangle)\ge N_{2nd} 
\end{equation}
For large $n$ in a $d$-dimensional graph, the volume $V$ of the graph scales as $V=c_dx^d$, where $c_d$ is a constant related to the shape of the graph, and $x$ is the length in one dimension. The surface area $A$ is the derivative of the volume with respect to $x$, $A=\partial V/\partial x$. Thus, we have:
\begin{equation}
\begin{aligned}
    n &\sim V=c_d x^d,\\
    \mathcal{P}_{\min} &\sim A=dc_dx^{d-1} \sim n^{(d-1)/d}
\end{aligned}
\end{equation}

\begin{figure}[ht]
\centering
\includegraphics[width=0.5\linewidth]{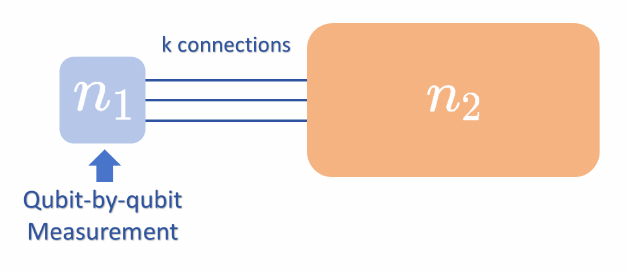}
\caption{
\textit{Illustration of how graph connectivity affects magic transport in a bipartite graph state.}
Subsystems $n_1$ (left) and $n_2$ (right) are connected by $k$ edges. Measuring all qubits in $n_1$ can inject at most $k$ magic states into $n_2$.
}
\label{Fig: MQC-connectivity}
\end{figure}

For the commonly used 2D graph, the potential magic resources scale at least as $\mathcal{P}\sim \sqrt{n}$, which already represents a gap greater than the linear gap.

For higher-dimensional graphs, we find that $\mathcal{P}\rightarrow O(n)$, leading to an exponential gap between MQC and classical computation. At this point, the effect of the graph's shape becomes negligible, and the scaling is primarily determined by the number of qubits.

Also, other structural features, like the connectivity of the graph, may affect the potential magic resources. To further clarify this point, we provide a brief supplementary discussion.

As illustrated in Fig.~\ref{Fig: MQC-connectivity}, we consider a toy model where the graph is bipartitioned into two subsystems containing $n_1$ and $n_2$ qubits, respectively, connected by $k$ edges. Within the MQC protocol, when all qubits in the $n_1$ subsystem are measured, at most $k$ T-type magic states can be injected into the $n_2$ subsystem. This observation suggests that higher connectivity between subsystems enables greater potential for magic transport.

However, there exists an important trade-off. While increasing connectivity typically facilitates magic transfer, in the extreme case of a fully connected graph, which is equivalent to a GHZ state up to local unitaries, the potential magic resource $\mathcal{P}$ may paradoxically decrease due to excessive delocalization of correlations.

This raises several intriguing questions for future investigation: 1. Does an optimal intermediate connectivity exist that maximizes $\mathcal{P}$? 2. Could spectral gap properties of the graph govern the rate of magic diffusion? 3. Can hierarchical graph designs balance localization and transport more effectively?

We leave these structural considerations for future theoretical explorations, as they may reveal deeper connections between graph topology and the distribution of magic resources.

\section{Experimental Setup}\label{Methods_Exp}

\begin{figure*}[htp]
    \centering
    \includegraphics[width=0.9\textwidth]{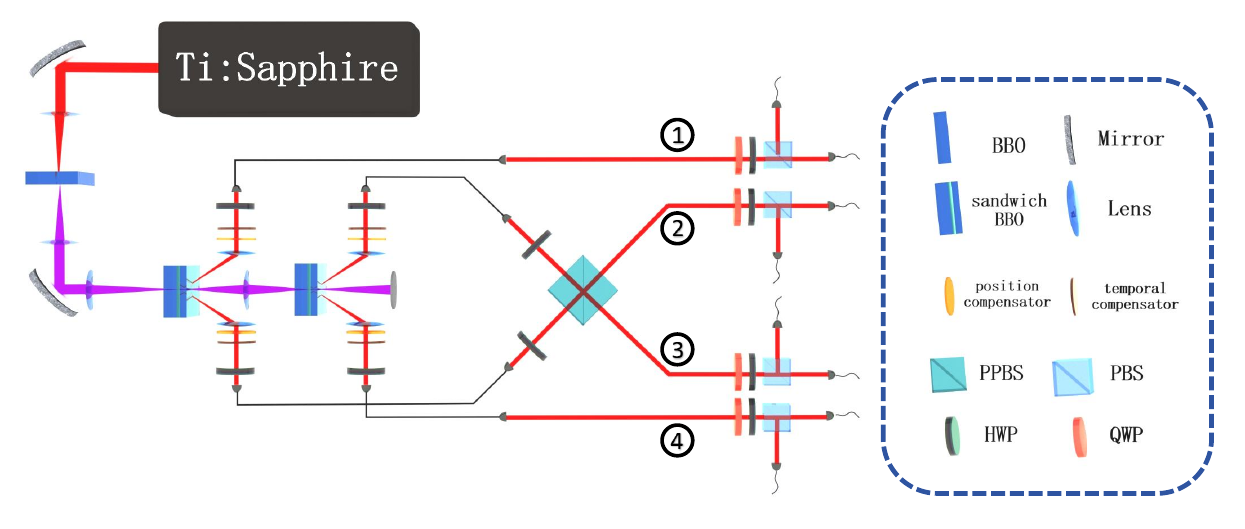}
    \caption{\textit{Scheme of the Experimental Setup.}
    A femtosecond laser centered at 780 nm (produced by a Ti: Sapphire laser) pumps a $\beta$-barium borate (BBO) crystal, generating violet light centered at 390 nm. This violet beam then pumps two separate BBO crystals in a "sandwich-like" configuration to create polarization-entangled photon pairs via type-II spontaneous parametric down-conversion (SPDC). Each sandwich BBO consists of a 2mm-thick BBO crystal, a zero-order half-wave plate, and a 1mm-thick BBO crystal. Here, the two extraordinary photons (e-photons) from both sources are directed to meet at a partial polarization beam splitter (PPBS) with $t_H = 1$ and $t_V = 1/3$. Hong-Ou-Mandel (HOM) interference occurs at the interface of PPBS. By post-selecting events where exactly one photon is detected in each output path, a cluster state is generated. This state is then measured using local single-qubit projective measurements, which involve a quarter-wave plate (QWP), a half-wave plate (HWP), and a polarization beam splitter (PBS).
    }
    \label{fig:setup}
\end{figure*}

A cluster state can be generated using an optical CNOT gate acting on two maximally entangled states. However, the success rate and fidelity of this method are not satisfactory. Therefore, we adopted the setup described in \cite{zhang2016experimental} to produce a high-quality cluster state from two non-maximally entangled states and a partial-polarization beam splitter (PPBS). The setup to produce the cluster state is shown in Fig.~\ref{fig:setup}. A femtosecond violet light sequentially pumps two sandwich-like $\beta$-barium borate (BBO) crystals. Each BBO sandwich comprises a 2mm-thick BBO crystal, a zero-order half-wave plate, and a 1mm-thick BBO crystal. The BBO crystals are beam-like cut \cite{kurtsiefer2001generation} to enhance brightness. The 2mm-thick BBO triples the brightness of the SPDC photons compared to the 1mm-thick BBO. This, along with position and temporal compensation and a half-wave plate (HWP) on one side, enables the generation of non-maximally polarization-entangled photons in the state $(|HH\rangle+\sqrt{3}|VV\rangle)/\sqrt{2}$. By adjusting the delay lines between the two sources, the two extraordinary photons (e-photons) from both sources are directed to meet at a PPBS. This PPBS has transmission coefficients of $t_H = 1$ for horizontally polarized light and $t_V = 1/3$ for vertically polarized light, and it introduces an $i$ phase shift for the reflected photon. When two vertically polarized photons meet at the PPBS interface, Hong-Ou-Mandel interference occurs. Consequently, a 4-photon polarization cluster state emerges after the PPBS, described by $|\text{cluster}\rangle=(|HHHH\rangle+|HHVV\rangle+|VVHH\rangle-|VVVV\rangle)/\sqrt{2}$, with a success rate of $1/4$. Following the definition of the cluster state, the qubit labels are assigned as shown in Fig.~\ref{fig:setup}. Photons from the first source represent qubits 1 and 2, while photons from the second source represent qubits 3 and 4. To optimize performance, 2nm filters are placed on the branches of qubits 2 and 3, and 3nm filters are placed on qubits 1 and 4, to eliminate the influence of frequency correlations. Further setup details can be found in \cite{zhang2016experimental}.

\begin{figure}
    \centering
    \includegraphics[width=0.6\textwidth]{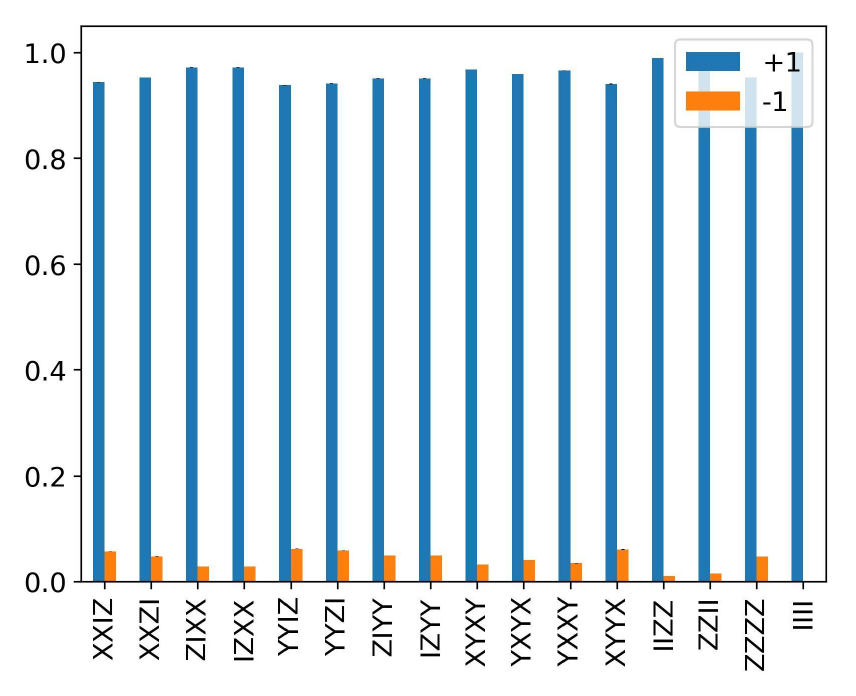}
    \caption{\textit{Stabilizer of the cluster state.} Experimental results on testing the 16 stabilizers of the cluster state, which are calculated from 9 bases in the experiments: ZZZZ, ZZXX, XXZZ, YYZZ, ZZYY, XYXY, YXYX, XYYX, YXXY. ``+1" and ``-1" are for the eigenvalues of the stabilizer, and the bar is for the percentage per eigenstate in experiments.  
    }
    \label{fig:stabilizer}
\end{figure}

Fig.~\ref{fig:stabilizer} shows the performance of the cluster state. A cluster state can be defined by 16 stabilizer bases, $B_i$, which can be calculated from 9 bases in experiments. To avoid multi-photon errors, the violet beam power was set to 40mW. By collecting 200 incidents per basis, we obtained the results shown in Fig.~\ref{fig:stabilizer}. The overall fidelity is $\langle \text{cluster}|\psi|\text{cluster}\rangle= \sum_{i=1}^{16} \langle B_i\rangle = 0.923\pm 0.001$. The error is calculated using Bootstrap over 80 incidents.

\section{Estimating Reserved Magic Resources from Randomized Measurement}\label{Mth:fewshot}
One experimental task involves quantifying the remaining magic resources $\R$ within a quantum state after undergoing MQC measurement. Since $\R (\ket{G})= \M_2([M]|G\rangle)$, estimating $\R$ involves estimating $\M_2$ in the experiments. As shown in Ref.~\cite{oliviero2022measuring}, $\M_2$ can be effectively estimated on IBM's 5-qubit and 7-qubit quantum processors.

Following established methods, we experimentally determined $\M_2$ using randomized measurements. $\M_2$ is calculated as $-\log \sum_P \Xi_P^2 - \log d$, where $\Xi_P = \langle P\rangle^2 / d$ and $P$ represents elements within the Pauli group $\mathcal{P}(n)$ excluding the identity and global phase factors. Interestingly, $\M_2$ can also be understood as a fourth-order property of a quantum state, expressible as $\M_2 = -\text{tr}(Q\psi^{\otimes 4}) - \log d$ for pure states $\psi$. Here, $Q = d^{-2} \sum_P P^{\otimes 4}$ involves the fourth tensor power of Pauli operators. While the Clifford group does not form a 4-design necessary for direct estimation of this fourth-order property, we can still leverage randomized Clifford measurements to estimate $\M_2$ using the following expression with every 4-incidents pair:
\begin{equation}
\begin{aligned}
    \M_2(\psi) =& -\log \mathbb{E}_{\vec{s}}\left[(-2)^{-s_1\oplus s_2\oplus s_3\oplus s_4}\mathbb{E}_{C \in \mathcal{C}l(n)}\prod_{i=1}^4\mathrm{Pr}(s_i|C)\right]- \log d,
\end{aligned}
\end{equation}
where $s_i$ is an n-bit string as outcomes of the experiments and $\oplus$ denotes bit-wise binary addition, $s_1 \oplus s_2 \oplus s_3 \oplus s_4 = \sum_i s_1^i \land s_2^i \land s_3^i \land s_4^i$.

For mixed states $\rho$, the Pauli spectrum $\Xi_P$ requires normalization since $\sum_P \Xi_P = \text{tr}(\rho^2)$. We introduce $\tilde{\Xi}_P = \Xi_P / \text{tr}(\rho^2)$ and modify $\M_2$ to incorporate the 2-R\'{e}nyi entropy of the state, $\mathbb{S}_2(\rho) = -\log \text{tr}(\rho^2)$. We have $\M_2 = -\text{tr}(Q\rho^{\otimes 4}) - \log d - \mathbb{S}_2(\rho)$. Both $\M_2$ and $\mathbb{S}_2(\rho)$ can be estimated from the same randomized Clifford measurements, exploiting the fact that the Clifford group forms a 3-design \cite{zhu2017multiqubit,li2023experimental}. $\mathbb{S}_2(\rho) = -\log \sum_{(s_1, s_2)}(-2)^{s_1 \oplus s_2}\mathbb{E}_{C \in \mathcal{C}l(n)}\prod_{i=1}^2\mathrm{Pr}(s_i|C) - \log d$ \cite{brydges2019probing, elben2019statistical, elben2023randomized}. The resulting expression for mixed states becomes:
\begin{equation}
\begin{aligned}
    \M_2(\rho) = &- \log \mathbb{E}_{\vec{s}}\left[(-2)^{-s_1\oplus s_2\oplus s_3\oplus s_4}\mathbb{E}_{C \in \mathcal{C}l(n)}\prod_{i=1}^4\mathrm{Pr}(s_i|C)\right] \\
    & - \log \mathbb{E}_{\vec{s}}\left[(-2)^{s_1 \oplus s_2}\mathbb{E}_{C \in \mathcal{C}l(n)}\prod_{i=1}^2\mathrm{Pr}(s_i|C)\right]. 
\end{aligned}
\label{eq:S2mixed_RM}
\end{equation}

Indeed, since the fourth-order operator $Q$ and the second-order Swap operator can both be written as the product of single-qubit ones, the previous estimators can be evaluated from the random single-qubit Clifford unitary, say the Pauli measurement \cite{oliviero2022measuring}. Moreover, employing the few-shot treatment \cite{singlezhou, my_p3ppt_paper}, we arrive at a concise few-shot estimator of $\M_2$ used in the experiment:
\begin{equation}
\begin{aligned}
    \M_2(\rho) =& -\log \mathbb{E}_P \mathbb{E}_{\vec{s}}\left[(-2)^{-s_1 \oplus s_2 \oplus s_3 \oplus s_4}\right]\\
    &\quad - \log \mathbb{E}_P \mathbb{E}_{\vec{s}}\left[(-2)^{-s_1 \oplus s_2}\right],
\end{aligned}\label{eq:S2mixed_FSRM}
\end{equation}
where $P$ represents Pauli measurement setting chosen from the set $\{X, Y, Z\}^{\otimes n}$ with equal possibility of $1 / 3^n$. The estimation process is as follows: (1) randomly select a Pauli measurement $P$ from $\{X, Y, Z\}^{\otimes n}$, (2) collect $N_M$ outcomes, each an n-bit string, (3) for each 4-string pair in $N_M$ outcomes, calculate $(-2)^{-s_1 \oplus s_2 \oplus s_3 \oplus s_4}$ and for each 2-string pair in $N_M$ outcomes, calculate $(-2)^{-s_1 \oplus s_2}$, and (4) calculate the means of the calculated values for all Pauli measurements using Eq.\eqref{eq:S2mixed_FSRM}.

\begin{figure}
    \centering
    \includegraphics[width=0.9\textwidth]{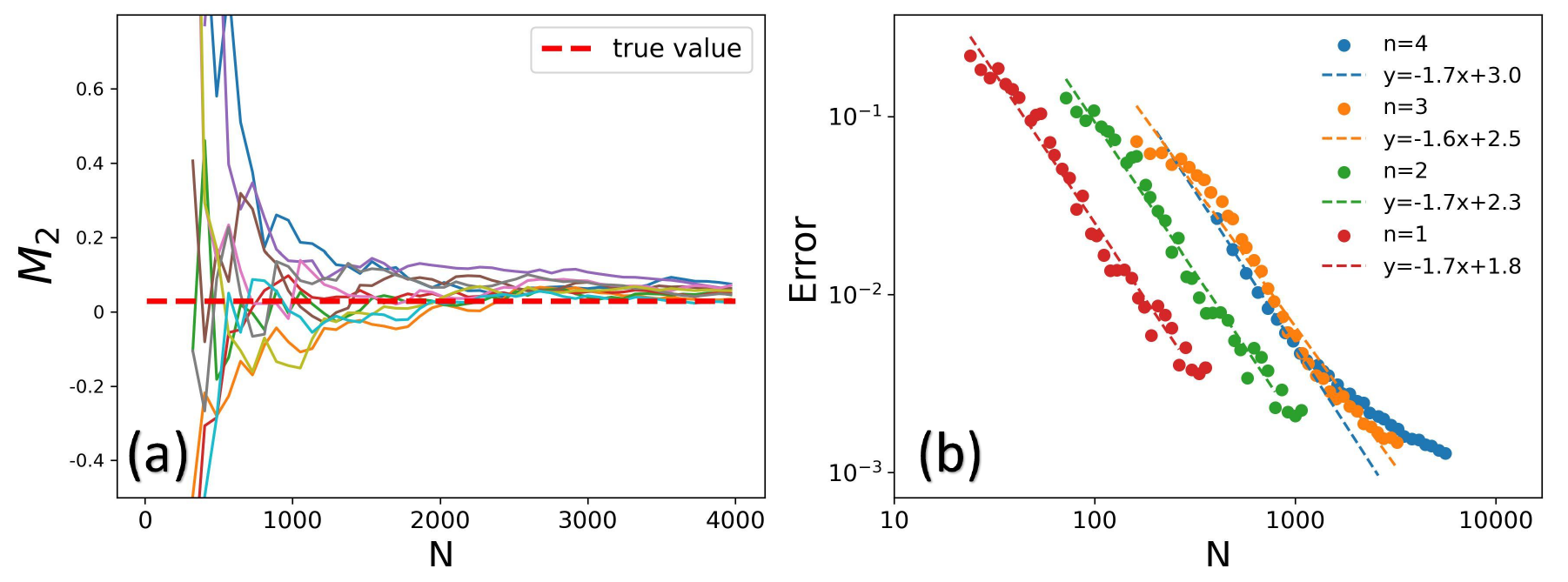}
    \caption{
    \textit{The experimental demonstration of measuring $\M_2$ with randomized measurement.} (a) Estimation of $\M_2$ of cluster state over the number of collected photons. (b) Scaling of estimation $\M_2$ over the number of collected photons for $n=1$ to $n=4$ linear cluster states in a log-log diagram. Dots represent the experimental results with uncertainty calculated from Bootstrap over the total photon number used. The dashed line represents the linear regression results.
    }
    \label{fig:exp_RM}
\end{figure}

The experimental implementation and results of the randomized measurement (RM) method for estimating $\M_2$ are presented in Fig.~\ref{fig:exp_RM}. This figure showcases an example of using RM to quantify the magic within a 4-qubit cluster state (a specific entangled state) and illustrates the performance. A graphics processing unit (GPU), Nvidia GeForce RTX 3070, is used to accelerate the post-processing. Estimating $\M_2$ involves calculating the value of $(-2)^{s_1 \oplus s_2 \oplus s_3 \oplus s_4}$ for every possible combination of four measurement outcomes within each measurement setting. With $N$ measurement outcomes, the number of such 4-incident-pairs $N' = \binom{N}{4}$ scales as $N^4$. Consequently, for a small number of measurement outcomes, the estimation error of $\M_2$ exhibits a $1/N^2$ dependence. However, as $N$ increases further, the uncertainty reduction slows and approaches $1/\sqrt{N}$ scaling. This phenomenon has been previously noted in similar studies in estimating entanglement \cite{elben2020mixed}. In the experiments, the linear regression in Fig.~\ref{fig:exp_RM}(b) shows that $\log \text{Err} = -1.7 \log N + b$, validating the calculation. The experimental data align with this theoretical understanding. A linear regression analysis on the log-log diagram in Fig.~\ref{fig:exp_RM}(b) observes a slope of $-1.7$. This result validates the expected scaling behavior and confirms the effectiveness of the RM method for estimating $\M_2$.

\section{Experimental Results for Linear graph and Box graph}

We scrutinize two typical processes: single-qubit rotation and quantum Fourier transformation, from which we demonstrate how the magic responses toward single-qubit measurement for 1-D graph and 2-D graph states. Here, the 1D graph state is the linear state, and the 2D graph state is the BOX state. Both of them can be from the 4-qubit cluster state, $|\text{cluster}\rangle = (|0000\rangle+|0011\rangle+|1100\rangle-|1111\rangle)/2$ \cite{walther2005experimental}. A linear state is equivalent to the cluster state with local operations $H\otimes I\otimes I\otimes H$. The BOX state is equivalent to the cluster state with local operations $H\otimes H\otimes H\otimes H$ and a SWAP operator between qubit 2 and qubit 3, which is equivalent to relabeling the qubits. 
The experimental estimation of the reserved magic resources comes from the few-shot randomized measurement\cite{my_p3ppt_paper} of each reserved state after one step of MQC, details see Methods.

\subsection{Arbitrary rotation from 1-D linear state}
Firstly, we demonstrate how the magic resources are altered in the Linear graph. For a linear graph, an arbitrary single-qubit rotation is feasible. For a single qubit state, the one that contains the most magic resources is Kitaev's T-state $|T\rangle = \cos{\theta_m} |0\rangle + \sin{\theta_m}e^{i\pi/4}|1\rangle$ with $\cos2\theta_m = 1/\sqrt{3}$\cite{bravyi2005universal}, which boast the magic resources of $\M_2(|T\rangle)=\log 3/2$. Another commonly used state is the Kitaev's H state $|H\rangle = (|0\rangle+e^{i\pi/4}|1\rangle)/\sqrt{2}$, whose magic resource is $\M_2(|H\rangle)=\log 4/3$. Compared with the T-state, the H state failed to be the maximum magic single-qubit state. Thus, for fairness, we use the T state magic resource as a unit to scale the invested and the reserved magic resources, this is we will use expressions like $\M_2(|T\rangle^{\otimes n}) = n \mathrm{T}$ to describe the magic resource, denoting as \textit{T-count}. 

\begin{figure}[htbp]
    \centering
    \includegraphics[width=0.8\textwidth]{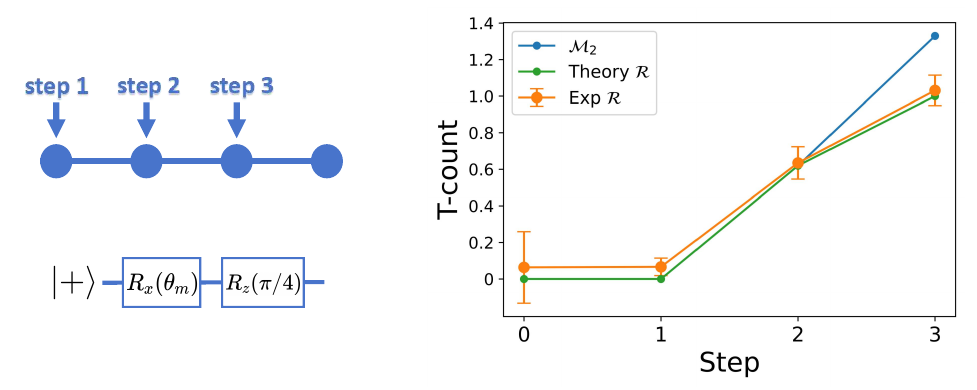}
    \caption{
        \textit{Experimental generation of the $|T\rangle$ state using a 1D graph.} The figure illustrates the three-step process used to generate a $|T\rangle$ state with a linear graph. The left side visually depicts the process, starting from an initial cluster state (Step 0) and progressing through three measurement steps with a standard MQC pattern. The measurement angles are ${0,\theta_m,\pi/4}$ for steps 1, 2, and 3, respectively. The plot on the right displays the total invested magic resources $\M_2$ and the reversed magic resource $\M_2$ at each step. Error bars are estimated using the bootstrap method. For step 0, an average of 50 incidents per measurement basis was used, and for other steps, an average of 80 incidents per measurement basis was used. 
    }
    \label{fig:exp_1D}
\end{figure}
Our experiment utilizes a linear graph and MQC to generate the $|T\rangle$ state, which serves as the maximum magic state with a single qubit. This process involves three sequential measurements: first, a measurement in the computational basis ${|0\rangle, |1\rangle}$ on the first qubit; then, a measurement in the basis $(|0\rangle\pm e^{i\theta_m}|1\rangle)/\sqrt{2}$ on the second qubit; and finally, a measurement in the basis $(|0\rangle\pm e^{i\pi/4}|1\rangle)/\sqrt{2}$ on the third qubit. The process can be mathematically expressed with a cluster state in the CME pattern as:
\begin{equation}
    |T\rangle = X^{s_2}Z^{s_1+s_3}[M_{3}^{\pi/4}]^{s_2}[M_2^{\theta_m}]^{s_1}[M_1^0]|\psi_{cluster}\rangle.
\end{equation}
Here, $s_i$ represents the measurement outcome (0 or 1) on the $i$th qubit. While the measurement outcomes on each qubit are random and influence subsequent measurement settings and corrections, our previous analysis established that they do not affect the overall magic resources within the system. Therefore, we average the final magic resources over all possible single-qubit measurement outcomes to obtain an estimation. 

The comparison between the invested magic resources and the reserved magic resources within the final state is illustrated in Fig.\ref{fig:exp_1D}. Due to inherent experimental imperfections and state preparation limitations, the measured magic resources may slightly exceed the theoretical values.
As Fig.\ref{fig:exp_1D} demonstrates, the invested magic and the reserved magic are equal during the first and second steps of the process. However, the third step inevitably results in a loss of magic resources. This is because a single qubit has a limited capacity of 1T to hold magic, and exceeding this limit results in a dissipation of the excess magic.

\subsection{QFT from 2-D Box state}
\begin{figure}[htbp]
    \centering
    \includegraphics[width=0.8\textwidth]{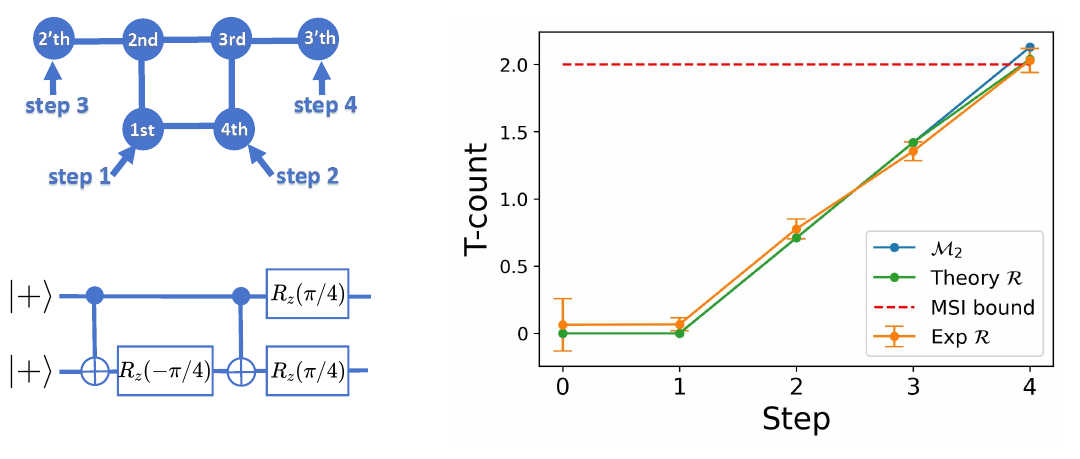}
    \caption{
    \textit{Experimental realization of Quantum Fourier Transform (QFT) using a 2D graph.} The left side of the figure depicts the 2D graph structure employed for implementing QFT and indicates the order of measurements performed on the qubits (steps 1 to 4). Additionally, it shows the equivalent circuit representation for generating the desired output state, $|CS\rangle$, with $n=2$ qubits. The right side of the figure presents the evolution of invested and reserved magic resources throughout the QFT process. The blue line represents the invested magic, quantified by $\M_2$, while the green line depicts the theoretical expectation for the reserved magic resources ($\M_2$). Steps 0 and 1 are the same as Fig.\ref{fig:exp_1D} and the estimated $\M_2$ and error bars are calculated from an average of 80 incidents per measurement basis. The red dashed line indicates the upper bound for magic state injection, which is 2T in this case, as only two qubits are retained in the final state. The orange dots represent the experimentally measured values of $\M_2$, demonstrating good agreement with the theoretical predictions.
    }
    \label{fig:exp_2D_QFT}
\end{figure}

QFT is a crucial component in quantum algorithms but has seen limited experimental demonstrations, even for the simplest case of $n=2$ qubits. From a circuit perspective, QFT requires controlled-rotation gates with arbitrary angles, denoted as $CR_k = \text{diag}[1,1,1,e^{i2\pi/2^k}]$. While $CR_1$ corresponds to a CZ gate, which doesn't require magic resources, implementing QFT for $n=2$ necessitates a special two-qubit gate known as the CS gate ($CS = CR_2 = \text{diag}[1,1,1,i]$). As illustrated on the left side of Fig.\ref{fig:exp_2D_QFT}, the CS gate can be decomposed into three T gates ($T = R_z(\pi/4) = R_3 = \text{diag}[1,e^{i\pi/4}]$), each of which can be implemented using MQC with the depicted graph structure and measurement pattern.

The power of QFT for $n=2$ is evident in the resulting QFT state, defined as $|\text{QFT}_n\rangle = \text{QFT}_n |+\rangle|0\rangle^{\otimes(n-1)}$, which possesses a nullity of $\nu(|\text{QFT}_n)\rangle = n-2$\cite{beverland2020lower}. Specifically, for $n=2$, the QFT state is $|\text{QFT}_2\rangle = |CS\rangle = (|00\rangle+|01\rangle+|10\rangle +i|11\rangle)/2$, representing the state with maximum magic content for a two-qubit system. As demonstrated in Fig.\ref{fig:exp_2D_QFT}, the $|CS\rangle$ state can be efficiently generated using a box-shaped graph structure, which is equivalent to a cluster state. For experimental convenience, the measurements in steps 3 and 4 are implemented as single-qubit rotations on qubits 3 and 4, respectively. The resulting CME pattern for generating the $|CS\rangle$ state is then given by:
\begin{equation*}
    |CS\rangle = X_3Z_3^{s_4}Z_2^{s_1} \hat{T}_3 \hat{T}_2 [M_4^{(\pi/8, 0)}][M_1^0]|\text{cluster}\rangle,
\end{equation*}
where $M_4^{(\pi/8,0)}$ represents a measurement on the 4th qubit in the basis $\cos\frac{\pi}{8}|0\rangle\pm \sin\frac{\pi}{8}|1\rangle$, while $M_1^0$ denotes a measurement on the 1st qubit in the basis $(|0\rangle\pm |1\rangle)/\sqrt{2}$.

The invested magic resources increase by approximately $\log\frac{4}{3}/\log\frac{3}{2}\approx$0.71T with each step of the process. Consequently, the total invested magic resources for steps 1 to 4 are 0T, 0.71T, 1.42T, and 2.13T, respectively. The corresponding reserved magic resources are 0, 0.71T, 1.41T, and 2.03T. In the experiments, we get $0.066\pm0.024 T, 0.777\pm 0.036 T, 1.354\pm 0.036 T, 2.030\pm 0.045 T$. Notably, all the invested magic resources are successfully injected into the system except for the last step, where a small amount of magic is lost. This loss occurs because 2.03T represents the maximum magic capacity of a two-qubit state, similar to how water spills over when a glass is filled to the brim. This experiment highlights the efficiency of MQC in utilizing and transferring magic resources.

To demonstrate that our conclusions are robust to the choice of magic quantifier, we repeated our analysis using the GKP-based magic measure proposed by Hahn et al. \cite{hahn2022quantifying}. This measure is equivalent to half of the 1/2-R\'{e}nyi entropy of the Pauli spectrum, $\frac{1}{2}\mathbb{M}_{1/2}(\psi)$. For a consistent comparison, we define a new unit of magic, $1T'= \frac{1}{2}\mathbb{M}_{1/2}(|T\rangle)$, and re-calculate the Invested, Potential, and Reserved magic resources accordingly.
The theoretically expected values for the cumulative invested magic are $[0T', 0T', 0.603T', 1.207T', 1.810T']$, while the corresponding reserved magic values are  $[0T', 0T', 0.603T', 1.207T', 1.794T']$. This confirms that the invested magic is almost perfectly converted into reserved magic, with only a negligible loss of $\mathcal{W}=\mathcal{M}-\mathcal{R}=0.016T'$, occurring in the final step. This outcome demonstrates that our analysis and its conclusions are not dependent on a specific magic measure.

\section{Arbitrary Measurement-Induced Magic Resources}\label{Methods_arbitrary}
\begin{figure*}[htbp]
    \centering
    \includegraphics[width=0.8\textwidth]{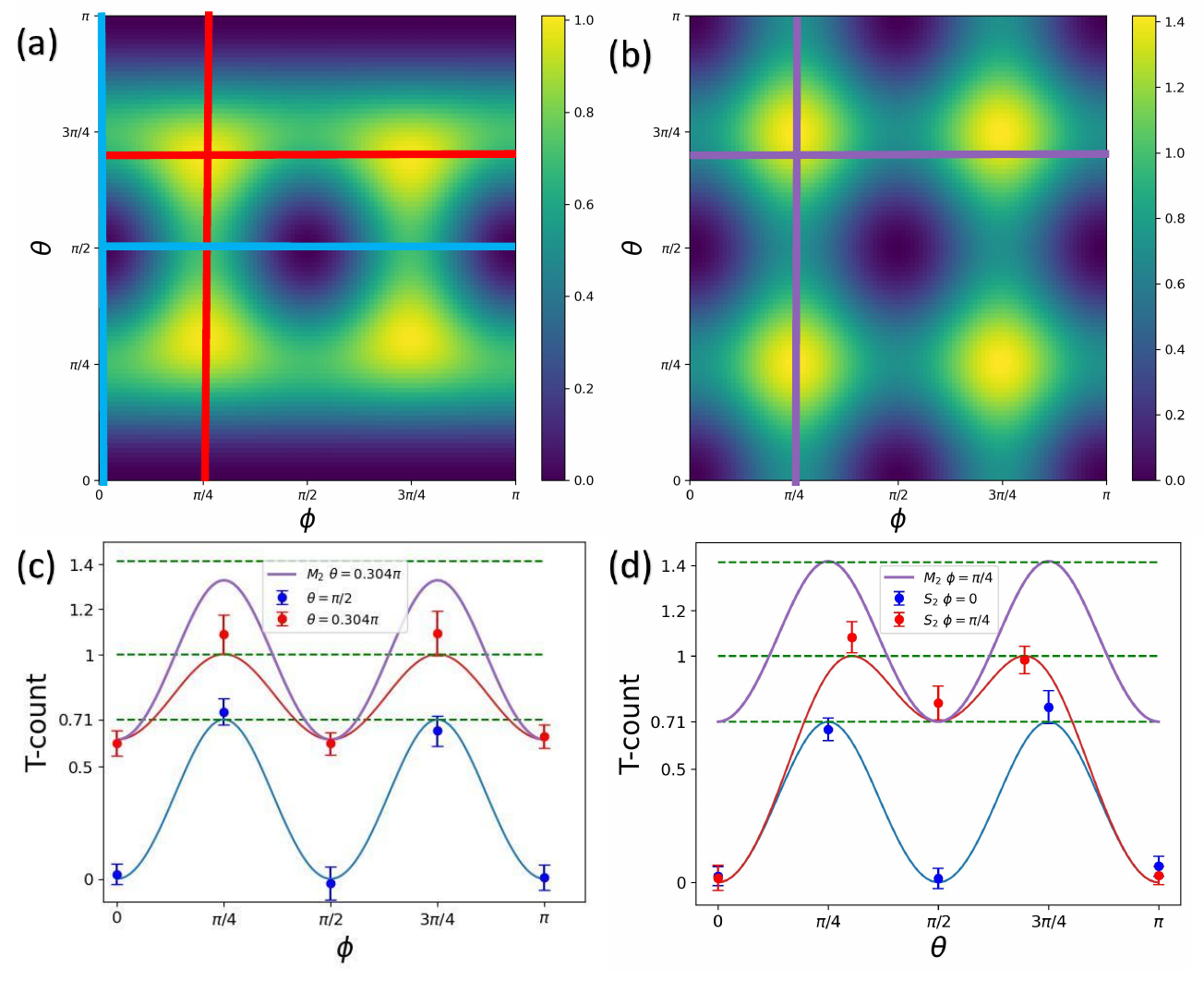}
    \caption{
    \textit{Magic resources under arbitrary single-qubit measurements.} This figure explores the relationship between invested and reserved magic resources when performing arbitrary single-qubit measurements of the form $|\psi\rangle = \cos\theta |0\rangle +\sin\theta e^{i\phi}|1\rangle$ on a maximally entangled state. (a) illustrates the reserved magic resources after the measurement, while (b) depicts the invested magic resources following the standard MQC pattern, calculated as $\M_2(|\psi\rangle) = \M_2(M^{\theta}) + \M_2(M^{\phi})$. 
    (c) and (d) focus on specific cases with fixed angles. (c) shows the invested magic resources for fixed values of $\theta$, comparing the arbitrary measurement case (red line for $\theta=\theta_m$) with the standard MQC pattern (grey line for $\theta=\theta_m$) and a measurement with $\theta=\pi/2$ (blue line). 
    Similarly, (d) presents the invested magic resources for fixed values of $\phi$, comparing the arbitrary measurement case (red line for $\phi=\pi/4$) with the standard MQC pattern (grey line for $\phi=\pi/4$) and measurement with $\phi=0$ (blue line). Dots in (c) and (d) represent experimental results with error bars from Bootstrap with 80 incidents per Pauli basis, which demonstrate good agreement with the theoretical predictions. Importantly, the invested magic resources for the arbitrary form measurements always match the reserved magic resources, as indicated by the alignment of the blue and red lines in subfigures (a) and (c)-(d) with the corresponding lines in (b).
    }
    \label{fig:exp_arbitrary_meas}
\end{figure*}

Measurement-based quantum computation (MQC) with linear graphs allows for arbitrary single-qubit rotations, enabling versatile quantum operations. However, the standard MQC approach, which employs measurements of the form $(|0\rangle + e^{i\phi}|1\rangle)/\sqrt{2}$, can lead to unnecessary consumption of magic resources, as demonstrated in Fig.~3(a). To address this issue, we explore the use of arbitrary single-qubit measurements to minimize magic resource waste. Consider an arbitrary single-qubit measurement described by the basis states $|\psi_\pm\rangle = \cos\theta|0\rangle \pm \sin\theta e^{i\phi}|1\rangle$. We can show that $|\psi_+\rangle$ can be obtained by a maximally entangled state ($E_{12}|+\rangle^{\otimes 2}$) followed by a specific single-qubit measurement $M_1^{(\theta,\phi)}$ and a Pauli Z correction ($Z_2^s$) depending on the measurement outcome: $|\psi_+\rangle = Z_2^s M_1^{(\theta,\phi)} E_{12}|+\rangle^{\otimes 2}$. In this case, the invested magic resources are equal to the $\M_2$ of the outcome state $|\psi\rangle$. 

Compared to the standard MQC approach, where the invested magic resources are given by $\M_2(M^{(\theta,\phi)}) \le \M_2(M^{\theta}) + \M_2(M^{\phi})$, the arbitrary measurement scheme offers a significant advantage. The standard MQC approach generally requires more magic resources than the amount preserved in the final state $\M_2(|\psi_\pm\rangle)$, leading to unnecessary waste.

This analysis reveals that any MQC computation on a linear graph can be efficiently implemented using just a single auxiliary qubit and an appropriately chosen arbitrary measurement, effectively eliminating magic resource waste. Fig.~\ref{fig:exp_arbitrary_meas} illustrates the impact of arbitrary measurements on magic resource consumption. The grey line, representing the invested magic resources for the standard MQC approach, is noticeably higher than the red line and dots, which depict the invested and reserved magic resources, respectively, for the arbitrary measurement case. The blue line and dots represent a similar comparison for a different set of measurement angles, further demonstrating the advantage of arbitrary measurements in conserving magic resources.

\section{Discussion about MSI and MQC}

It is important to emphasize that a direct, fully quantitative comparison between MQC and MSI is intractable. Implementing an arbitrary unitary via MSI typically requires decomposing the target operation into a Clifford+magic-gate sequence, where the number of consumed magic states depends heavily on the chosen synthesis protocol. This process is indirect and, in many cases, the minimal required number of magic states is unknown or uncertain. By contrast, MQC offers a constructive and direct realization of a given unitary through the preparation of an appropriate graph state and a prescribed sequence of single-qubit measurements. Because of this asymmetry, a one-to-one resource comparison between MQC and MSI is inherently intractable in full generality.

Moreover, there is currently no universal benchmark for MSI that specifies the ancillary qubit overhead, gate depth, and connectivity requirements for arbitrary unitaries. Without such a standardized baseline, it is difficult to align MSI’s resource costs with those of MQC in a rigorous and universally fair manner. Addressing this limitation is an important open problem for future work.

In our discussion, we quantify magic resources using the \textit{T-count}. Other works may quantify magic resources in terms of the state $|H\rangle=\frac{1}{\sqrt{2}}(|0\rangle+e^{i\pi/4}|1\rangle)$, referred to as $1H$. However, the key difference is that the $|T\rangle$ state contains the maximum magic resources in a single qubit, which allows us to identify special cases. For instance, in our experiments, MQC effectively generates the $|CS\rangle$ state, which contains $\mathbb{M}_2(|CS\rangle)=2.03T$ -- greater than its qubit number $n=2$. This suggests that such a state cannot be generated by injecting two $T\rangle$ states at the outset and Clifford gates. A similar case arises with the Hogger state, $|\text{Hogger}\rangle = \big((1+i)|000\rangle-|010\rangle+|011\rangle-i|100\rangle+|101\rangle\big)/\sqrt{6}$, which has $\mathbb{M}_2(|\text{Hogger}\rangle)=3.6T$,  also greater than its qubit number $n=3$. The MQC method can always effectively achieve the state that contains the maximum magic resources (no matter the measures), which is certainly greater than its qubit numbers. 

In the MQC framework, we can adopt a \textit{‘borrow-first-measurement-later’} approach to generate a target state. For example, if the target is an n-qubit state containing $xT$ magic resources $(x>n)$, except for the n qubits for the final state, we start by borrowing $m-n$ qubits and use Clifford operations to generate an appropriate graph state $|G\rangle$. By following the MQC protocol, single-qubit measurements on the $n-m$ qubits remaining qubits will deterministically generate the target state, without any approximation.

In contrast, the MSI framework requires the preparation of magic states at the beginning, followed by Clifford gates to generate either the target state or an approximation within arbitrary precision. From the previous discussion, we know that MSI cannot generate states like $|CS\rangle$ and $|\text{Hogger}\rangle$ without discarding auxiliary qubits. Furthermore, MSI lacks the flexibility to employ the ‘borrow-first-measurement-later’ strategy, requiring the preparation of more magic states than the final number of qubits in the target state.

Thus, the key difference lies in how magic resources are distributed and concentrated in the two paradigms. MSI typically injects standardized magic states uniformly at the outset, without an inherent mechanism for concentrating non-stabilizerness on specific subsystems. By contrast, MQC invests magic progressively through adaptive measurements on an entangled graph state, enabling a more structured allocation of resources. In certain cases --- such as the Quantum Fourier Transform (QFT) example discussed in the main text—the invested magic closely matches the reserved magic, indicating exceptionally efficient usage. This illustrates a distinctive advantage of MQC: it can concentrate the required magic onto a compact subset of qubits, achieving a genuine form of space-saving. Our notion of magic efficiency ---quantifying usable magic per qubit --- captures this resource concentration property.

In summary, a fully comprehensive, case-by-case resource comparison between MQC and MSI remains an open and challenging task. Nonetheless, the framework developed in this work provides a clear and operational perspective on magic efficiency and space utilization, offering a first step toward understanding the true resource trade-offs between different quantum computational models.
 
Therefore, we believe that MQC offers significant advantages over MSI, particularly in terms of space efficiency.

\bibliography{reference}